\documentclass[structabstract]{aa}
\usepackage{graphicx}
\usepackage{float}
\usepackage{txfonts}
\usepackage{natbib}
\usepackage{longtable}
\usepackage{supertabular}

\bibpunct{(}{)}{;}{a}{}{,}

\begin{document}

\newcommand{\neon}{[Ne~II]}
\newcommand{\lne}{L$_{\rm [Ne~II]}$}
   \title{Searching for gas emission lines in \emph{Spitzer} Infrared Spectrograph (IRS) spectra of young stars in Taurus}


   \author{Baldovin-Saavedra, C.
          \inst{1,}
          \inst{2}   
   		  \and Audard, M.
          \inst{1,}
          \inst{2}
          \and
          G\"udel, M.
          \inst{3}
          \and Rebull, L.~M.
          \inst{4}
          \and Padgett, D.~L.
          \inst{4}
          \and Skinner, S.~L.
          \inst{5}
          \and Carmona, A.
          \inst{1,}
          \inst{2}   
          \and Glauser, A.~M.
          \inst{6,}
          \inst{7}
          \and Fajardo-Acosta, S.~B.
          \inst{8}
          }

   \institute{ISDC Data Centre for Astrophysics, Universit\'e de Gen\`eve, 16 Chemin d'Ecogia, CH-1290 Versoix, Switzerland
         \and
             Observatoire Astronomique de l'Universit\'e de Gen\`eve, 51 Chemin de Maillettes, CH-1290 Sauverny, Switzerland
		\and 
			University of Vienna, Department of Astronomy, T\"urkenschanzstrasse 17, A-1180 Vienna, Austria
		 \and
		 	Spitzer Science Center, California Institute of Technology, 220-6 1200 East California Boulevard, Pasadena, CA 91125 USA 
		  \and
		  	Center for Astrophysics and Space Astronomy, University of Colorado, Boulder, CO 80309-0389, USA	
		 \and 
			ETH Z\"urich, 27 Wolfgang-Pauli-Str., CH-8093 Z\"urich, Switzerland					
		  \and
		  	UK Astronomy Technology Centre, Royal Observatory, Blackford Hill, EH3 9HJ Edinburgh, UK
		  \and
		  	IPAC, California Institute of Technology, 770 South Wilson Avenue, Pasadena, CA 91125, USA 	
             }

   \date{Received August 2010; accepted January 2011}

 
  \abstract
   {  Our knowledge of circumstellar disks has traditionally
            been based on studies of dust. However, gas dominates 
            the disk mass and its study is key to our understanding
            of accretion, outflows, and ultimately planet formation.
            The \emph{Spitzer} Space  Telescope provides access to 
            gas emission lines in the  mid-infrared,  providing crucial new 
            diagnostics of the physical conditions in accretion 
            disks and outflows.}
   {  We seek to identify gas emission lines in mid-infrared spectra of
         64 pre-main-sequence stars in Taurus. Using line luminosities and
         other known star-disk-outflow parameters, we aim to identify correlations
         that will help to constrain gas heating, excitation mechanisms, and
         the line formation.}
   {We have based our study on \emph{Spitzer} observations using the Infrared Spectrograph (IRS), 
   mainly with the high-resolution modules.
   Line luminosities (or $3\sigma$ upper limits) have been obtained by fitting Gaussian profiles to the lines. 
   We have further searched for correlations between the line luminosities and different parameters related to the star-disk system.}    
   {We have detected H$_2$ (17.03, 28.22~$\mu$m) emission in 6 objects, [Ne~II] (12.81~$\mu$m) emission in 18 objects, and [Fe~II] (17.93, 25.99~$\mu$m) emission in 7 objects. 
   [Ne~II] detections are found primarily in Class~II objects. 
   The luminosity of the [Ne~II] line (L$_{\rm{Ne II}}$) is in general 
         higher for  objects known to drive jets than for those without 
         known jets, but the two groups are not statistically distinguishable.
         L$_{\rm{Ne II}}$ is correlated with X-ray luminosity, but for Class II
         objects only. L$_{\rm{Ne II}}$ is also correlated with disk mass and
         accretion rate when the sample is divided  into high and low accretors.
         Furthermore, we find  correlations of L$_{\rm{Ne II}}$ with  mid-IR
         continuum luminosity and with luminosity of the [O I] (6300 \AA) line,
         the latter being an outflow tracer. L$_{\rm [Fe II]}$ correlates with
         $\dot{\rm M}_{\rm acc}$. No correlations were found between
         L$_{\rm H_{2}}$ and several tested parameters.}
   { Our study reveals a general trend toward accretion-related phenomena as the origin of the gas emission lines. Shocks in jets and outflowing material are more likely 
   to play a significant role than shocks in infalling material. The role of X-ray irradiation is less prominent but still present for [Ne II], in particular for 
   Class II sources, while the lack of correlation between [Fe II] and [Ne II] points toward different emitting mechanisms.  
   }

	
	 \authorrunning{Baldovin-Saavedra et al.}
	 \titlerunning{Searching for gas lines in the young population of the TMC}

   \maketitle
%
\section{Introduction}
\label{intro}

Protoplanetary disks are formed by a mixture of gas, dust and ices.
The bulk of their mass is in gaseous form (mainly H$_2$) but the opacity is dominated by dust at infrared (IR) and optical wavelengths. 
Ro-vibrational transitions of H$_2$ in the IR are difficult to stimulate due to the combination of low temperatures found in protoplanetary disks and the lack of a permanent electric dipole moment of the H$_2$ molecule.
For this reason, our current knowledge of protoplanetary disks is largely based on the study of dust (e.g., \citealt{beckwith:1990aa,hartmann:2008aa,mann:2009aa}). 

One of the key questions in star formation is the disk lifetime, since it provides a strong constraint on the planet formation timescale.
Many studies have been dedicated to this problem (e.g., \citealt{carpenter:2009aa,meyer:2009aa}).
The signatures of protoplanetary disks, namely thermal emission from dust grains in the millimeter and the infrared, and molecular line emission from CO in the sub-millimeter, typically disappear after a couple of Myr (e.g., \citealt{duvert:2000aa,andrews:2005aa}). 
However, it is not clear whether the material has been completely cleared out or if present observations are not sensitive enough to allow the detection of its remnant. 
Recent evidence indicates that the gas and dust lifetimes are not the same (e.g., \citealt{pascucci:2009ab,fedele:2010aa}).
Detections of quiescent emission of molecular H$_2$ in weak-lined T Tauri stars (WTTS) that do not show infrared excess, i.~e., those that are not expected to have a disk, support the hypothesis that gas lives longer than dust in disks (e.g., \citealt{bary:2002aa,bary:2008ab}). 
To address the question of disk dissipation timescales, studies of the gas component of protoplanetary disks independent of dust are particularly needed.

To study gas in protoplanetary disks, molecular spectroscopy, in particular emission lines, can be used. 
As the temperature and density of the gas vary as function of the radius in the disk and the mid-plane distance, 
spectral lines at different wavelengths probe different disk regions, and gas at different physical conditions.
In the sub-millimeter and millimeter regimes, observations probe cold gas (T $\sim 20-50$~K) at large radii ($>$ 20 AU) through emission lines such us CO species and HCO$^+$ (e.g., \citealt{mannings:1997aa,pietu:2007aa}). 

In the near-IR, the H$_2$ emission arises from ro-vibrational transitions, probing gas at T$\sim1000$~K (e.g., \citealt{weintraub:2000aa,carmona:2007aa,bary:2008aa}).
After H$_2$, CO is the most abundant species found in disks. 
The different excitation conditions of this molecule probe a wide range of disk radii;
the first overtone band at 2.3~$\mu$m traces hot ($\sim4000$ K) and dense material (n $>10^{10}$~cm$^{-3}$) of the innermost part of protoplanetary disks. 
 Observations show that the emission has the characteristic shape suggestive of a rotating disk \citep{carr:1989aa}.
The CO ro-vibrational band at 4.7 $\mu$m is estimated to arise within 1-10 AU from the star (e.g., \citealt{najita:2003aa,blake:2004aa,rettig:2004aa,salyk:2007aa,carmona:2008ab}) with characteristic temperatures between 100 to a few 1000~K.

The mid-IR traces the warm gas at radii of a few AU up to several tens of AU (the giant planet formation region).
At these wavelengths, H$_2$ emission is observed, arising from pure rotational transitions at $12.28$, $17.03$, and $28.22$~$\mu$m (e.g.,~\citealt{van-den-ancker:1999aa,lahuis:2007aa,martin-zaidi:2007aa,bitner:2008aa}). 
Furthermore, many detections of forbidden atomic lines have been reported, such as [Ne~II]  (12.81 $\mu$m), [Fe~I] (24.0 $\mu$m), [Fe II] (17.93 and 25.99 $\mu$m), and [S~III] (18.71 $\mu$m), opening up a new window into the diagnosis of gas in protoplanetary disks. 

When studying the gas emission from protoplanetary disks, it is important to discriminate among emission produced by (i) gas in jets or outflows, (ii) accreting gas falling onto the star, and (iii) gas in the disk itself.
High spectral and spatial resolution observations enable the study of the line profiles and possible velocity shifts with respect to the stellar rest velocity.
However, the heating mechanism of gas and the ionization of atomic species is not well understood.
\citet{glassgold:2007aa} predicted that stellar X-rays heat the upper layers of protoplanetary disks within 25~AU from the star, exciting atomic species.  
\citet{gorti:2008aa} considered both stellar X-rays and ultraviolet (UV) irradiation, proposing several emission lines as possible diagnostics of the gas in disks irradiated by high energy photons: 
forbidden lines of [Ne~II] (12.81~$\mu$m), [Ne~III] (15.55~$\mu$m), [Fe~II] (25.99~$\mu$m), [O~I] (63~$\mu$m), and pure rotational lines of H$_2$.
Moreover, \citet{alexander:2008aa} successfully reproduced the observed [Ne~II] line luminosities by modeling the line profile using a photoevaporative wind model,
which considered the UV flux from the central star as the only ionizing source. 
In addition, far-uv (FUV), extreme-UV (EUV), and X-ray photons may be produced by accretion shocks as the material falls into the central star \citep{hollenbach:2009aa}.
Forbidden emission lines in the optical and near-IR have been used for a long time to probe the shocked gas from T~Tauri stars (e.g.,~\citealt{mundt:1990aa,hirth:1997aa,hartigan:2009aa,agra-amboage:2009aa}). 
\citet{hollenbach:1989aa} studied the infrared emission from interstellar shocks and proposed that for low pre-shock densities (n~$\leq 10^{4-5}$ cm$^{-3}$), forbidden  emission lines of [Fe~II], [Ne~II], [Si~II], and [O~I] are particularly strong.

Forbidden emission lines are used to trace the electron densities of the environment where they originate; typical densities of $10^3 - 10^7$~cm$^{-3}$ are probed.
At these densities, the origin is in low-density winds or shocks rather than in the infall of material through accretion funnels. On the other hand, permitted emission lines exhibiting large 
widths and inverse P-Cygni features are associated with infalling material \citep{muzerolle:2001aa}.  

Up to now, observations have not been conclusive in giving a broadly accepted explanation of the heating and ionizing mechanism exciting the gas component in protoplanetary disks. 
We aim to contribute to the understanding of this problem by studying a large sample of pre-main sequence stars of low and intermediate mass.
We have used observations done by the \emph{Spitzer} Space Telescope \citep{werner:2004aa} Infrared Spectrograph (IRS, \citealt{houck:2004aa}) in order to study the emission from gas lines. 
Although higher spectral resolution observations are needed in order to obtain kinematical information for the emitting gas,
 the high sensitivity of \emph{Spitzer} does enable us to access sources not detectable from the ground. 
Furthermore, the high number of sources observed 
allows the extraction of statistically significant information from a large sample. 
We have made use of archival spectra of 61 pre-main sequence stars plus 3 objects from our own \emph{Spitzer} program. 
This study is complementary to previous investigations; our work includes objects of different evolutionary stages, while the somewhat related previous study by \citet{gudel:2010aa}~concentrated on a pure Class~II sample. 
Additionally, in our analysis we have made a distinction based on the YSO class, in contrast to previous studies (e.g., \citealt{lahuis:2007aa,flaccomio:2009aa}).

Sections~\ref{sample} and \ref{datana} are dedicated to the description of the sample and of the data analysis, respectively.
Results are presented in Section~\ref{results} and  discussed in Section~\ref{discussion} together with the presentation of correlation tests. A brief summary of the results and conclusions are given in Section~\ref{conclusions}.
In Appendix~\ref{ind_objects} (available in the electronic version) we have included additional information on individual objects whenever one of the studied gas lines was detected.

\section{Sample}
\label{sample}

Our study is based on objects from the Taurus Molecular Cloud (TMC).
This star-forming region is known to form mainly low-mass stars, allowing the study of the star formation process without 
the influence of high-mass young stars, which are known to have strong UV radiation fields.
Its nearby distance (137 pc, \citealt{torres:2007aa}) enables us to access fainter sources.
All the objects have similar ages, lie at nearly the same distance, and are presumably formed out of the same parent cloud.

A further motivation to focus our study on the TMC region is the large amount of multi-wavelength observations available. 
The TMC was indeed been observed extensively in X-rays with \emph{XMM-Newton}: the XEST survey covered $\sim$5 square degrees of the TMC, \citealt{gudel:2007aa}. In the optical, the Canada-France-Hawaii Telescope (CFHT) 
covered 28 square degrees of the TMC (Monin et al.\ 2010, in preparation), and the Sloan Digital Sky Survey (SDSS) covered 48 square degrees \citep{padmanabhan:2008aa}. 
In the IR, \emph{Spitzer} mapped $\sim$43 square degrees (\emph{Spitzer} Taurus Legacy Project; \citealt{padgett:2007aa}, 2011, in preparation). 
Multiwavelength observations are crucial in order to test the influence of the stellar X-rays in the heating and excitation of the gas species.

Our study is based on publicly available archived data obtained by IRS with its high resolution modules (short-high -- SH -- and long-high -- LH), with a resolving power of R$\sim600$ ($\sim 500$~km s$^{-1}$) for wavelength coverage between $9$ and $37$~$\mu$m.  
Most of the targets studied belong to GTO (Guaranteed Time Observations) programs; several publications exist that focussed on the study of the silicate and ice features \citep[e.g.,][]{watson:2004aa,furlan:2006aa,furlan:2008aa}.
In particular, we used data from the GTO program number 2, from the Legacy programs 171, 179, and from the GO (General Observer) programs 3470, 3716, and 30300.
In addition, we included 3 low-mass pre-main sequence stars identified by the \emph{Spitzer} TMC survey for which we obtained IRS high and low-resolution spectra (program 50807, PI. Baldovin-Saavedra). 
For one object (SST~041412.2+280837) we only have the spectrum taken with the low-resolution modules (long-low -- LL -- and short-low -- SL). 

Our sample consists of 64 low and intermediate mass pre-main sequence stars.
Their infrared classification is taken from the literature and is based on the slope of the spectral energy distribution (SED) in the infrared.  
The sample includes 10 Class~I objects, 40 Class~II, 4 Class~III, and 6 intermediate mass Herbig AeBe stars.
We found 4 objects classified in the literature as both Class I and Class II. 
Table \ref{prop} (available in the electronic version) summarizes the properties of the stars in our sample, including the bibliographic references for the adopted values.  

\section{Data Analysis}
\label{datana}

\subsection{Initial Steps}
\label{initial_steps}

The data were retrieved from the \emph{Spitzer} archive; we started the analysis from the post-basic calibrated data, i.e., spectra extracted automatically by the pipeline (version S15.3.0 for SH and S17.2 for LH 
\footnote{We checked for a few sources that the latest version of the pipeline (S18.7.0) did not change the obtained line luminosities.}).
The spectra of the three targets from our own program were treated using the pipeline version S18.7.0.\footnote{Our observations were reduced from the basic calibrated data, correcting for the co-adding bug in S18.7.0 and for time and 
position-dependent dark currents on the LH array (DarkSettle 1.1).}

In a first step we used IRSCLEAN\footnote{Available from http://ssc.spitzer.caltech.edu} to create masks for the rogue pixels from the basic calibrated data images (rogue pixels have abnormally high dark current and/or photon responsivity). 
Once the images were free from obvious rogue pixels, we re-extracted the spectra using SPICE (Spitzer IRS Custom Extraction) regular extraction. Furthermore, in the cases where we detected emission lines or where the detections were not obvious, 
we also used optimal extraction (details on the regular and optimal extraction can be found in the SPICE User's Guide Version, 16.2.1+).

Some of the spectra also presented residual fringes; these interference patterns are typical for infrared detectors as their surface coating acts as a Fabry-P\'erot etalon, leading to a sinusoidal modification of the detector response of 
spectrally dispersed light. Although they were corrected during the pipeline processing, residual fringes remained in a few cases that we removed using the tool IRSFRINGE$^3$.

\subsection{Background}
\label{subsect:background}

The mid-IR background comes predominantly from zodiacal light at the IRS wavelengths, while diffuse emission from the Taurus cloud
contributes only little. Instrumental background (rogue pixels) is also significant for the high-resolution modules. Therefore,
sky observations are needed in order to correct for the sky and instrumental background. 
In this study, however, most of our source sample had no dedicated background observations. Good rogue pixel removal proved to be crucial
for our emission line study.

Some of the GTO observations were performed using a series of 6 pointings per object with a shift corresponding to approximately half of the PSF.
In these cases, for the on-source observations we used the 2 pointings with higher flux and averaged them; these pointings are equivalent to the standard staring mode observations, i.e., 2 slit  positions with the target 
located at 30 and 60\% of the slit length.
For each object, we checked using \emph{Spitzer} Infrared Array Camera (IRAC) 8~$\mu$m images that the pointings correspond to the SIMBAD position of the target, and that the flux levels of the 2 used pointings were equivalent. 

One central issue lies in understanding the origin of the emission. 
It could be that the lines detected are not really emitted by the star-disk system, but rather come from the background or a nearby source.
Our data are in the form of an echellogram, which is a two-dimensional representation of a spectrum containing both spatial and spectral information.  
As a first step, we searched each spectrum and searched for evident line emission; when found, we examined in the echellogram two spectral regions with no emission lines adjacent to the position of the line studied.  
These two regions were used as a measure of local background plus stellar continuum; they were averaged and then subtracted from the line region. 
In this way, any possible extended emission can be identified; moreover, by removing the continuum, the detection becomes more evident.  
This method did not result in the discarding of any detections. However, it is not sensitive to potential extended emission from the direct vicinity of the star.

\subsection{Line Fitting}
\label{fitting}

In order to obtain line luminosities, we used a function consisting of a Gaussian plus a linear component to account for the continuum within a range of typically $\pm 0.6~\mu$m of the line.
All the parameters of the function (height, center, and width) were left free to vary. 
The integrated flux was calculated as the area under the Gaussian fit once the continuum was subtracted,
and the line luminosity was calculated using a distance to the TMC of 137 pc from \citet{torres:2007aa}\footnote{The distance to the TMC has often been considered to be 140~pc. In comparison to the 137~pc used in this study, this can result 
in a difference of the order of 4\% in the line luminosities.}.
A detection was claimed when the peak of the line was higher than 3 times the level of the noise in the spectra.
The noise level is the standard deviation of the continuum flux ($\sigma$) in the wavelength range of interest.

Most of the detected lines were spectrally unresolved; the full width at half-max (FWHM) obtained from the fit was similar to the instrumental FWHM ($\sim 500$~km~s$^{-1}$).
In the cases where the lines were not been detected, we obtained 3$\sigma$ upper limits for their luminosities by multiplying the 3$\sigma$ noise value of the continuum by the FWHM.
  
\section{Results}
\label{results}

Our study is focused on the search for pure rotational emission lines from H$_2$ (12.28, 17.03, 28.22 $\mu$m), and forbidden emission lines from [Ne II] (12.81 $\mu$m), [Ne III] (15.56 $\mu$m), [Fe II] (17.94, 25.99 $\mu$m), 
and [S III] (18.71, 33.48 $\mu$m). We detected [Ne~II] in 18 objects in our sample, [Fe~II] in 7 objects, and H$_2$ in 6 objects. 
We did  not detect [Ne~III] or [S~III].
The line luminosities and upper limits are presented in Table \ref{linelum}, with the luminosities expressed in units of $10^{28}$~erg~s$^{-1}$; bold fonts indicate detected lines, and normal fonts are 3$\sigma$ upper limits. 
In Appendix~\ref{ind_objects} (available in the electronic version) we provide information on the single objects with detected emission lines.  

\begin{figure*}[tp!]
   \centering
   		\includegraphics[width=12.0cm]{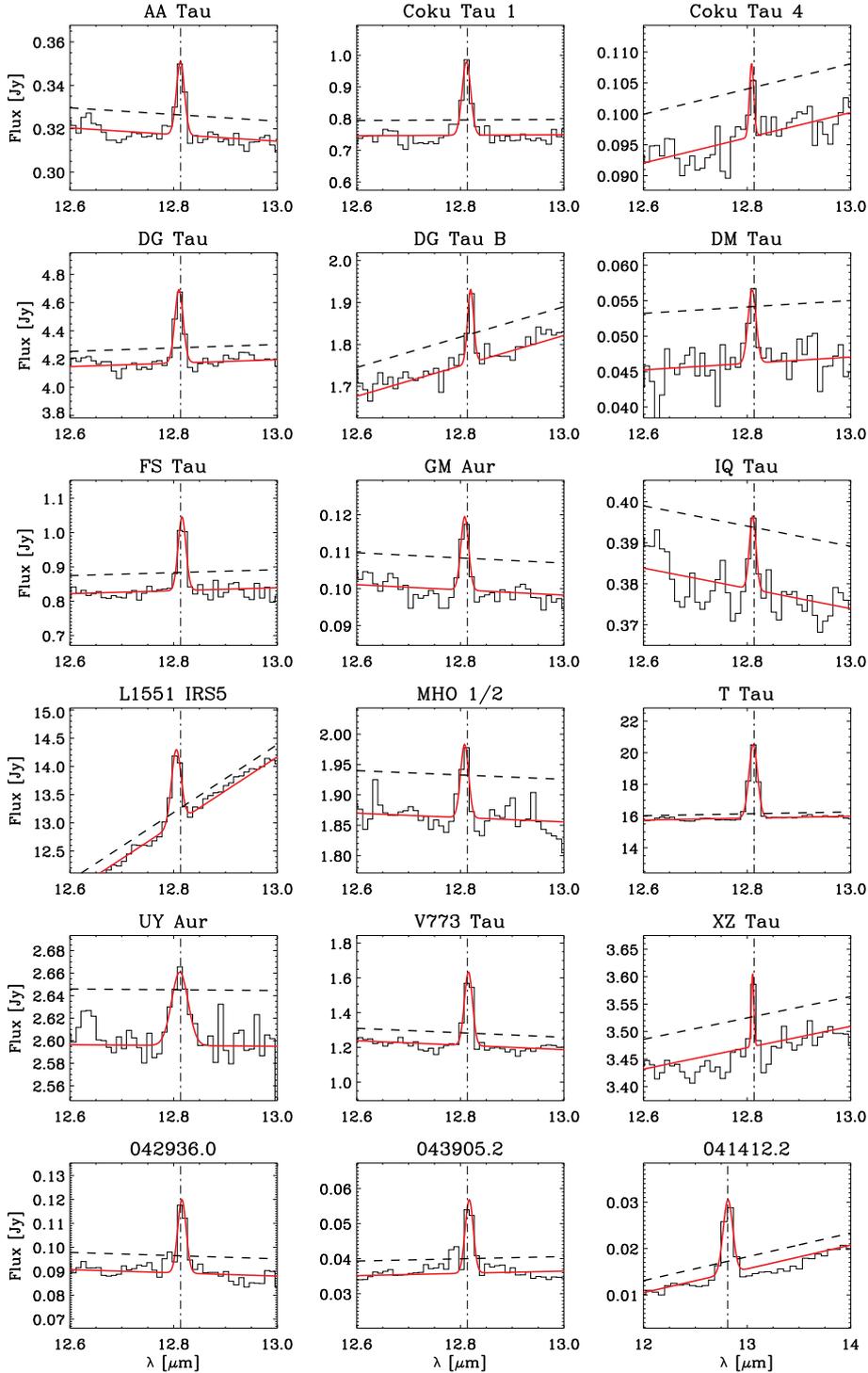}
    \caption{Detections of [Ne~II] at 12.81 $\mu$m. The spectra are plotted as solid histogram, the lines were fitted using a gaussian profile plus a polynomial component for the continuum, the fit is plotted using a solid line. The 3$\sigma$ detection threshold is plotted as a dashed line. The name of each star is labeled in the top of each plot. We recall that the spectrum of SST~041412.2+280837 was obtained using the low resolution modules (SL and LL).}      	 
     \label{NeIIdetections}
\end{figure*}

\begin{figure*}[t!]
   \centering
   		\includegraphics[width=12.0cm]{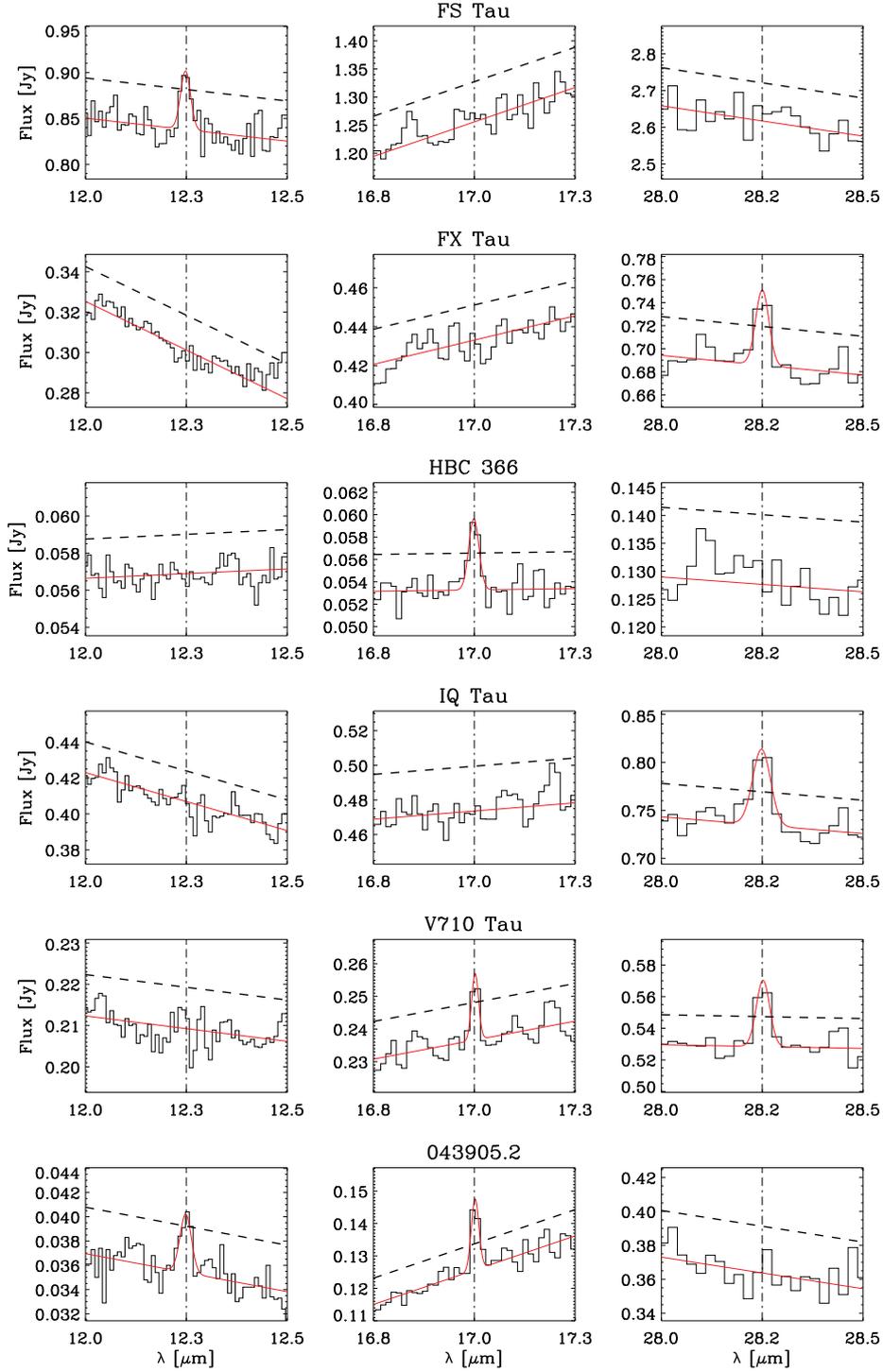}
    \caption{The plotted spectra show the regions around the H$_2$ lines at 12.28, 17.03 and 28.22~$\mu$m for the sources where we have detected at least one of the lines.}    	 
     \label{H2detplot}
\end{figure*}

\begin{figure*}[h]
	\centering
		\begin{minipage}[!b]{\textwidth}	
     		$\begin{array}{cc}
	    			\includegraphics[height=3.cm]{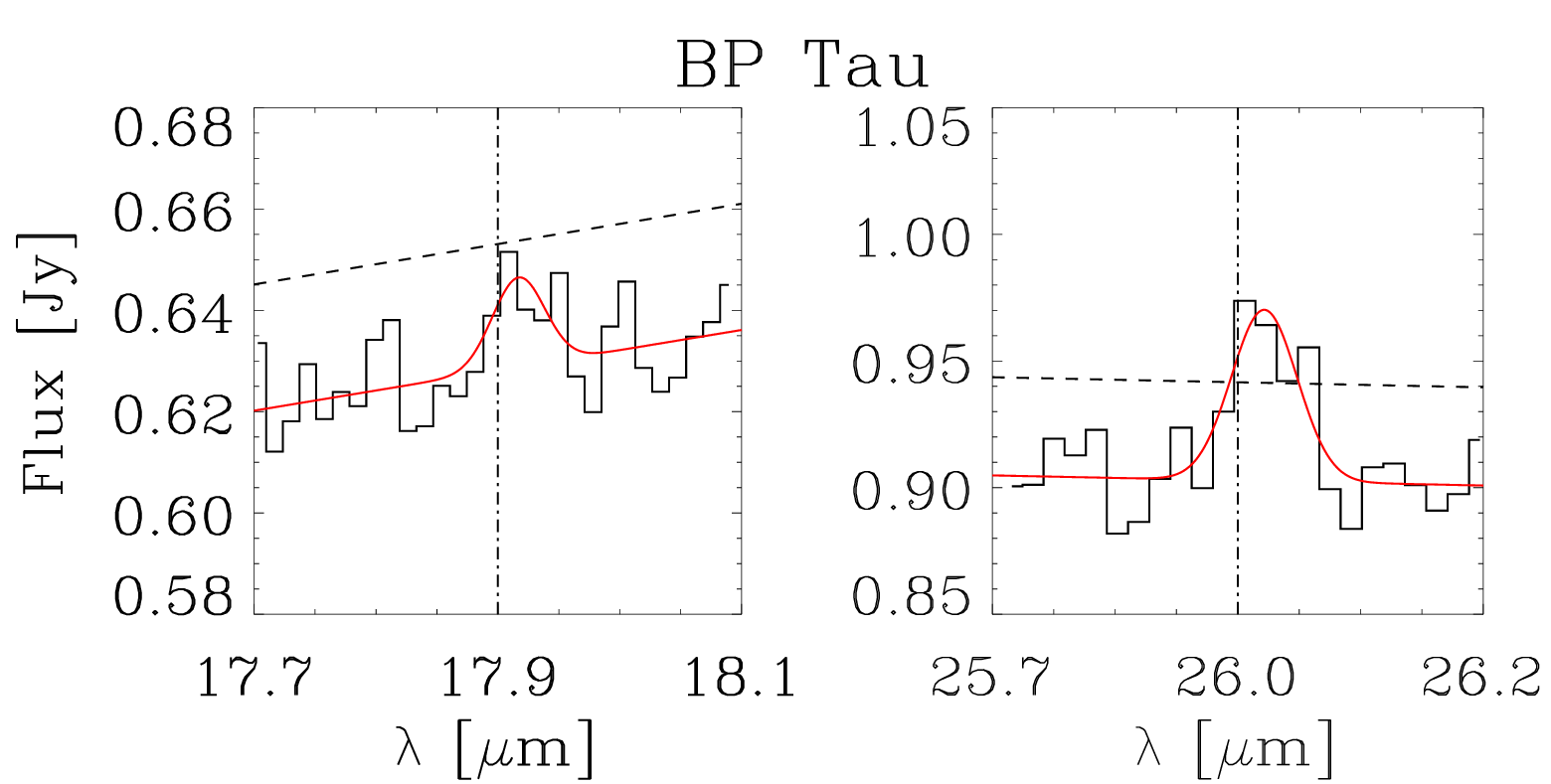} &
	    			\includegraphics[height=3.cm]{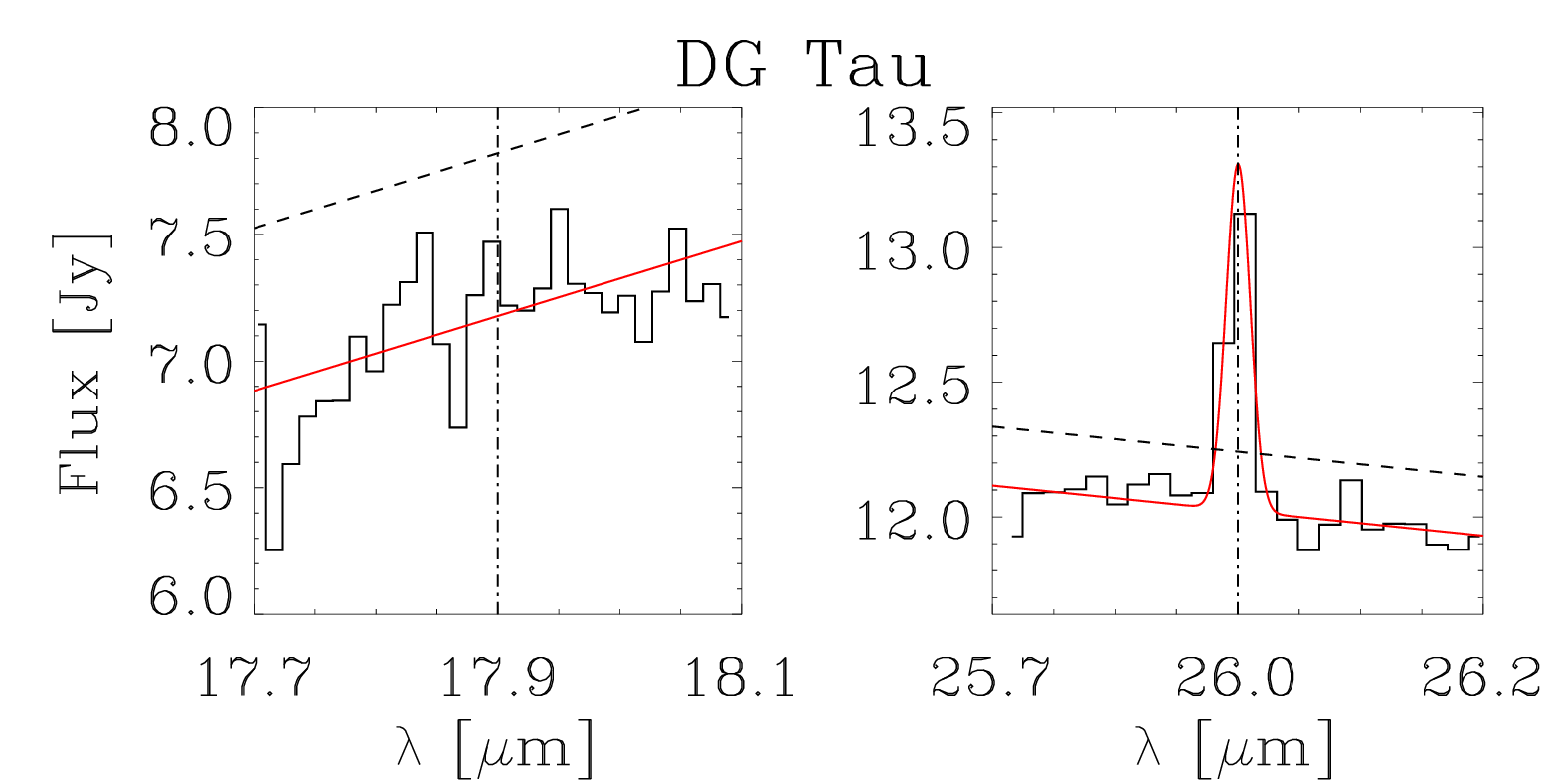} \\
	    			\includegraphics[height=3.cm]{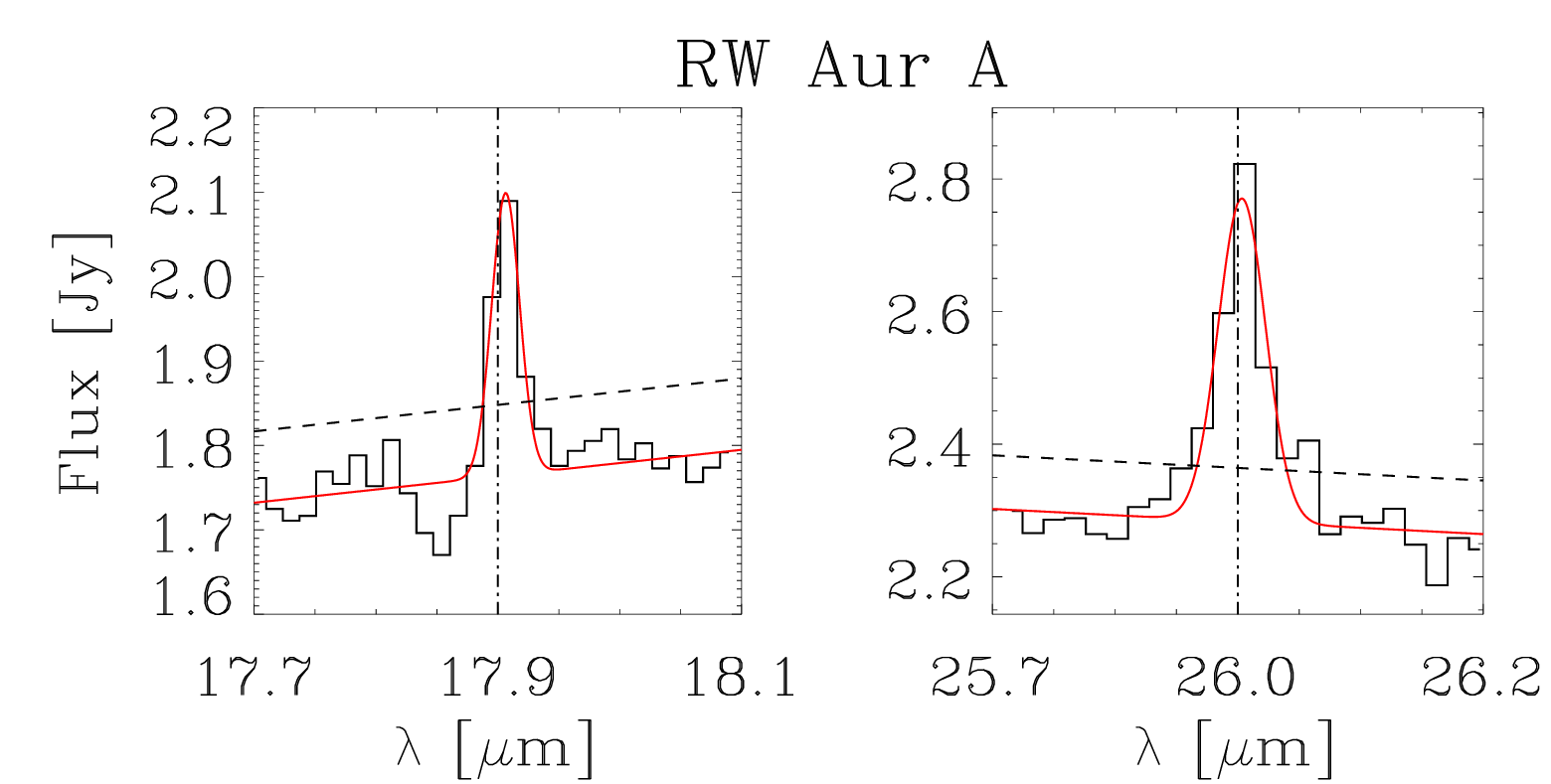} &
	    			\includegraphics[height=3.cm]{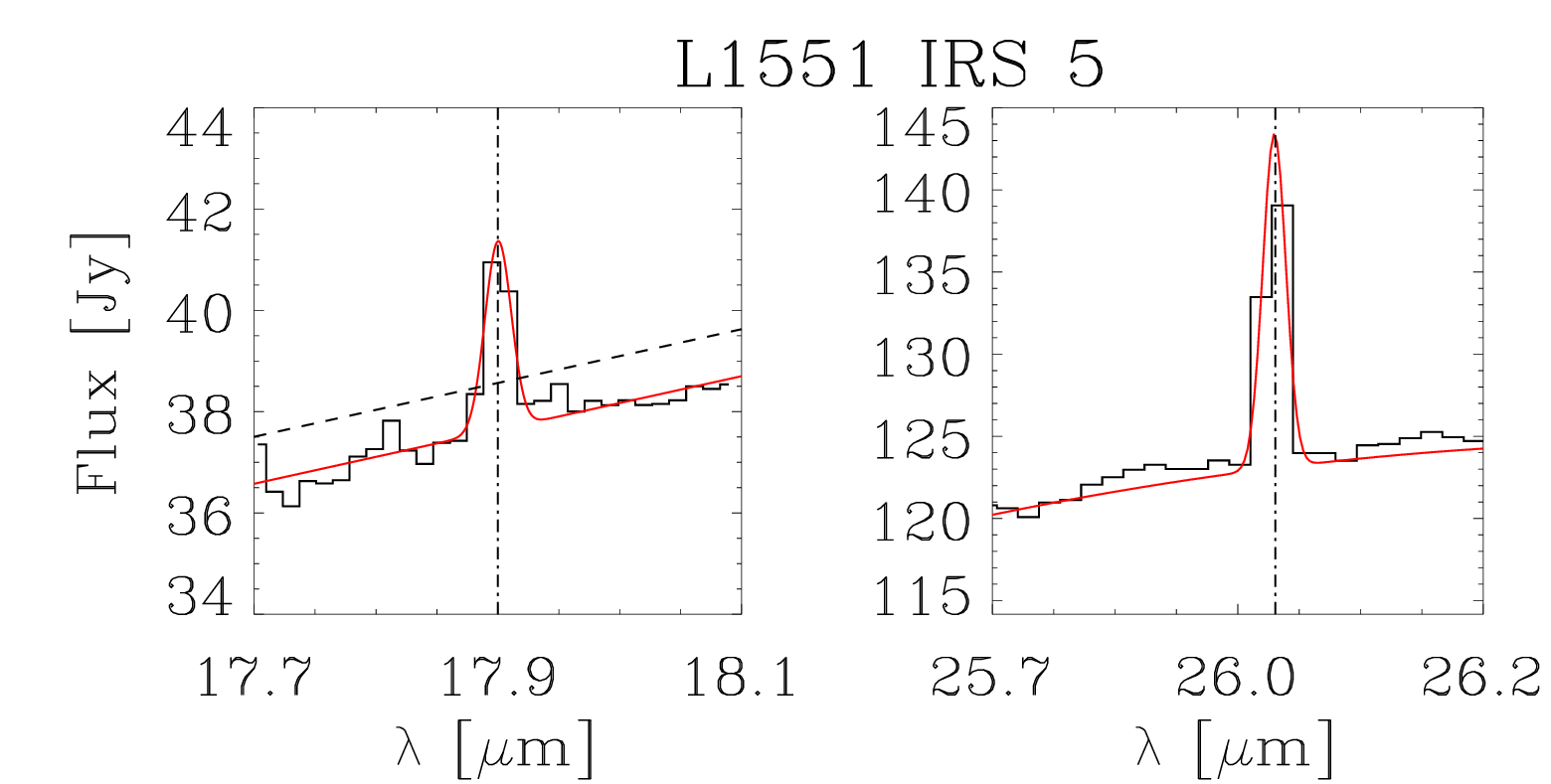} \\
				\includegraphics[height=3.cm]{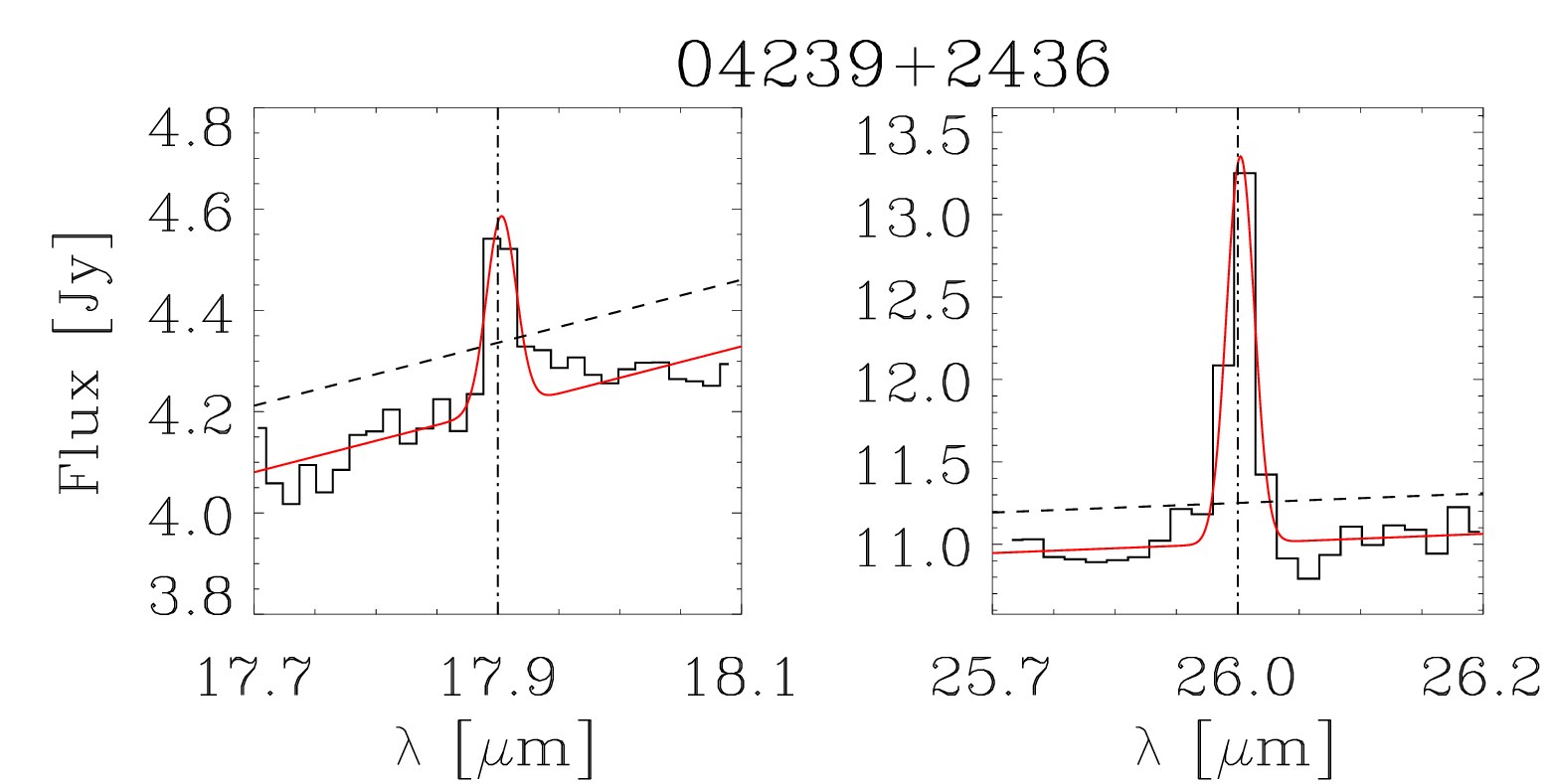} &
				\includegraphics[height=3.cm]{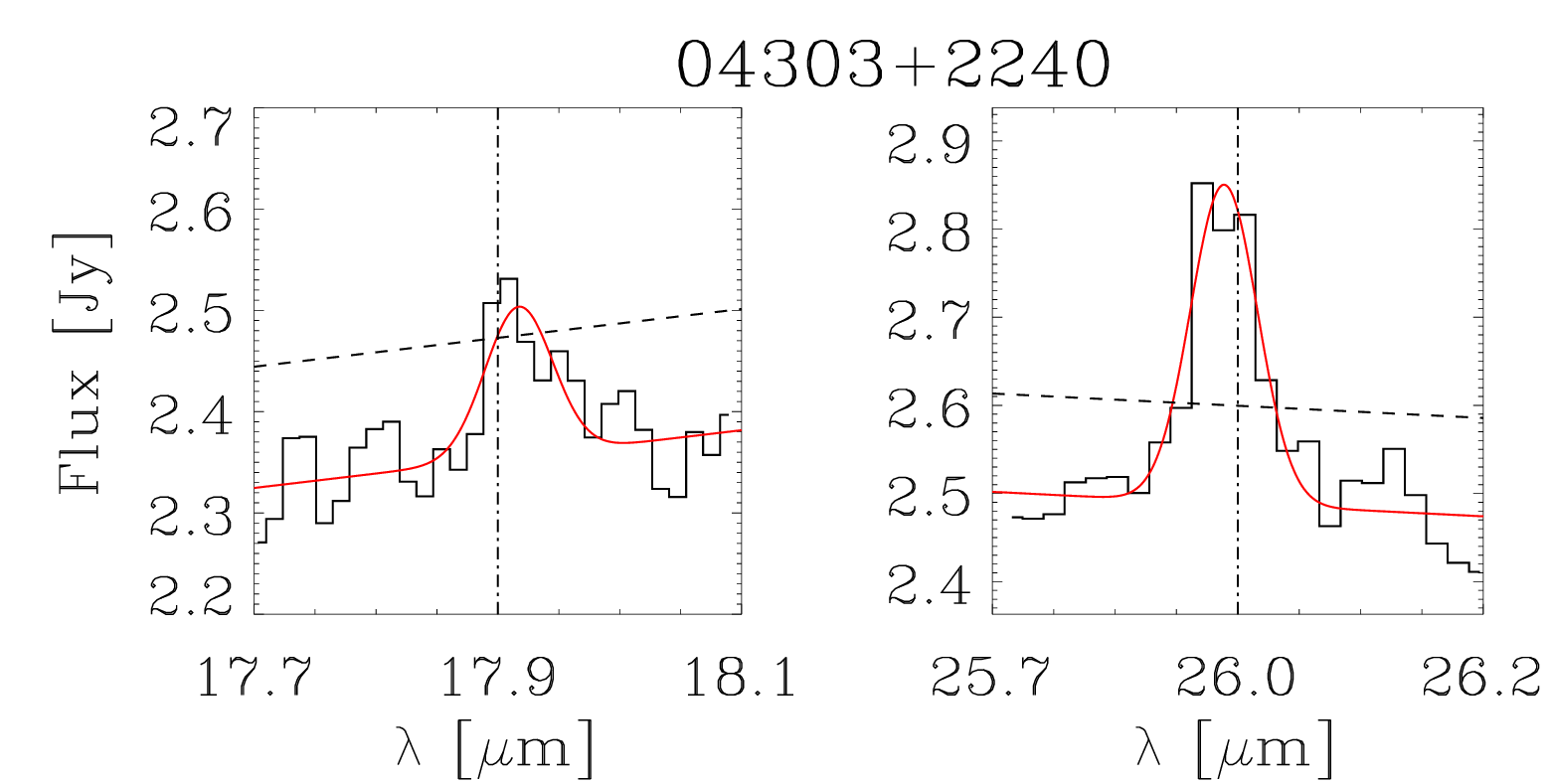} \\
				\includegraphics[height=3.cm]{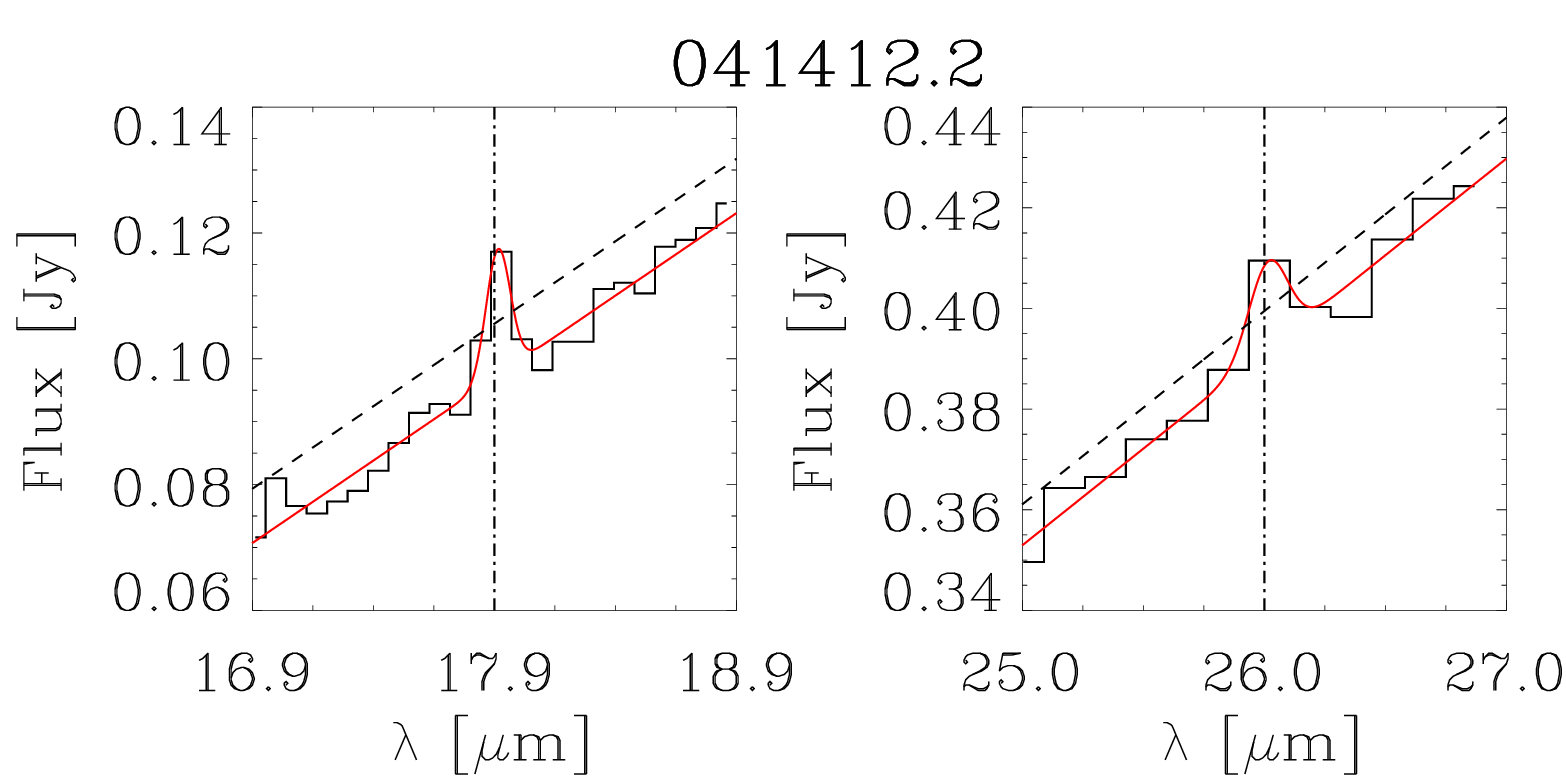}			
	  		\end{array}$  
		\end{minipage}
    \caption{Sources with [Fe~II] detections at 17.94 and 25.99 $\mu$m. We recall that the spectrum of SST~041412.2+280837 was obtained using the low resolution modules (SL and LL).} 
     \label{fig:FeIIdet}
      \end{figure*}

In Fig.~\ref{NeIIdetections} we present the spectra of the targets with detected [Ne~II] line emission, whereas we present the spectra surrounding the position of the [Ne~II] line for the sources with non-detections in Appendix~\ref{neon_non_det} (available in the electronic version). 
Figures~\ref{H2detplot} and \ref{fig:FeIIdet} show the line spectra of the sources with detected emission lines from H$_2$ and [Fe~II]. 
In all figures, the red line shows the Gaussian fit to the line, the dashed line shows 3$\sigma$ detection threshold, and the vertical dashed-dotted line shows the expected position of the [Ne~II] line. 
The title of each panel is the name of the target.


\section{Discussion}
\label{discussion}

\subsection{[Ne~II] emission}
\label{ne23}

The search for [Ne~II] emission has gained importance in the last few years thanks to an increasing number of reported detections in the spectra of young stars observed mainly with \emph{Spitzer} 
(e.g., \citealt{pascucci:2007aa,lahuis:2007aa,espaillat:2007ab,flaccomio:2009aa,gudel:2010aa}). The sample presented in \citet{pascucci:2007aa} included six targets showing evidence of optically thick disks. 
The [Ne~II] line was detected in four of them. Moreover, \citet{pascucci:2007aa}  reported a correlation between the luminosity of the [Ne~II] line and the stellar X-ray luminosity.
\citet{lahuis:2007aa}  detected [Ne~II] in 15 out of 76 low-mass pre-main sequence stars from different star forming regions, and  also reported one detection of [Ne~III]. 
Those authors concluded that the [Ne~II] detections and the [Ne~II]/[Ne~III] line flux ratio are consistent with excitation through X-ray heating of the circumstellar disk from the central star, although excitation through EUV might contribute as well.
In addition, \citet{flaccomio:2009aa} obtained \emph{Spitzer}~IRS spectra for 28 young stellar objects in the Ophiuchus region, obtaining 10 detections of [Ne~II] and one detection of [Ne~III]. 
They did not find a correlation between the luminosity of the [Ne~II] line and the stellar X-ray luminosity. Instead, the luminosity of the [Ne~II] line seemed to be correlated with disk mass accretion rate, consistent with \citet{espaillat:2007ab}. 
\citet{gudel:2010aa} based their study on 92 sources observed with \emph{Spitzer} IRS, obtaining detections of [Ne~II] in 58 of them.
A weak correlation was found between the luminosity of the [Ne~II] line and the stellar X-ray luminosity, whereas
a good correlation was also found between L$_{\rm [Ne II]}$ and jet/outflow parameters, such as the luminosity of the [O~I] line.

We have detected the [Ne II] line in 18 objects, representing 28\% of our sample. 
This detection rate is similar to the results of \citet{flaccomio:2009aa}. 
That study was based only in the Ophiuchus region, obtaining a detection rate of $36 \%$.
\citet{lahuis:2007aa} and \citet{gudel:2010aa} used a mixed sample with stars from different star-forming regions, obtaining $\sim 20$ and $\sim 60$~\% detections respectively. 
The sample from \citet{gudel:2010aa} did not contain any Class~I or Herbig AeBe stars. 
Furthermore, their data reduction differed from ours, as they used the method introduced in \citet{lahuis:2010aa} in which the compact and extended component of the line emission are separated during the spectral extraction procedure.
 
Among the objects in our sample with [Ne II] detections, the distribution in terms of evolutionary state is the following: 11\% of the [Ne~II] detections correspond to Class~I objects, 67\% of the detections correspond 
to pure Class~II, and 22\% of the detections correspond to objects classified  as Class I and II in the literature.

The highest luminosity of the [Ne~II] line in our study came from T~Tau, which is also one of the most luminous young star in the X-rays (L$_{\rm X}= 8 \times 10 ^{30}$~erg~s$^{-1}$). 
This system is known to drive an outflow responsible for the [Ne~II] emission \citep{van-boekel:2009aa}.  L1551~IRS~5 has also a [Ne~II] luminosity comparable to that of T~Tau, of order $10^{30}$~erg~s$^{-1}$.
Previous studies in the optical and near-IR have shown that L1551~IRS~5 drives a strong molecular outflow (see Appendix \ref{ind_objects}), so there is a suggestion in this case that the [Ne~II] could be produced in an outflow/jet environment.
We note that its X-ray luminosity is not well constrained, given that the source is very absorbed \citep{bally:2003aa}; the value available is associated with the jet. 
For this reason, in the correlation tests presented in Section~\ref{Corr:LX},  we do not consider the X-ray luminosity of L1551~IRS~5.

Other sources with [Ne~II] detections known to drive outflows are AA~Tau, DG~Tau, DG~Tau~B, UY~Aur, and CoKu~Tau~1.
In general, these sources show high [Ne~II] luminosities ($> 10^{29}$~erg~s$^{-1}$) except for AA~Tau (L$_{\rm [Ne~II]} \sim3 \times 10^{28}$ erg s$^{-1}$). 
We might conclude that for high [Ne~II] luminosities, the heating of gas and excitation of neon is caused by the outflow activity.
This trend is actually observed in the study of \citet{gudel:2010aa}; sources known to drive jets or outflows show an excess of [Ne~II] emission when compared with sources that do not present jets. 

To check the suggested difference between sources driving jets and sources without jets, we have divided our sample into sources showing jet emission (typically forbidden line emission in the optical and near-IR) and sources without known jets. 
Using the Kaplan-Meier (K-M) estimator available within the ASURV\footnote{Available from http://astrostatistics.psu.edu/statcodes/asurv} package (\citealt{feigelson:1985aa,isobe:1986aa}), we have extracted the cumulative distribution of [Ne~II] 
luminosities for these two groups. The K-M estimator is the basis of survival analysis. It is a generalized maximum-likelihood estimator for the population from which the sample is drawn. Although it was first used in right censoring problems (i.e., lower limits), it has been adapted for application to left censoring problems (i.e., upper limits). The basic concept underlying the K-M estimator is the survivor function, which is defined as the probability of a value being larger than 
or smaller than the upper or lower limit, depending on the type of censoring. The survivor function is a non increasing step function which only jumps at uncensored values. When the sample is large, it approaches the true survivor function of the population. For details we refer to \citet{feigelson:1985aa}.

The sub-sample of sources without jets has 38 objects; among them, 9 have [Ne~II] detections. 
The mean of the distribution is $\rm{L_{[Ne~II]}}=(0.8 ^{+0.5} _{-0.3}) \times 10^{28}$ erg~s$^{-1}$. 
The sub-sample of sources with jets has 26 objects; among them we have 9 detections of the [Ne~II] line. 
The cumulative distribution of [Ne~II] luminosities is extended towards higher values, with a mean of $\rm{L_{[Ne~II]}}=(1.6 ^{+0.9} _{-0.6})\times 10 ^{28}$~erg~s$^{-1}$, almost twice the mean value of the sub-sample of sources with no jets. 
However, both distributions overlap. Moreover, the median of the two sub-samples are very close, with $\rm{L_{[Ne~II]}}=4\times 10^{28}$~erg~s$^{-1}$ and $\rm{L_{[Ne~II]}}=5\times 10^{28}$~erg ~s$^{-1}$ for jet and non-jet sources, respectively.
Thus, we cannot distinguish between the two populations, in contrast with \citet{gudel:2010aa}.
This difference might be due to the sample selection criteria. In fact, \citet{gudel:2010aa} chose a sample composed only of Class~II objects.
Furthermore, their sub-sample of jet-driving sources, unlike ours, did not contain objects with indirect evidence of jets (i.e., high-velocity emission line components).

The number of detections of [Ne~II] in our sample (18 objects) has prompted us to search for possible correlations or trends between the luminosity of the line and parameters related to the star-disk system, and outflow/jet indicators. 
The parameters studied are: stellar mass (M$_{*}$), stellar luminosity (L$_{\rm star}$), X-ray luminosity (L$_{\rm X}$), H$\alpha$ equivalent width, mass accretion rate ($\dot{\rm M}_{\rm acc}$), wind mass loss rate ($\dot{\rm M}_{\rm wind}$), luminosity of the [O~I] line in the optical (L$_{\rm [OI]}$), and luminosity of the [Fe~II] line at 26~$\mu$m (L$_{\rm [Fe II]}$). If correlations exist, they would give new hints on the neon excitation mechanism. 

Due to the high number of upper limits for the luminosity of the [Ne~II] line in our sample, we have tested the correlations using survival analysis implemented in the ASURV package.
We have used 3 statistical tests: Cox's proportional hazard model, generalized Kendall's $\tau $ rank correlation, and Spearman's $\rho$. 
The Kendall and Spearman methods allow the use of upper limits in one or both of the variables tested. Therefore, we have included upper limits for all the parameters tested.
In contrast, the Cox hazard model allows only upper limits in the dependent variable. 
The Spearman method is not reliable for data sets smaller than 30 data points. Nevertheless, we report for completeness the probabilities obtained with this method even when the number of data points is smaller than 30. 
Calculations were done using the logarithm of values. Table~\ref{table:3} summarizes our results, in which we have included the values for the correlation parameters between parentheses for completeness. 
In the case of the Cox hazard method, the parameter reported is the $\chi^2$ of the regression fit. For the Kendall method, the value reported is the ratio between the correlation parameter $\tau$ and its standard deviation $\sigma_{\tau}$, 
a function of the correlation. For the Spearman's $\rho$ method, the parameter reported is the estimated correlation. 

\begin{figure*}[!h]
   \centering
   	\begin{minipage}[!b]{1.\textwidth}
		\centering
			$\begin{array}{ccc}
		    \includegraphics[height=6.2cm]{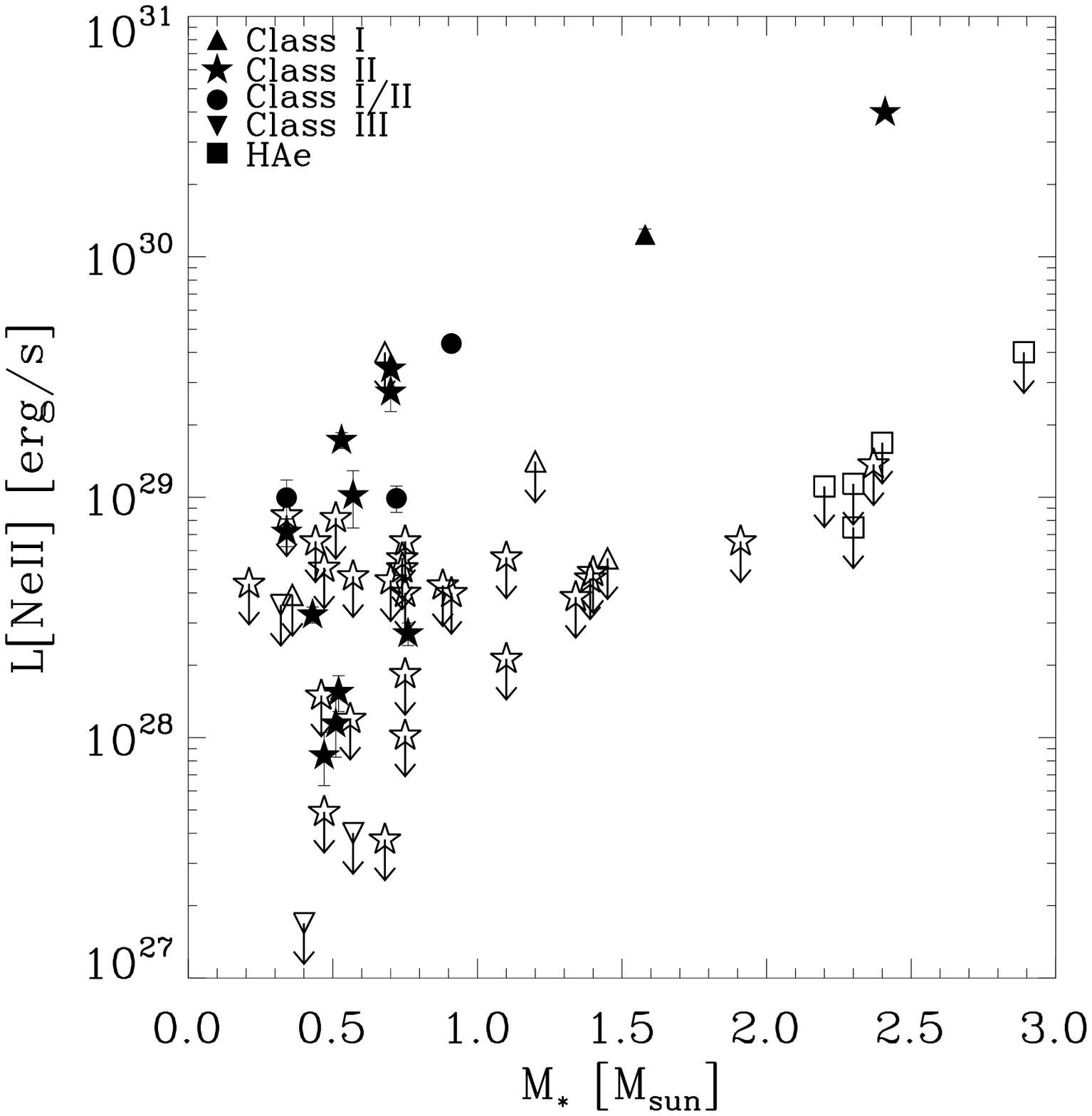} &
			\includegraphics[height=6.2cm]{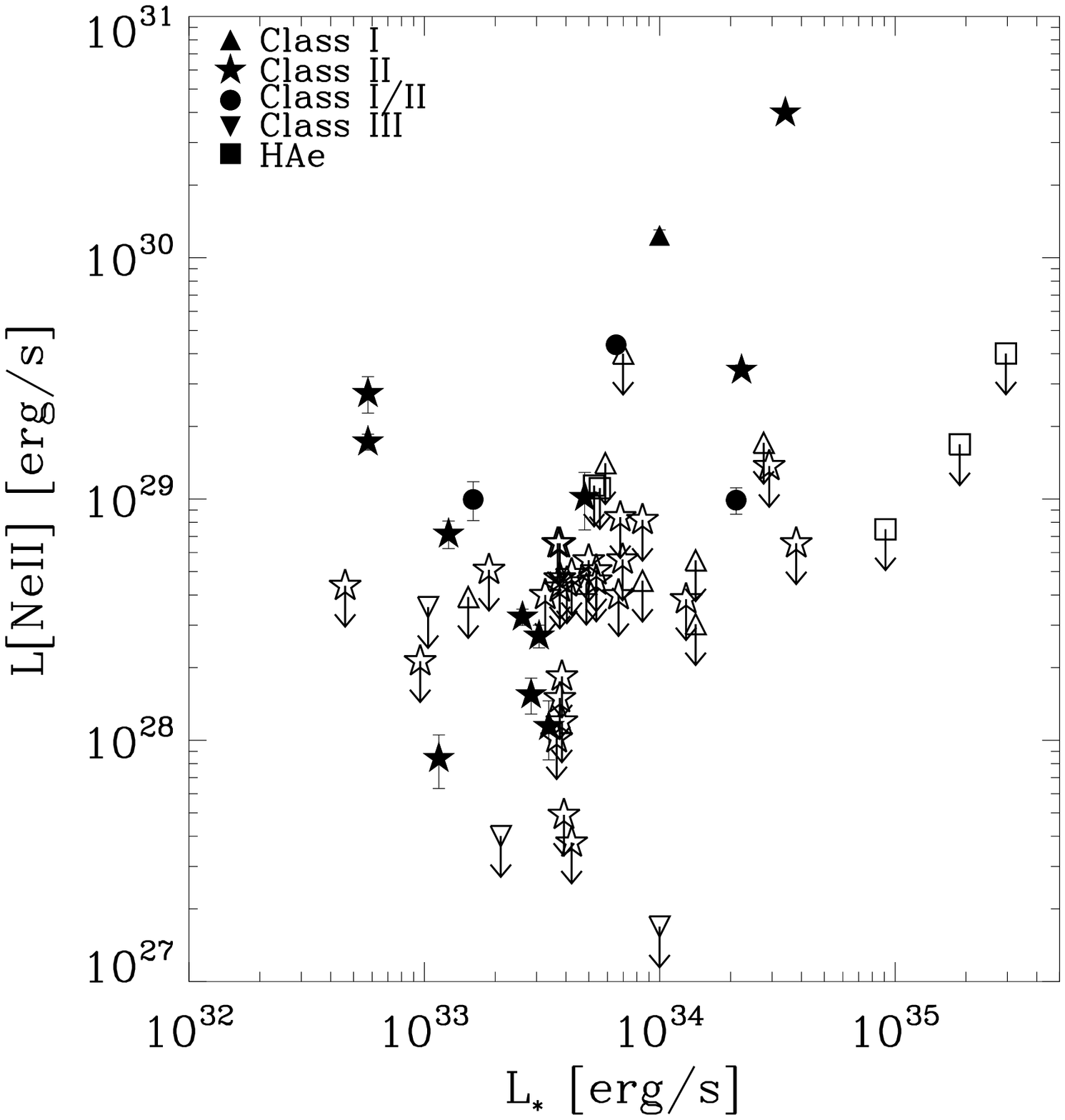}&
			\includegraphics[height=6.2cm]{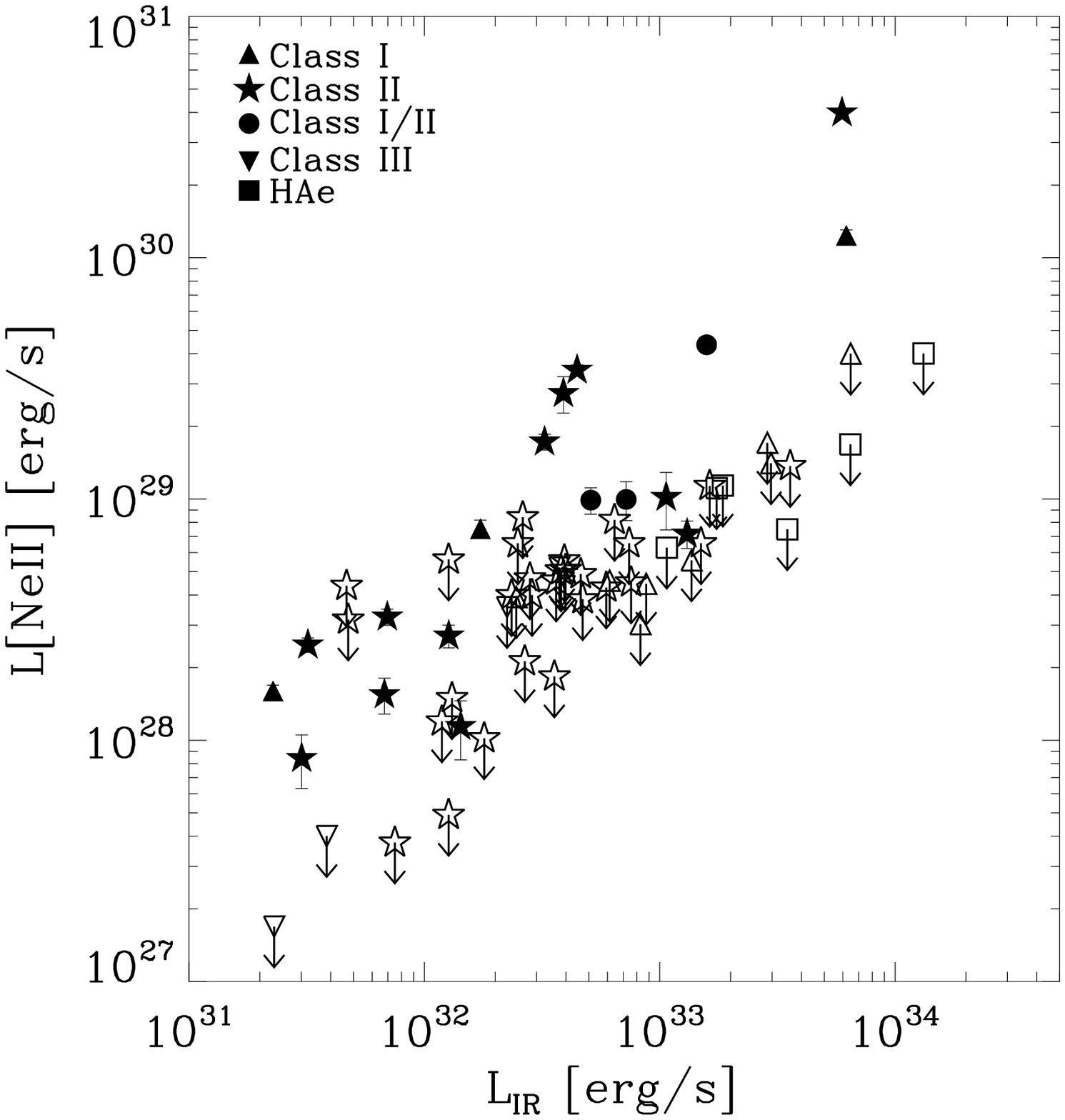}	   		 
	\end{array}$
		$\begin{array}{ccc}			 
			\includegraphics[height=6.2cm]{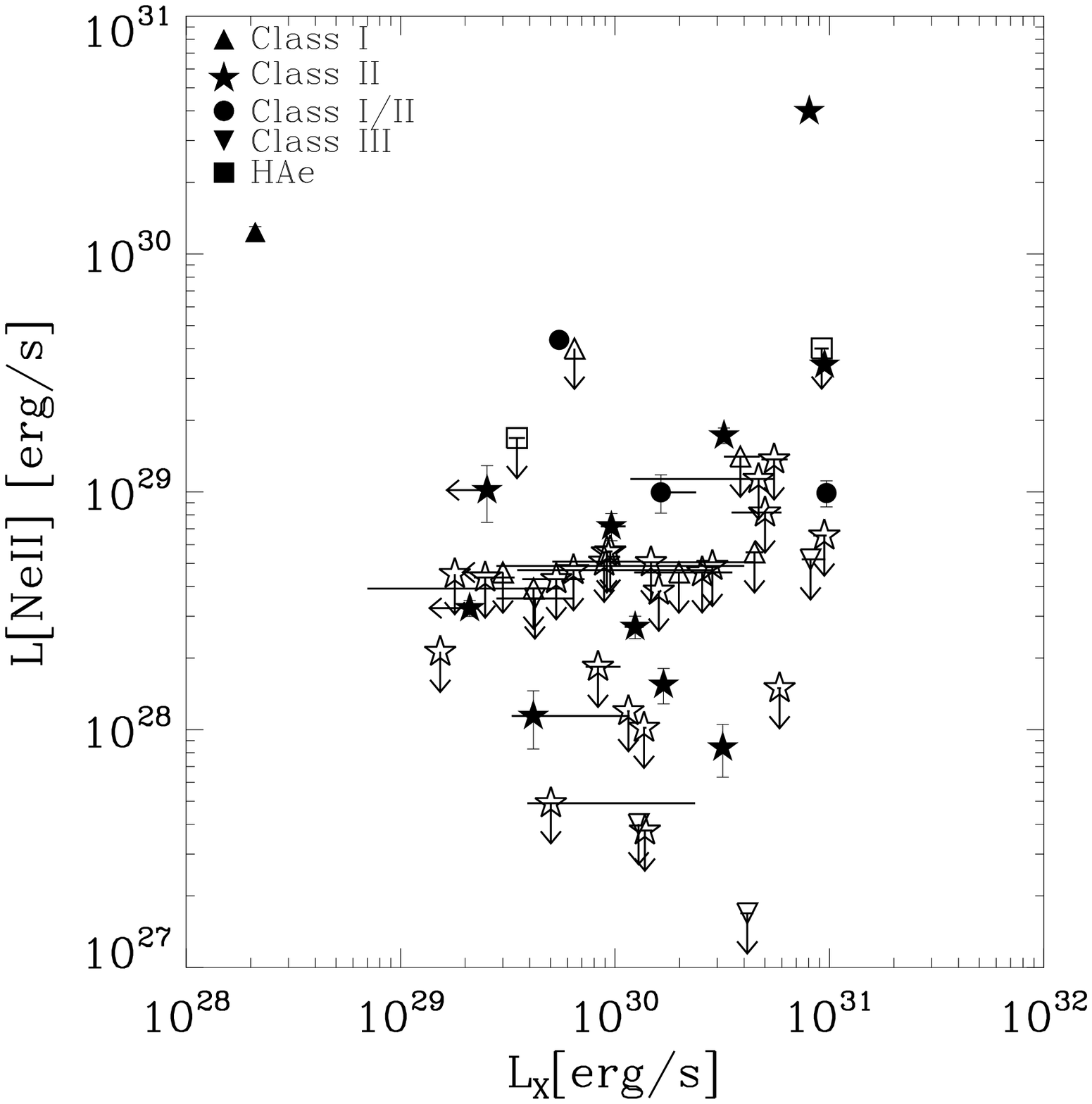} &
		    \includegraphics[height=6.2cm]{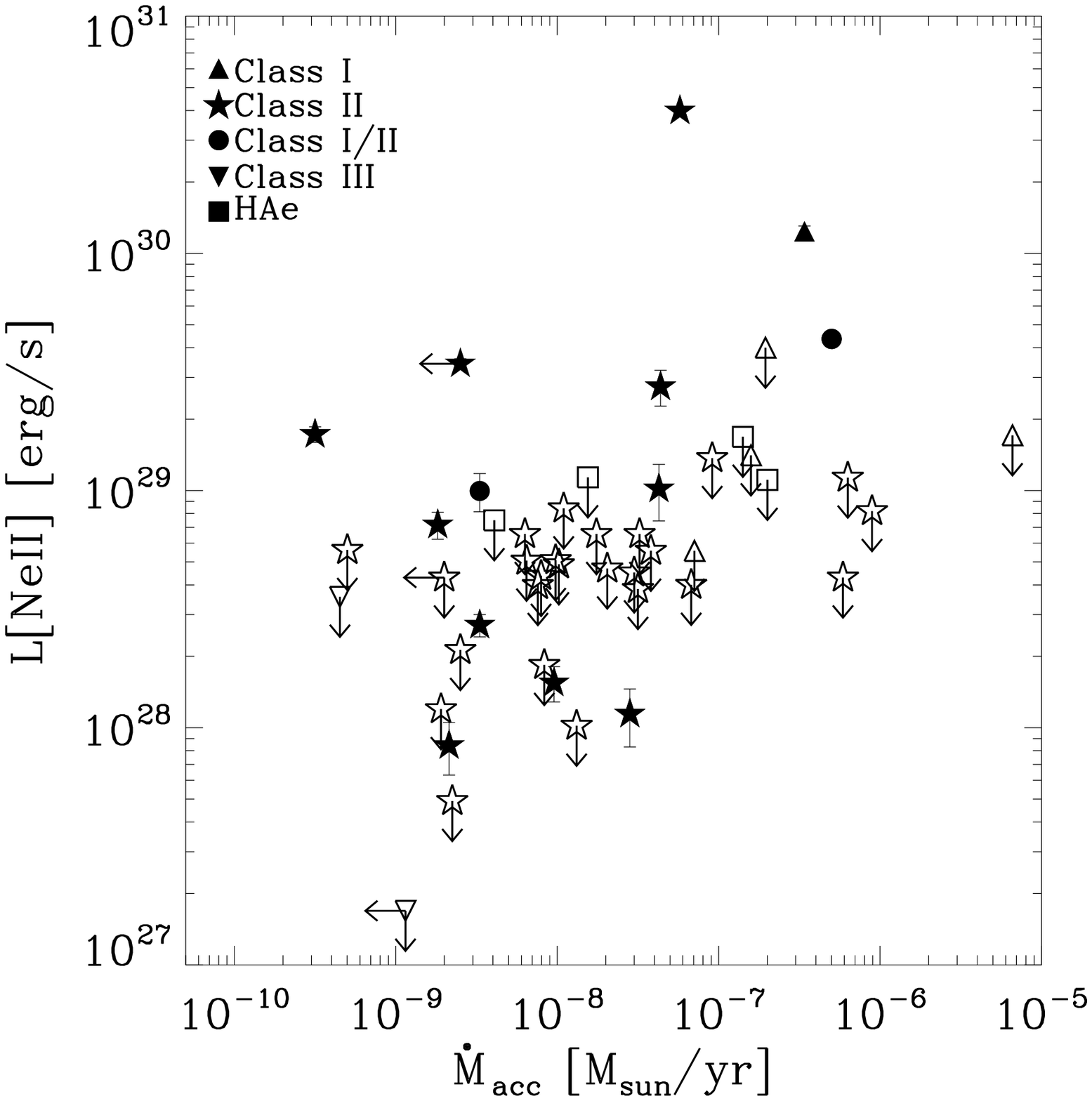} &
		    \includegraphics[height=6.2cm]{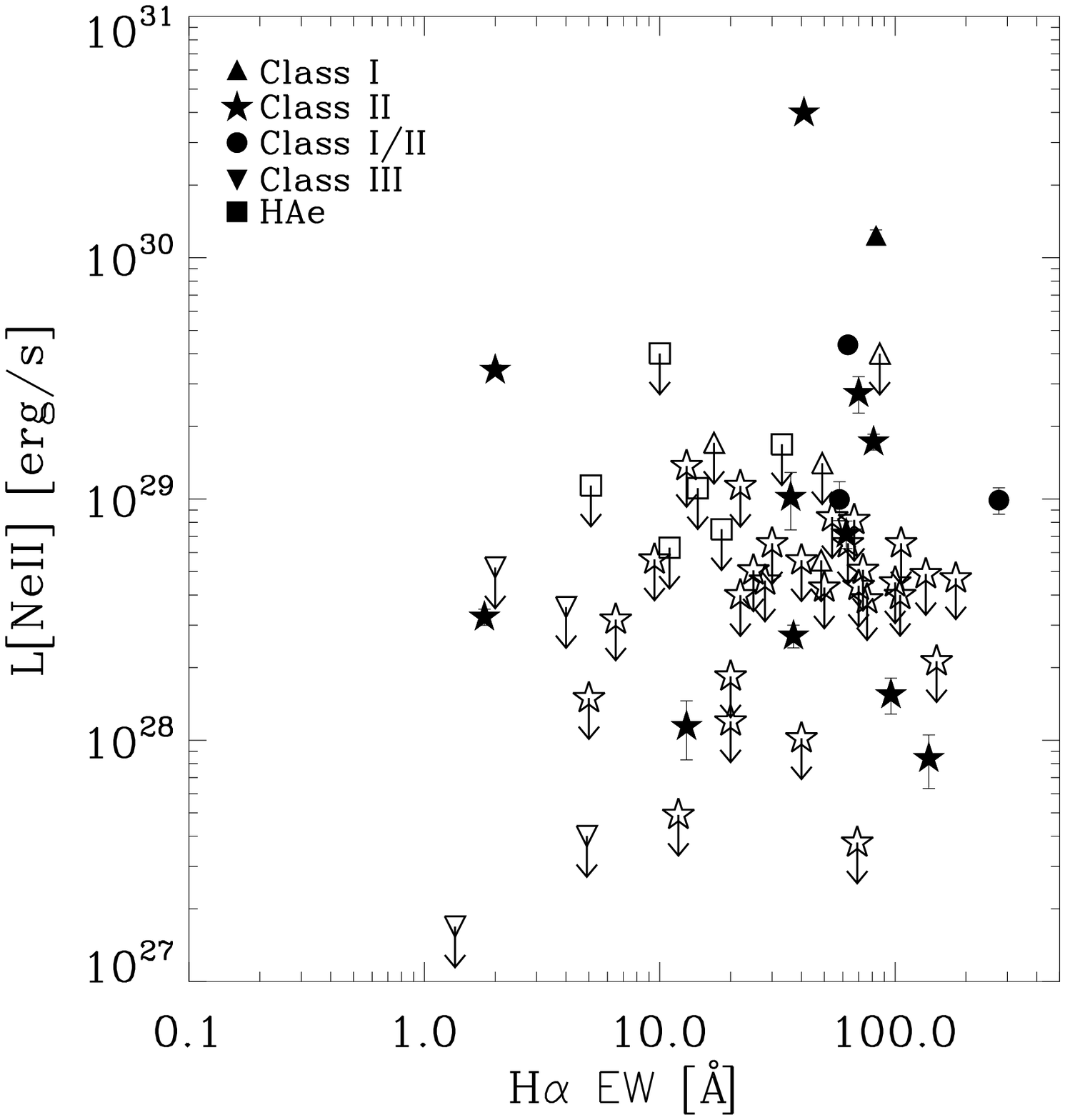} 
	\end{array}$	   
	$\begin{array}{ccc}
		    \includegraphics[height=6.2cm]{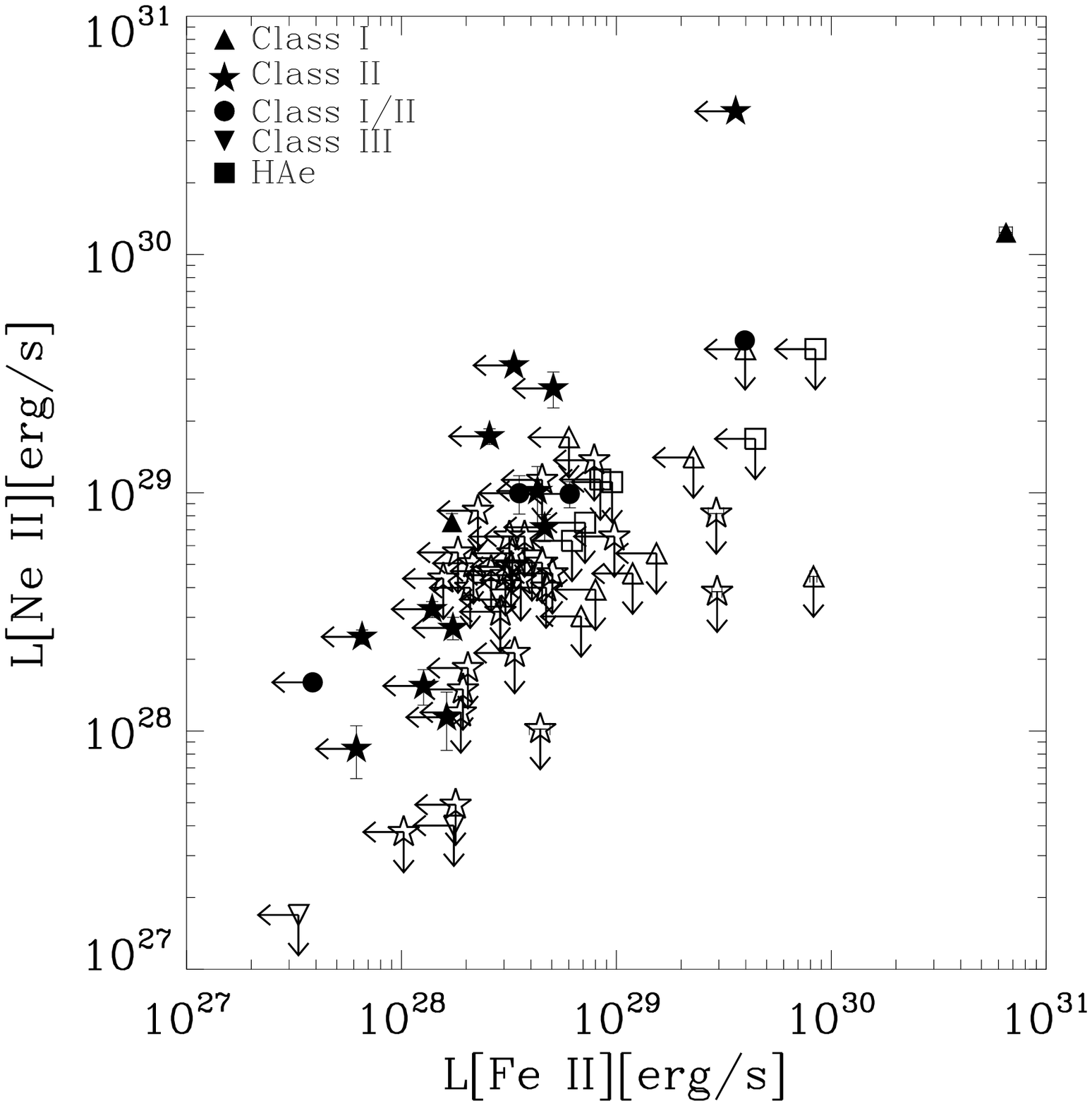}  &	
			\includegraphics[height=6.2cm]{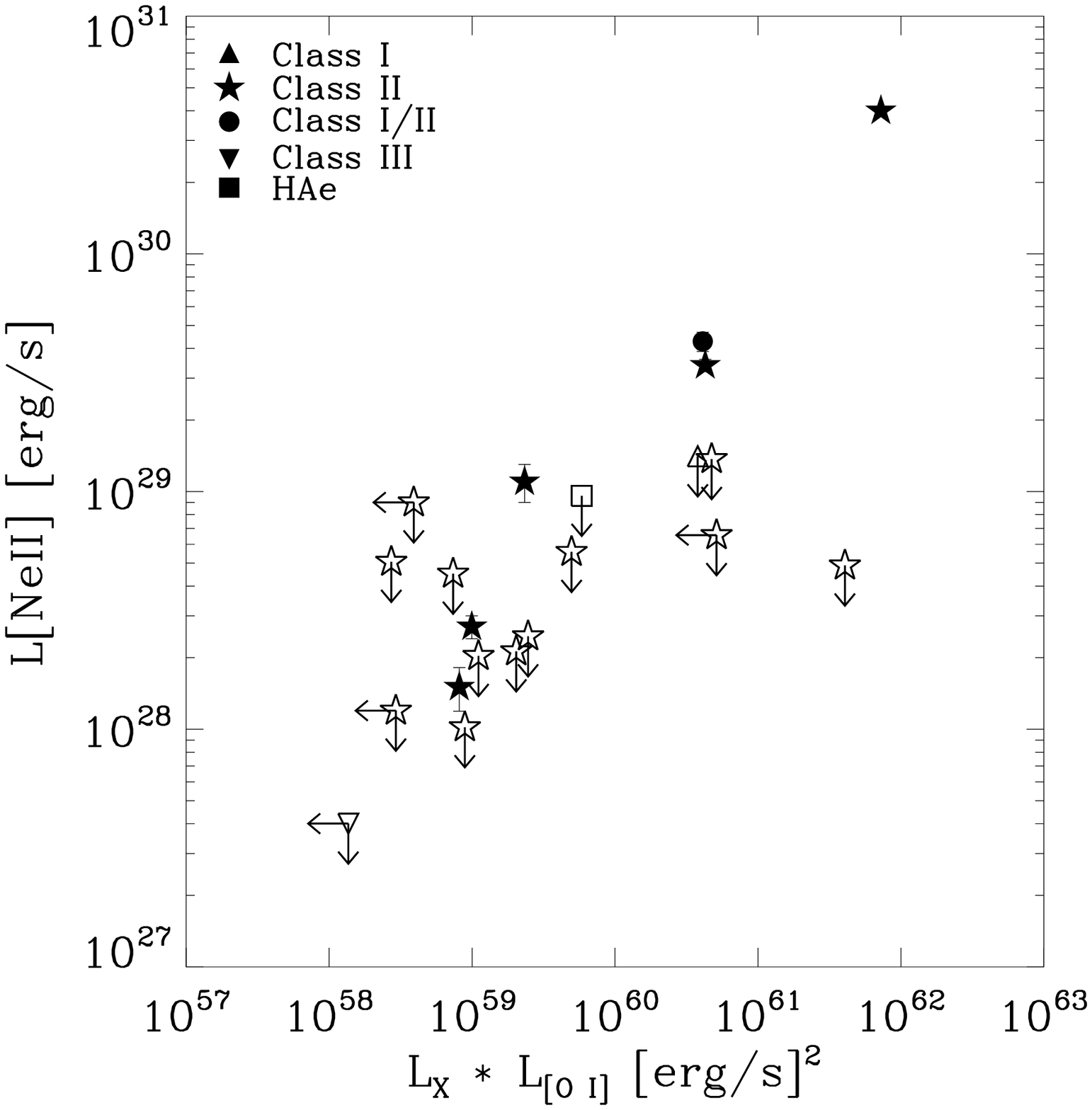}&
			\includegraphics[height=6.2cm]{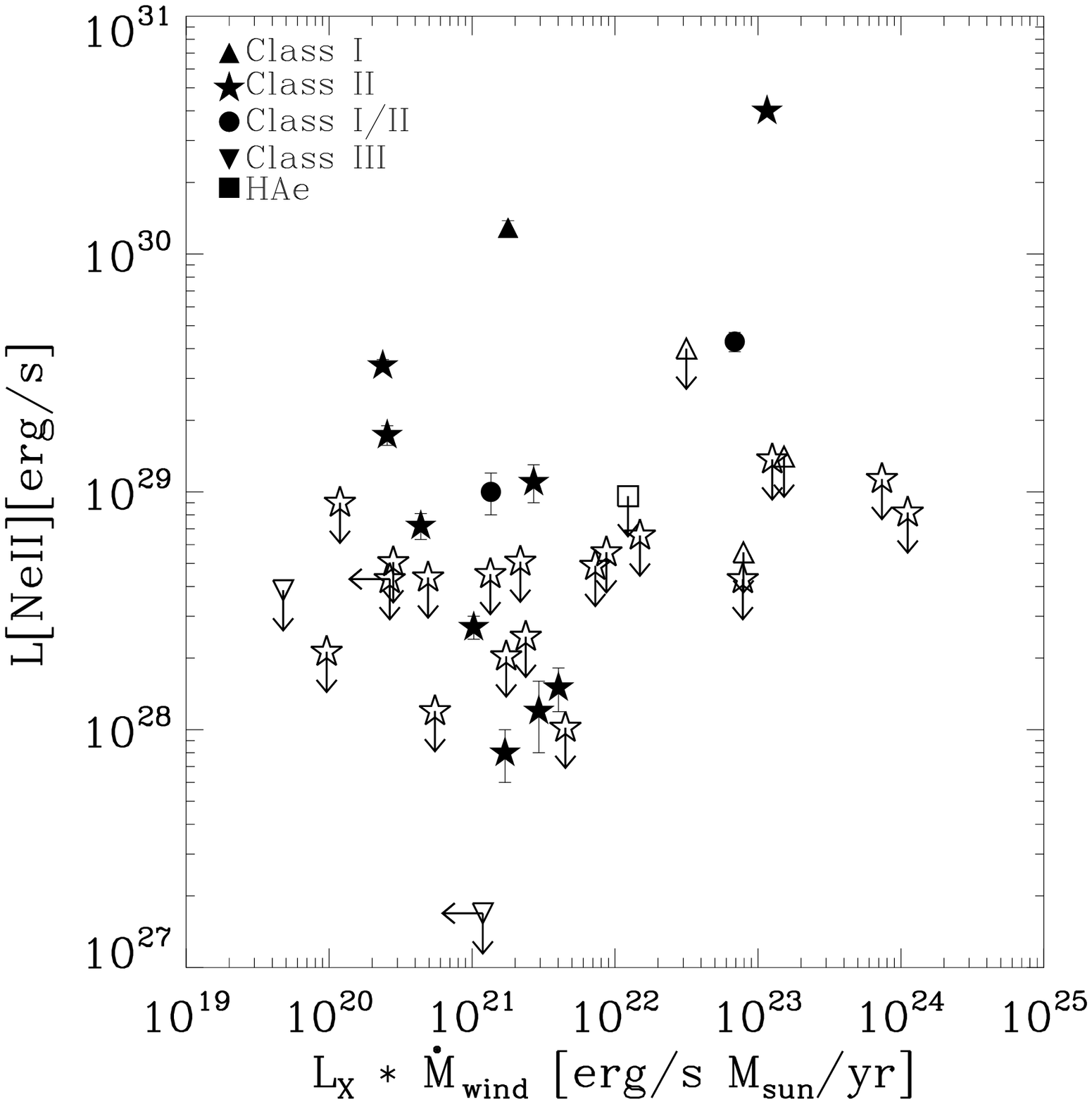}
		\end{array}$

	\end{minipage}
     \caption{ [Ne~II] luminosity versus the different parameters for which correlations have been tested. The different symbols indicate Class~I, II, III or Herbig stars as indicated in the legend. Filled symbols indicate [Ne~II] detections while empty symbols with arrows indicate upper limits.}
     \label{Ne2corr}
 \end{figure*}

\begin{figure*}[!h]
   \centering
   	\begin{minipage}[!b]{1.\textwidth}
		\centering
			$\begin{array}{ccc}
			    \includegraphics[height=6.3cm]{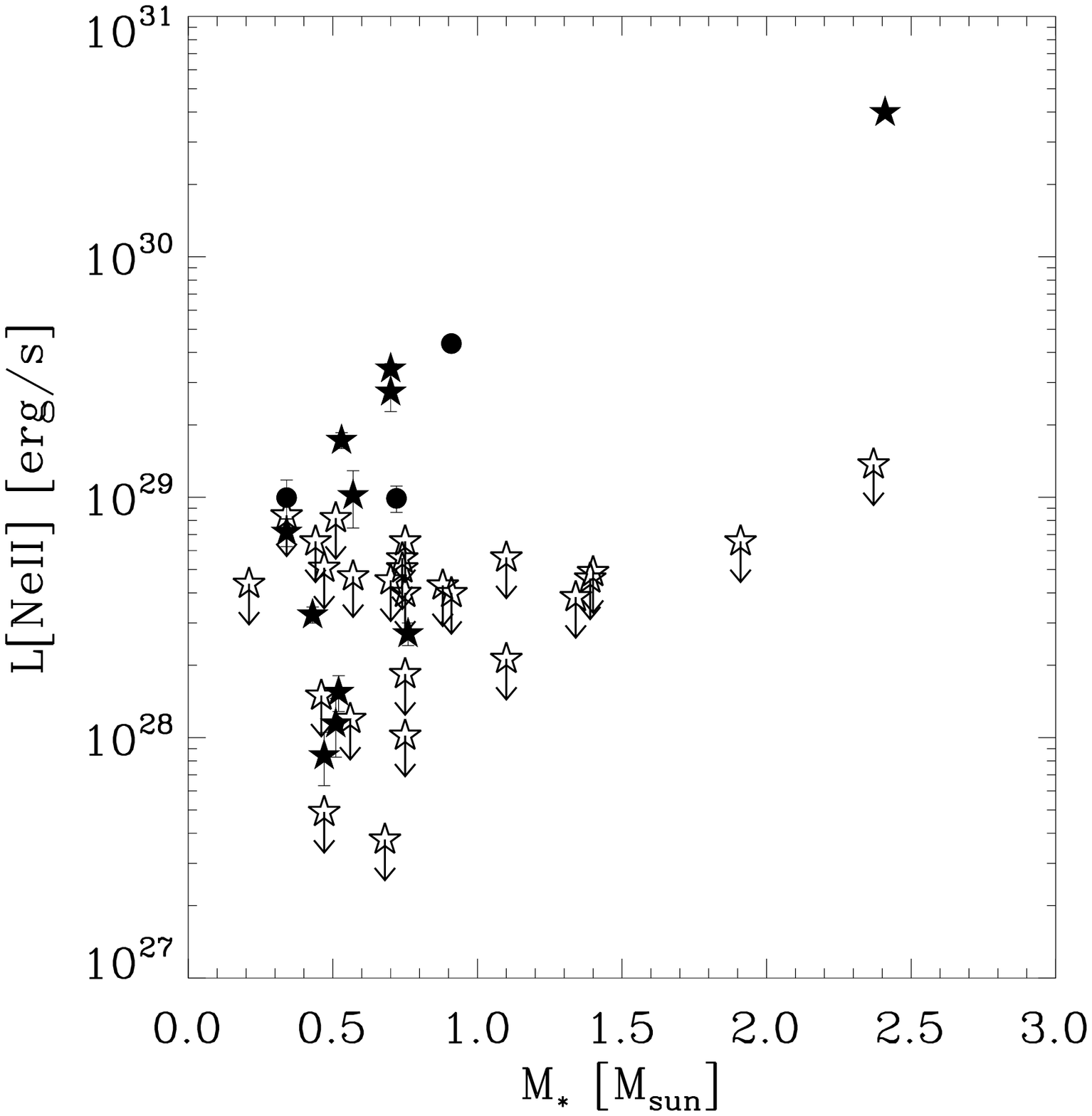} &
				\includegraphics[height=6.3cm]{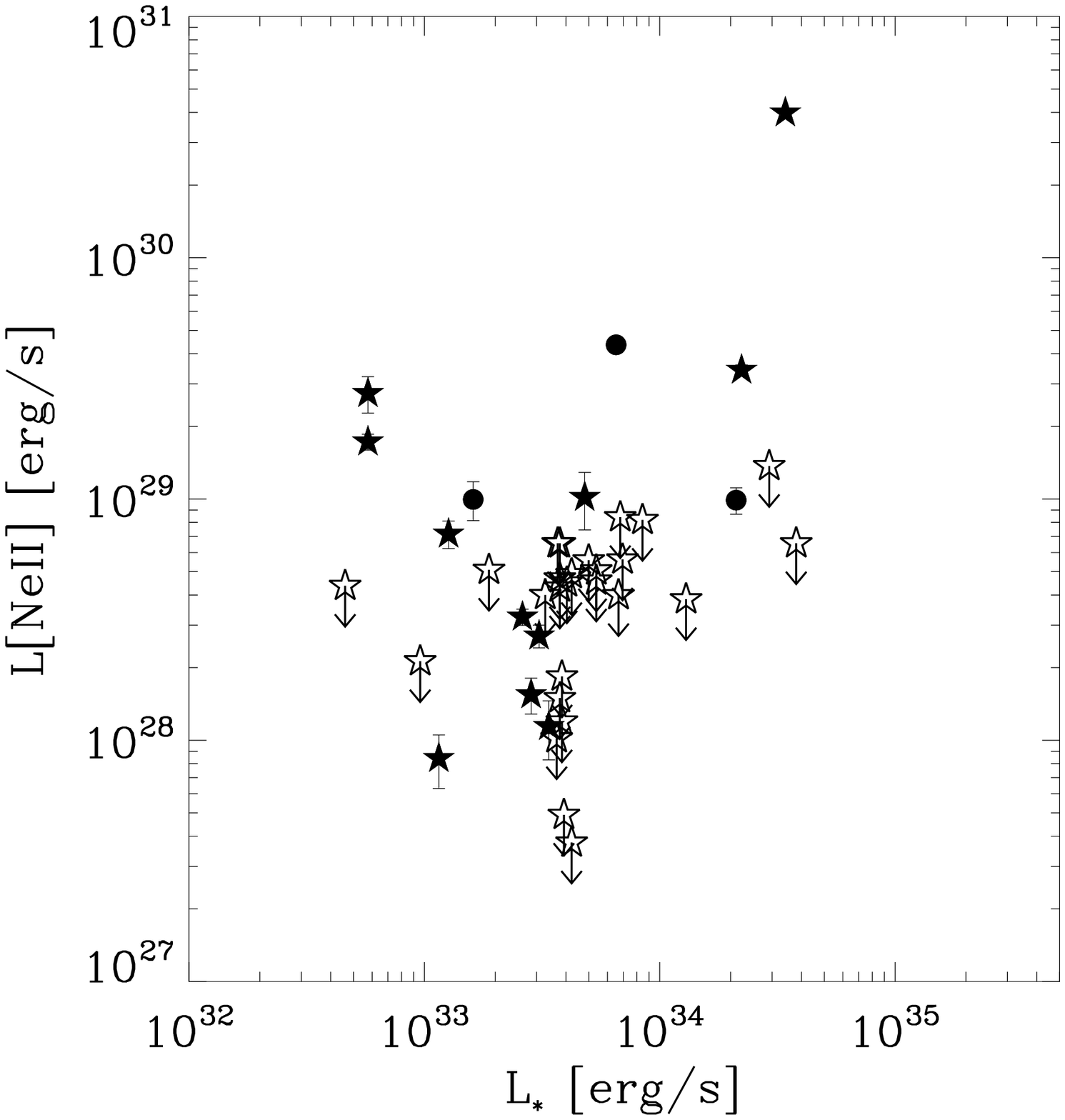} &
				\includegraphics[height=6.3cm]{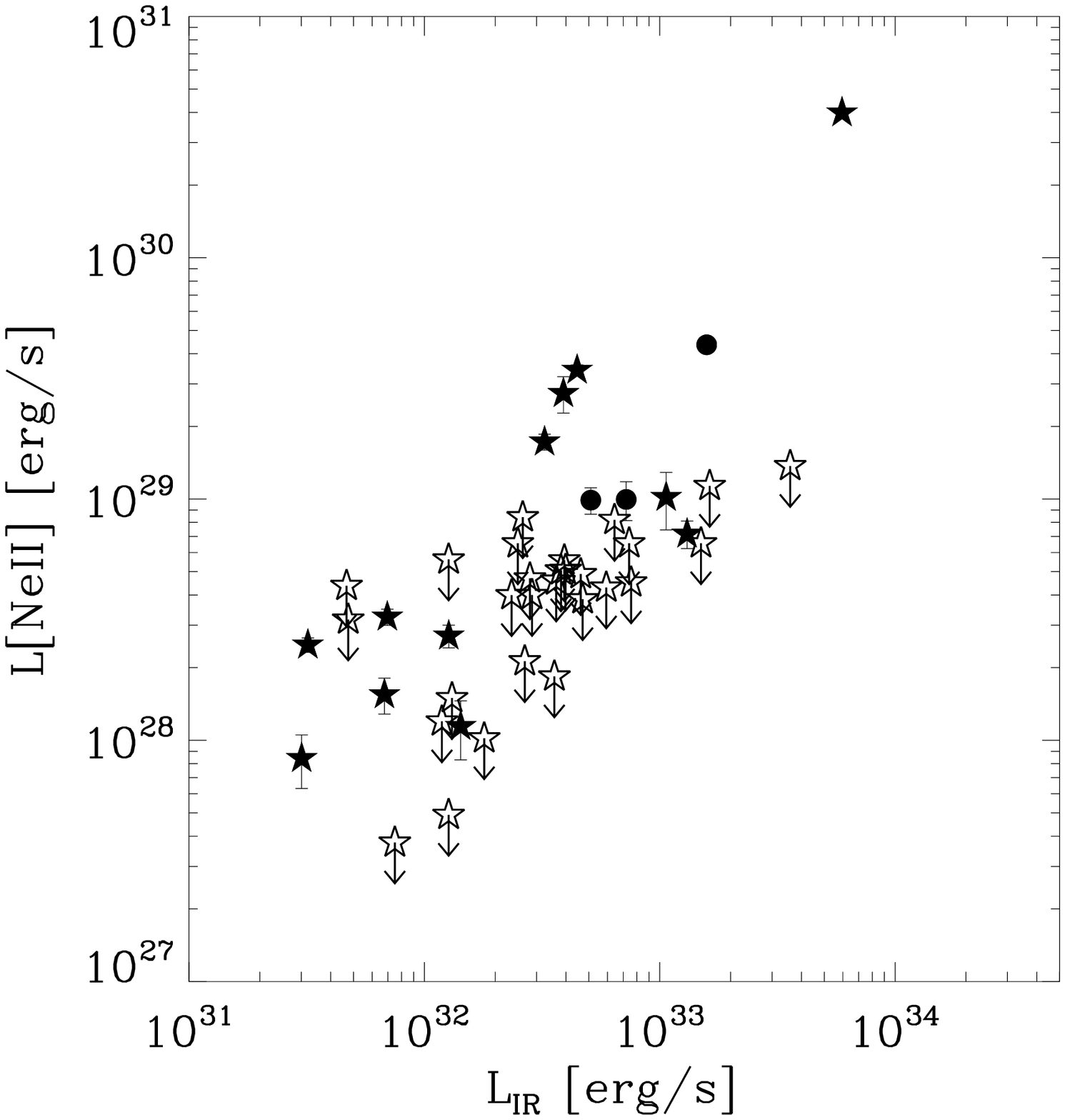}
			\end{array}$
		$\begin{array}{ccc}
				\includegraphics[height=6.3cm]{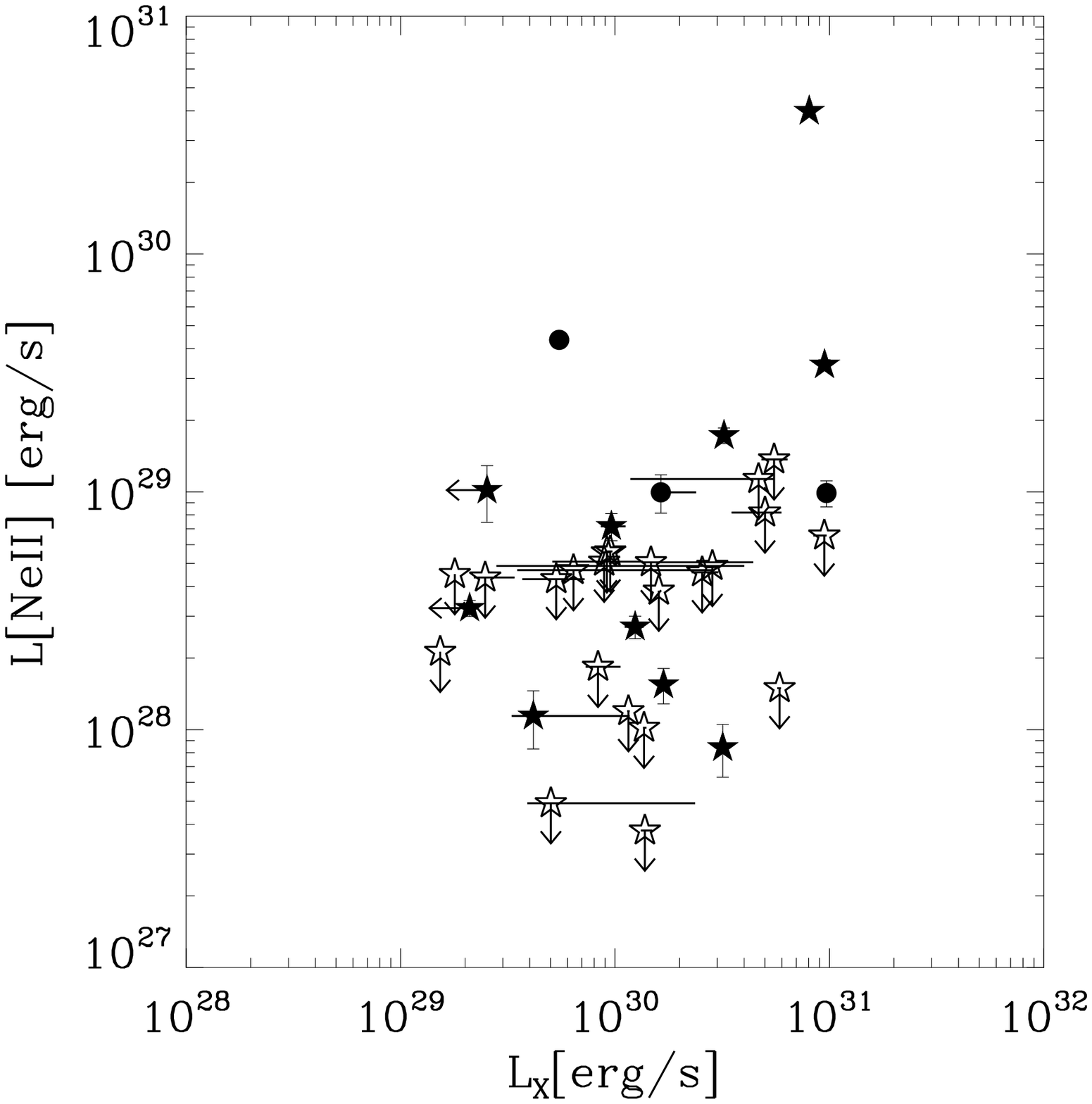} & 	    			 				
			    \includegraphics[height=6.3cm]{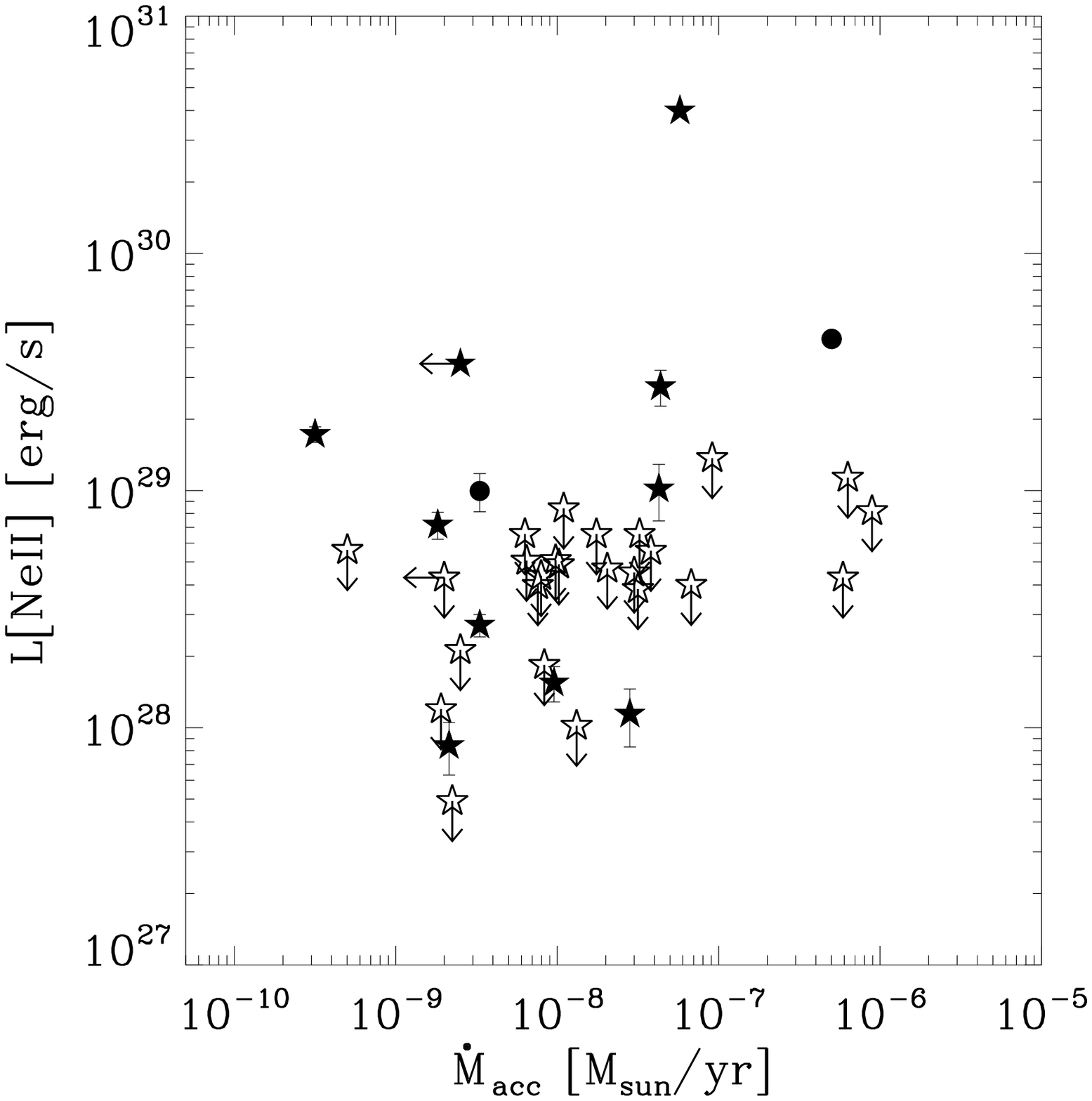} 
			    \includegraphics[height=6.3cm]{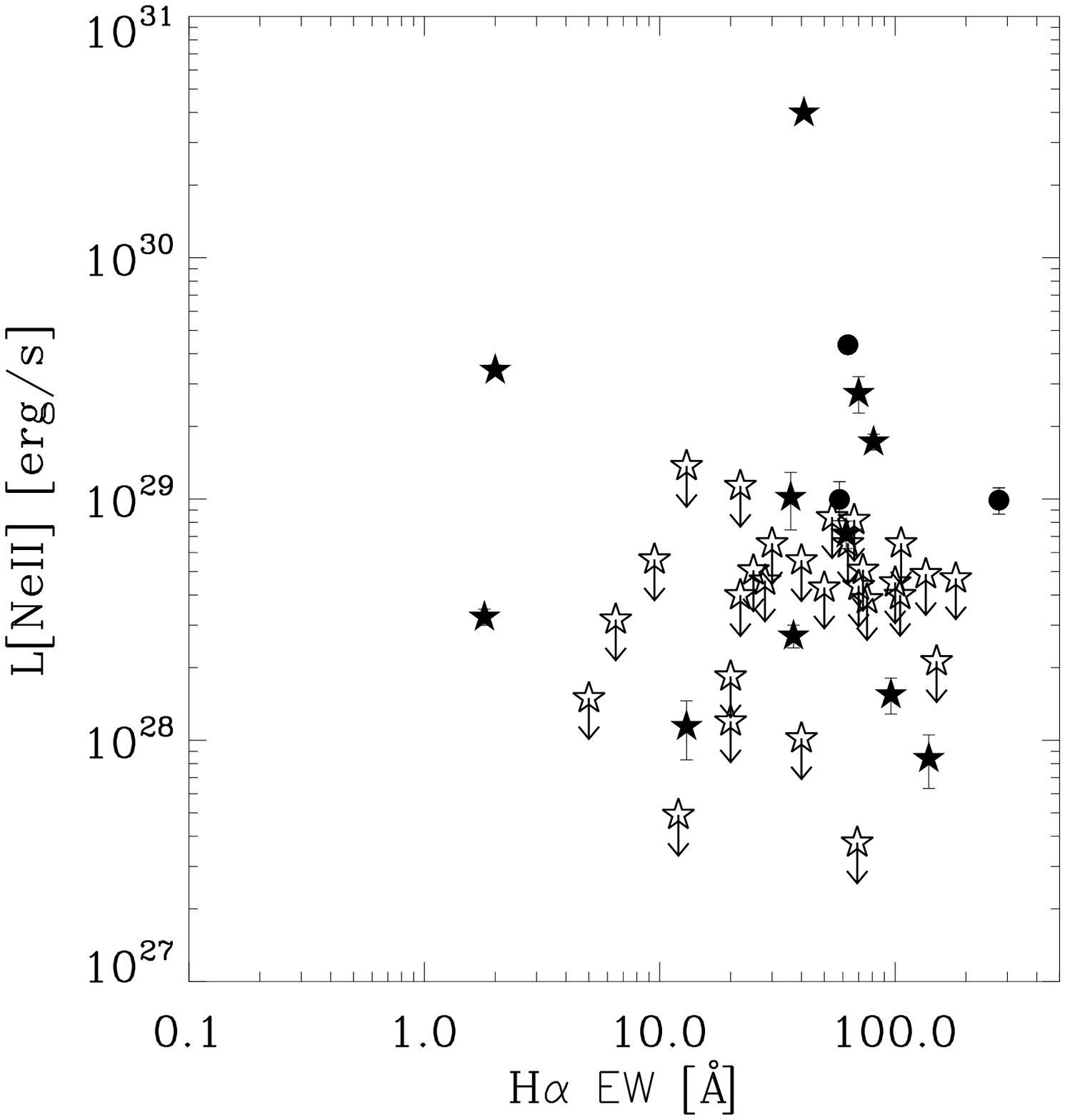} &
		\end{array}$
		$\begin{array}{ccc}				
			    \includegraphics[height=6.3cm]{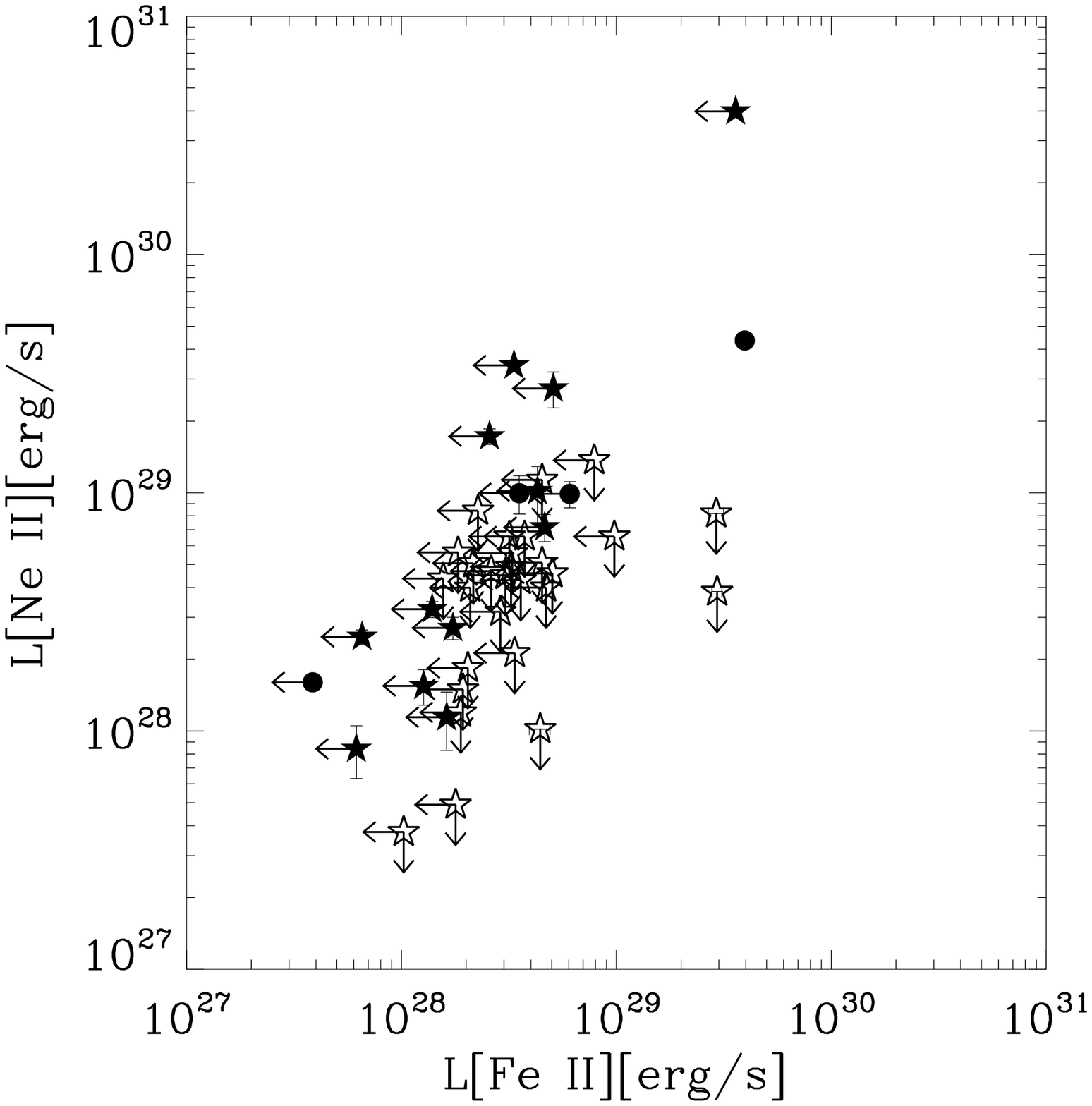} &
				\includegraphics[height=6.3cm]{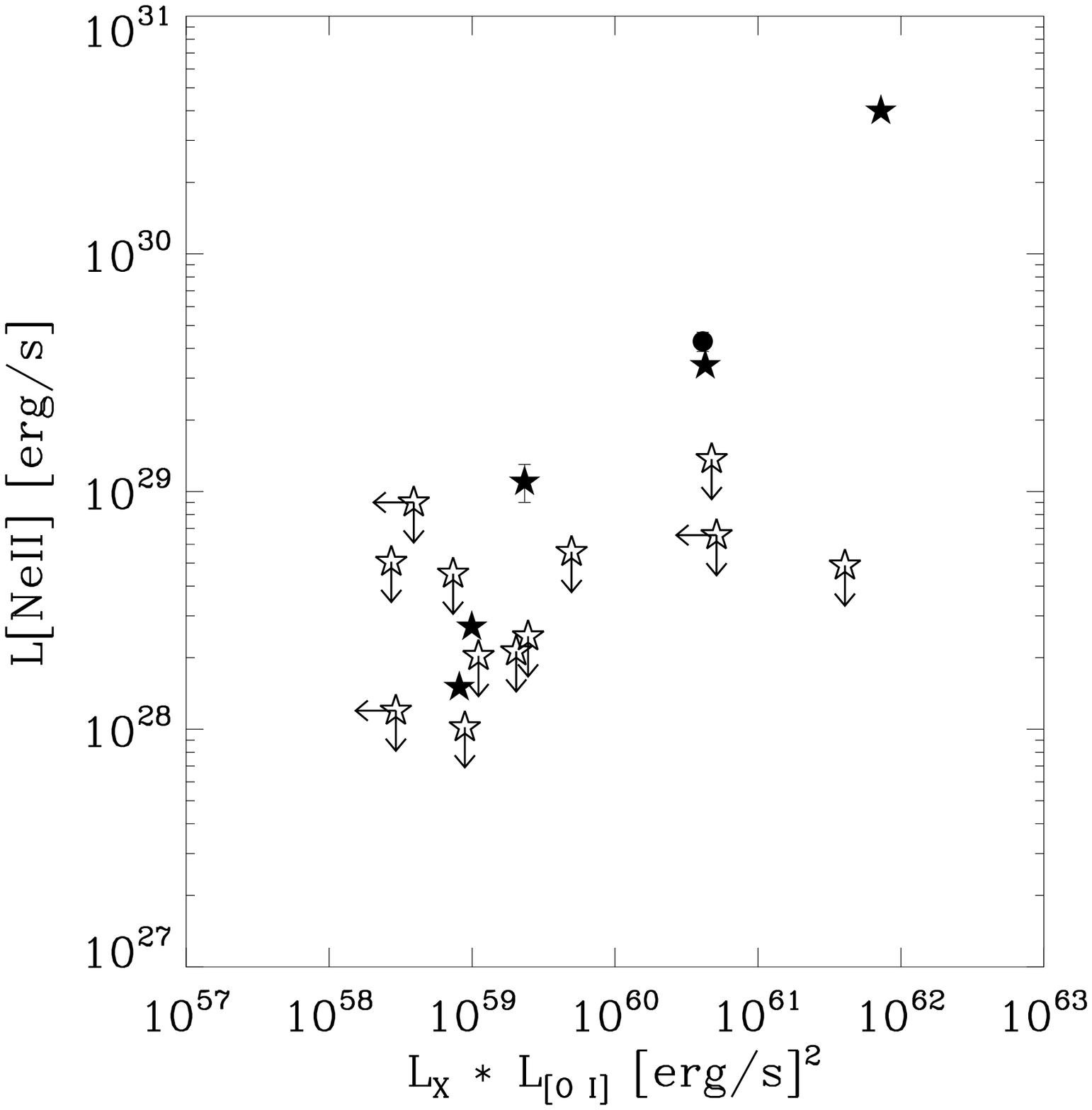} &
				\includegraphics[height=6.3cm]{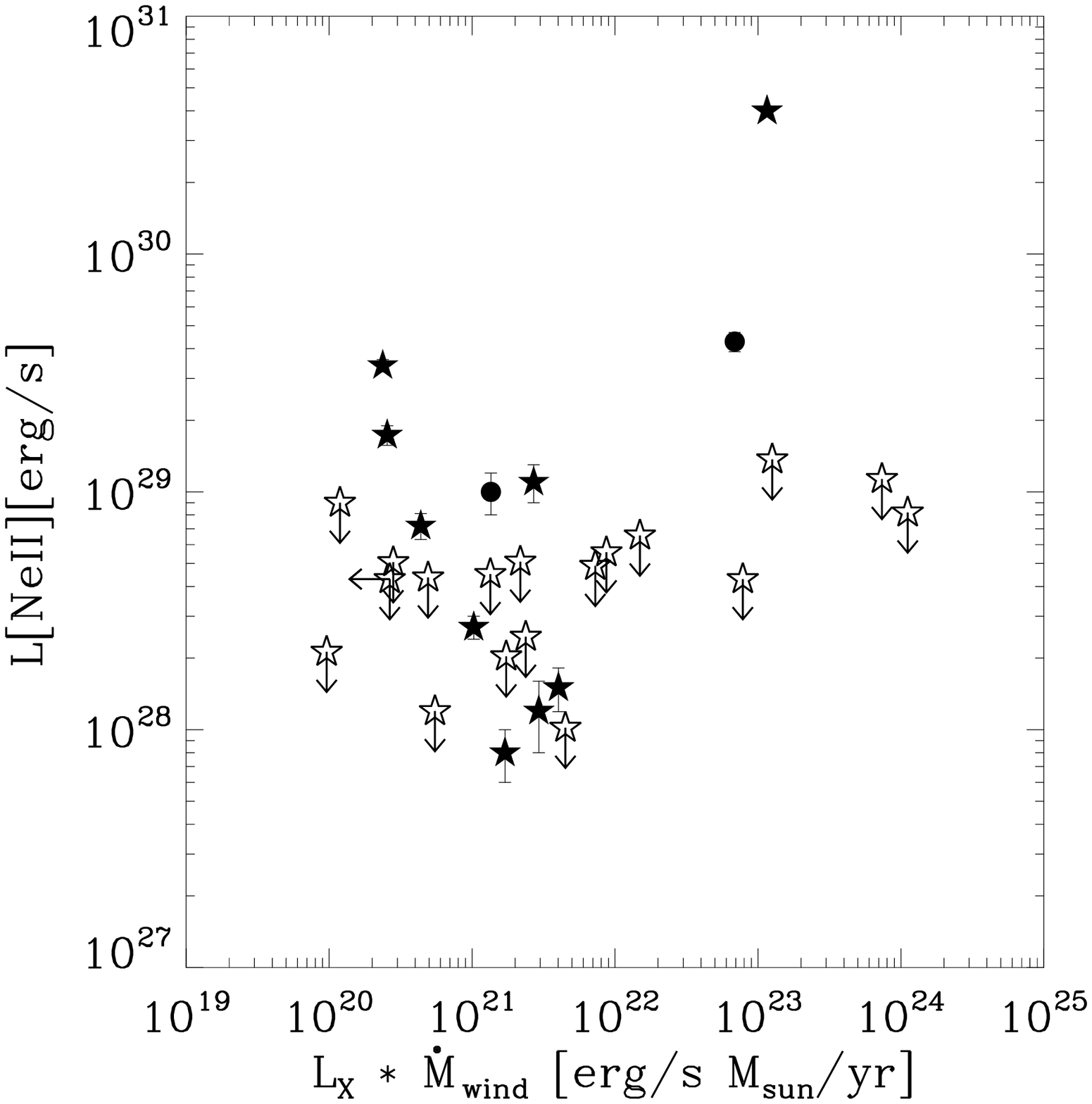}
		\end{array}$
	\end{minipage}
    \caption{ Same as Fig.~\ref{Ne2corr} for Class II (stars) and sources classified as I and II in the literature (circles).}
     \label{Ne2corr_CII}
        				
 \end{figure*}

We have tested the correlations using the whole sample (i.e., all infrared classes).
In addition, we have created a sub-sample with the Class~II sources, given that the majority of the [Ne~II] detections are from these objects.
In Figs.~\ref{Ne2corr} and \ref{Ne2corr_CII} we present L$_{\rm [Ne~II]}$ as a function of the different parameters studied for the complete sample and for the sub-sample of Class~II objects separately. Although we find significant correlations (at a 5\% threshold) based on several methods included in the ASURV package, there is a certain chance that the conclusion of correlation is incorrect. 
To this end, we have used a bootstrap method to determine the chance of finding a correlation by chance. Such a method implies the creation of a sampling distribution from the original data. In practice, we 
created $ 5\times 10^{4}$ random samples in each testing case by keeping the independent variable unchanged and permuting each time the dependent variable (L$_{\rm [Ne~II]}$ in this case) in order to break 
the links between each pair of points, keeping the number of detections and upper limits unchanged. We have applied the correlation tests using the ASURV package for each sample. Then we have constructed the distribution of the 
correlation parameters (i.e., $\tau$ for Kendall, $\rho$ for Spearman, and $\chi^2$ for Cox), and we have looked into the percentage of cases with correlation parameters larger than the ones obtained for the original sample. 
In all correlations tested, the probabilities were generally less than 4\%, thus confirming the correlations presented in Table \ref{table:3}.

\begin{table*}
\caption{Results from survival analysis correlation tests for the luminosity of the [Ne~II] line.}
\label{table:3}
\centering
\begin{tabular}{lcccccc}
\hline\hline
\multicolumn{7}{c}{ \bf All infrared classes}\\
 & & \multicolumn{2}{c}{N$^\circ$ Upper Limits}  & & Probability and Correlation Parameter& \\
\hline
Variable 	&	N$^\circ$ Points &	Total  & [NeII]	 &Cox &	Kendall &	Spearman \\
\hline
M${_*}$ 							&	53	&	38	&	38	&	0.92 (0.010)	&	0.62 (0.49)	&	0.55 (0.084)	\\
L${_*}$ 						 	&	57	&	42	&	42	&	0.56 (0.34)	&	0.82 (0.23)	&	0.93	 (0.012)\\
L$_{\rm X}$ 						&	47	&	35	&	33	&	...  	&	0.09	 (1.7) &	0.18 (0.18)	\\
L$_{\rm IR}$						&	64	&	46	&	46	&	{\bf0.04} (4.4)	&	{\bf0.01} (2.4)	&	{\bf0.01} (0.31)	\\
EW(H$\alpha$) 					&	56	&	42	&	42	&	{\bf0.04} (4.3)	&	{\bf0.03} (2.1)	&	{\bf0.01} (0.3)	\\
$\dot{\rm M}_{\rm acc}$ 		 	&	47	&	36	&	32	&	...  	&	0.62 (0.50)	&	0.37 (0.13)	\\
$\dot{\rm M}_{\rm acc-high}$ 		&	27	&	21	&	20	&	...	&	0.13 (1.5)	&	{\bf0.04} (0.39)	\\
$\dot{\rm M}_{\rm acc-low}$ 		&	19	&	15	&	12	&	...  	&	0.20 (-1.3)	&	0.55 (-0.14)	\\
L$_{\rm [OI]}$ 					&	27	&	21	&	16	&	...  	&	{\bf0.02} (2.2)	&	{\bf0.02} (0.46)	\\
L$_{\rm X}$L$_{\rm [OI]}$ 		&	22	&	17	&	12	&	...	&	0.06 (1.9)	&	{\bf0.03} (0.46)	\\
L$_{\rm X}\dot{\rm M}_{\rm wind}$&	36	&	28	&	23	&	...	&	0.67 (0.42)	&	1.00	 (0.00)\\
L$_{\rm [Fe II]}$ 				&	64	&	61	&	46	&	...  	&	0.09 (1.7)	&	0.10	 (0.21)\\
\hline
\multicolumn{7}{c}{ \bf Class II}\\
\hline 
M${_*}$ 							&	40	&	26	&	26	&	0.10	 (2.6) &	0.46 (-0.74)	&	0.27 (-0.17)	\\
L${_*}$ 						 	&	40	&	26	&	26	&	0.94 (0.005)	&	0.85 (0.19)	&	0.88 (0.02)	\\
L$_{\rm X}$ 						&	35	&	24	&	22	&	...	&	{\bf0.04} (2.0)	&	{\bf0.05} (0.34)	\\
L$_{\rm IR}$						&	44	&	28	&	28	&	{\bf0.03} (4.7)	&	{\bf0.004} (2.8)	&	{\bf0.01} (0.39)	\\
EW(H$\alpha$) 					&	41	&	28	&	28	&	0.26	 (1.3)&	0.29 (1.1)	&	0.23	 (0.19)\\
$\dot{\rm M}_{\rm acc}$ 		 	&	36	&	26	&	24	&	...	&	0.90 (0.12)	&	0.99 (-0.002)	\\
$\dot{\rm M}_{\rm acc-high}$ 		&	19	&	14	&	13	&	...  	&	{\bf0.05} (1.9)	&	{\bf0.04} (0.49)	\\
$\dot{\rm M}_{\rm acc-low}$ 		&	17	&	12	&	10	&	...	&	{\bf0.02} (-2.2)	&	{\bf0.05} (-0.49)	\\
L$_{\rm [OI]}$ 					&	24	&	18	&	18	&	...  	&	{\bf0.02	} (2.3)&	{\bf0.03	} (0.45)\\
L$_{\rm X}$L$_{\rm [OI]}$ 		&	19	&	14	&	13	&	...	&	0.10 (1.9)	&	0.09	 (0.44)\\
L$_{\rm X}\dot{\rm M}_{\rm wind}$	&	30	&	22	&	19	&	...  	&	0.61 (-0.51)	&	0.63	 (-0.90)\\
L$_{\rm [Fe II]}$ 				&	44	&	42	&	28	&	...  	&	0.62 (0.50)	&	0.41 (0.12)	\\
\hline
\hline
\end{tabular}
\tablefoot{The first column shows the variable for which the correlation is tested, the second column shows the total number of points considered, the third column shows the total number of upper limits used (dependent and independent variable), 
the fourth column shows the number of upper limits for the luminosity of the [Ne~II] line (dependent variable), and the probabilities (P) of a correlation achieved by chance are presented in columns 5, 6, and 7, where P~$\leq 5 \%$ are shown in bold fonts.
The value of the correlation parameters are shown in parentheses. We have defined  $\dot{\rm M}_{\rm acc-high}$ and $\dot{\rm M}_{\rm acc-low}$ as follows: $\dot{\rm M}_{\rm acc-high} \geq 10^{-8}$M$_{\odot}$yr$^{-1}$ and 
$\dot{\rm M}_{\rm acc-low} < 10^{-8}$M$_{\odot}$yr$^{-1}$. The Cox method cannot be used when upper limits are present in both of the tested variables. See text for more details.
}
\end{table*}

\subsubsection{Correlations with stellar properties}
\label{Corr:Mstar}

We have tested correlations with the stellar mass and luminosity.
In a previous study, \cite{flaccomio:2009aa} found that the stellar mass does not influence the [Ne~II] luminosity.
In our sample, the detected [Ne~II] lines  are concentrated in sources at the low end of the mass range (with stellar masses typically $<$1~M$_{\odot}$); only two objects with M$_*\geq$~1~M$_{\odot}$ show [Ne~II] emission. 
In \citet{flaccomio:2009aa}, the detected [Ne~II] lines are spread across a larger range of stellar masses. 
Despite the different mass ranges probed by the two samples, the result is the same: L$_{\rm{[Ne~II]}}$ is not dependent on the stellar mass.
Since the stellar mass and luminosity are related through the relation L$_* \propto$~M$_* ^2$ \citep{hartmanBook:2009}, we
expect to obtain a similar lack of correlation between L$_{\rm[Ne~II]}$ and the stellar luminosity, which is indeed confirmed.

\subsubsection{Correlations with high-energy irradiation}
\label{Corr:LX}

The stellar X-ray luminosity is one of the key parameters driving the [Ne~II] emission proposed in several theoretical studies, e.g., \cite{glassgold:2007aa}. 
The first observational study suggesting such a correlation was \citet{pascucci:2007aa}, although this result was based on a sample of just four objects, spanning a small range in L$_{\rm X}$. 
\citet{espaillat:2007ab} increased the sample to seven objects but did not find any evident correlation between L$_{\rm[Ne II]}$ and L$_{\rm X}$. 
The same result was obtained recently by \citet{flaccomio:2009aa} based on a larger sample and spanning a larger range in L$_{\rm X}$.
Nevertheless, \citet{gudel:2010aa} have found a statistically significant dependence on the stellar L$_{\rm X}$ for their sample confined to Class~II objects and spanning a broad range of L$_{\rm X}$.

Our sample spans a range of X-ray luminositis between $10^{29}$ and $\sim 10^{31}$~erg~s$^{-1}$, similar to the sample presented by \citet{gudel:2010aa}.
The correlation between L$\rm{_{Ne II}}$ and L$\rm{_{X}}$ is non-existent for the complete sample (all infrared classes), i.e., consistent
with \citet{flaccomio:2009aa}. But we observe a correlation with L$_{\rm X}$ for the sub-sample of Class~II objects, consistent with \citet{gudel:2010aa}. 
These results are not contradictory: in fact, the difference between the previous studies lies in the sample selection.
\citet{flaccomio:2009aa} used a mixed sample, including Class~I to Class~III objects, similar to our sample (although we have included also Herbig AeBe stars), whereas
the sample used in \citet{gudel:2010aa} contains only Class~II objects.
Therefore, we do expect to find a lack of correlation like in \citet{flaccomio:2009aa} when using our whole sample, and the same result as in \citet{gudel:2010aa} when restricting the sample to Class~II objects. 

\subsubsection{Correlations with accretion parameters}
\label{Corr:Mdot}

The mass accretion rate ($\dot{\rm M}_{\rm acc}$) is an indirect measurement of the evolutionary state of the sources.
We expect $\dot{\rm M}_{\rm acc}$ to decrease as the star evolves toward the main sequence.
\citet{espaillat:2007ab} and \citet{flaccomio:2009aa}  found a positive correlation between L$_{\rm{[Ne~II]}}$ and $\dot{\rm M}_{\rm acc}$.
\citet{hollenbach:2009aa} studied the irradiation of the disk by high-energy photons (EUV and soft X-rays). 
They showed that in order for the high-energy photons to penetrate the strong protostellar winds and illuminate the disk beyond 1~AU, low wind mass-loss rates were needed (\.M$_{\rm wind} < 10^{-9}$~M$_{\odot}$yr$^{-1}$). 
This critical value was translated in \citet{hollenbach:2009aa} into a critical disk mass accretion rate of \.M$_{\rm acc} <10^{-8}$~M$_\odot$~yr$^{-1}$.
Sources with low mass accretion rates would be more suitable to study the effects of high-energy photons.
Following this idea, we have split our sample between high and low accretors, setting the limit at 
$\dot{\rm M}_{\rm acc}$ = $10^{-8}$~M$_\odot$~yr$^{-1}$, as proposed in \citet{hollenbach:2009aa}. 
We have, therefore, defined two sub-samples: 
\.M$_{\rm acc-high}\geq 10^{-8}$~M$_\odot$~yr$^{-1}$, and \.M$_{\rm acc-low}< 10^{-8}$~M$_\odot$~yr$^{-1}$.
 For the sub-sample of Class~II objects, we find a positive correlation between L$_{\rm{[Ne~II]}}$ and \.M$_{\rm acc}$ for the group \.M$_{\rm acc-high}$ and a negative correlation with  L$_{\rm{[Ne~II]}}$ for the group \.M$_{\rm acc-low}$. 
This result is not found when considering the sub-sample of Class~II objects with no separation in the mass accretion rate range.
We observe a difference between low and high accretors; the anti-correlation found for low-accretors is statistically more significant.
But we are cautious about this result: this  correlation has been discussed in \citet{gudel:2010aa} in the sense that jet driving sources are younger and, in general, have higher mass accretion rates than sources without jets. 
They argued that the segregation between both samples might produce an apparent correlation. Indeed, in our study, the sub-sample of low-accretors ($\dot{\rm M}_{\rm acc-low}$) is dominated by non-jet sources; 18 out of the 21 sources in 
this sub-sample do not have evidence of jets.On the other hand, in the sub-sample of high accretors ($\dot{\rm M}_{\rm acc-high}$), which accounts for 27 sources, 16 present observational evidence of jets/outflows. 

The H$\alpha$ line (seen usually in emission in young stars) is another tracer of accretion. 
Its equivalent width is commonly used to differentiate between accreting and non-accreting stars. 
Typically, EW(H$\alpha$)~$> 10$~\AA~ for accreting stars, and EW(H$\alpha$)$< 10$~\AA ~for non-accreting stars \citep{martin:1997aa}.
The result for the test between L$_{\rm{[Ne~II]}}$ and EW(H$\alpha$) is not clear in our sample. 
Although we found a correlation for the whole sample (i.e., all infrared classes), this result is not repeated for the sub-sample of Class~II objects.

Finally, we have tested a correlation of L$_{\rm{[Ne~II]}}$ with the luminosity in the mid-IR, in order to check if bright emission lines are related to a bright continuum emission. 
The mid-IR luminosity has been obtained for each object by integrating the continuum flux in the SH range (9 to 19~$\mu$m).
Since in most cases we do not have background observations, we have over-plotted the SH and SL spectra and checked that the continuum emission is dominated by the source and not by the background.
Given the small aperture and wavelength coverage of the SH module, the background contribution is expected to be small (e.g., \citealt{furlan:2006aa}). 
 We have obtained a correlation between L$_{\rm{[Ne~II]}}$ and L$_{\rm{IR}}$. This result is confirmed by the bootstrap technique. 
The origin of this correlation is not clear, although not likely to be the product of bias in the observations.
The rms of the continuum increases with L$_{\rm{IR}}$, but the signal-to-noise ratio for the detected [Ne~II] lines is distributed uniformly over the L$_{\rm{IR}}$ spanned by our sample, with no preference for either high or low values of
 the signal-to noise ratio. This correlation was also found by \citet{lahuis:2010aa} in a sample of Class~I sources. 

\subsubsection{Correlations with jet/outflow parameters}

Forbidden atomic lines of [O~I] (6300 \AA) and [Fe~II] (1.64~$\mu$m) are used to trace the outflow activity in young stars (e.g., \citealt{cabrit:1990aa,agra-amboage:2009aa}).
\citet{gudel:2010aa}  investigated this correlation to test the relation between [Ne~II] emission and outflow activity.
\citet{shang:2010aa} modeled the flux and shape of the [Ne~II] and [O~I] lines and found a clear correlation between both lines, indicating that neon would also trace jets.

We have tested the correlation between the luminosity of the [Ne~II] line and the luminosity of the [O~I] line at 6300~\AA~ compiled from the literature, obtaining a positive result.
We have also tested a correlation with the product of L$_{\rm X}$ and L$_{\rm [O~I]}$, as proposed in \citet{gudel:2010aa}, but
a correlation was not found.  We have further used the luminosities of the [Fe~II] line at 25.99~$\mu$m obtained in our sample to study the dependence of [Ne~II] on this tracer, and no correlation was found.
 We emphasize, however, that the tests were done with a large number of upper limits due to the lack of detections in the tested parameters. 

Three objects with detected [Ne~II] emission also show [Fe~II]: L1551~IRS~5 (Class~I), DG~Tau (Class~I/II), and SST~041412.2+280837. 
Since [Fe~II] is a known tracer of shocks, the origin of the [Ne~II] emission in these particular cases might be explained by the shock scenario. 
However, this is not clear for the rest of the objects with [Ne~II] emission for which there is no detected emission from [Fe~II].

\subsubsection{Correlations with {\rm L}$_{\rm X}\times${\rm \.M}$_{\rm wind}$}

According to the model presented in \citet{shang:2010aa}, [Ne~II] is produced in jets from young stars.
Based on the X-wind model theory, they obtained line profiles for [Ne~II], [Ne~III], and [O~I].
They tested the dependency of the line luminosities on critical parameters of this model such as L$_{\rm X}$ and the wind mass loss rate (\.M$_{\rm wind}$). 
The best correlation was found for the product of both parameters L$_{\rm X} \times$ \.M$_{\rm wind}$.
The relation between wind mass loss rate and accretion rate is \.M$_{\rm wind} =f \times$ \.M$_{\rm acc}$, with $f\approx 0.25$ \citep{shang:2004aa}.  
However, in our sample, no such correlation is found.

\subsection{[Fe~II] emission}
\label{fe2}

[Fe~II] is a tracer of shock activity typically observed in the near-IR at 1.53 and 1.64~$\mu$m (\citealt{davis:2003aa,hartigan:2009aa}). 
Its excitation mechanism can also be stimulated by the irradiation of protoplanetary disks by high-energy photons \citep{gorti:2008aa}.
We have detected [Fe~II] emission in 7 sources. In five cases, we have detected both iron lines (at 17.93 and 25.99~$\mu$m).
Typically, whenever both lines have been detected, the luminosity of the [Fe~II] line at 25.99~$\mu$m is roughly a factor of 3 higher than the luminosity of the line at 17.93~$\mu$m.
This result is consistent with \citet{hollenbach:1989ab}, where in a shock environment the [Fe~II] line at 25.99~$\mu$m is a more effective coolant than the line at 17.93~$\mu$m. 
Four out of the seven sources with detected [Fe~II] lines (L1551~IRS5, DG~Tau, RW~Aur, IRAS~04239+2436, and IRAS~04303+2240) have previous indications of jet activity based on near-IR and/or optical observations (see Appendix~\ref{ind_objects}).
We have tested the correlation between the luminosity of the [Fe~II] line (25.99~$\mu$m) obtained in this study with different parameters: stellar mass, luminosity, X-ray luminosity, infrared luminosity, mass accretion rate, 
and luminosity of the [O~I] line (see Table~\ref{table:4}). 
In some cases (mass and luminosity) the results from the different correlation tests for L$_{\rm [Fe~II]}$ give largely different results. 
We remind that mostly upper limits are available for the [Fe~II] line; indeed, we have only 7 detections for 53 to 64 points, depending on the parameter tested.
Furthermore, most correlations were found with the Spearman method, but the Kendall and Cox methods are in general more robust and no correlations were found with these methods.
In summary, except for mass accretion rate, no correlation is found for the [Fe~II] line luminosity.

\begin{table*}
\caption{Correlations tested for the luminosity of the [Fe~II] line.
}
\label{table:4}
\centering
\begin{tabular}{lcccccc}
\hline\hline
\multicolumn{7}{c}{ \bf All infrared classes}\\
 & &\multicolumn{2}{c}{N$^\circ$ Upper L. }& \multicolumn{3}{c}{Probability } \\
\hline
Var. 	&	N$^\circ$ Points &Total  & [FeII] &Cox &	Kendall  &	Spearman \\
\hline
M$_{*}$		&	53	&	48	& 48	 & 0.55 (0.45)	 &	0.17 (1.3)	&	{\bf 0.001} (0.44)\\
L$_{*}$   	&	57	&	51	& 51	 & 0.77 (0.08) 	 &	0.14 (1.5)	&	{\bf0.001} (0.46)	\\
L$_{\rm X}$ &	47	&	43	& 40 & ...	 &	0.66 (0.44)	&	0.25 (0.17)	\\
L$_{\rm IR}$	&	64	&	57	& 57 & 0.81 (0.05)	 &	0.62 (0.49)	&	0.49 (0.09)	\\
$\dot{\rm M}_{\rm acc}$  &	47	&	42	 & 39 	&...	&	{\bf 0.01} (2.6)	&	{\bf 0.0004} (0.52)	\\
L$_{\rm[O I]}$   	&	27	&	24	& 24 &...	&	0.16 (1.4)	&	0.09 (0.33)	\\

\hline
\hline
\end{tabular}
\tablefoot{Probabilities (P) of a correlation achieved by chance are presented in columns 5, 6, and 7, where P~$\leq 5 \%$ are shown in bold fonts.
The value of the correlation parameters are shown between parentheses. 
}

\end{table*}

\subsection{H$_2$ emission}
\label{H2}

Three pure rotational lines of H$_2$ are covered by the \emph{Spitzer} IRS: 12.28~$\mu$m (S2), 17.03~$\mu$m (S1), and 28.22~$\mu$m (S0).
Detections of H$_2$ in the near-IR (S(1) transition at 2.12~$\mu$m) were reported in a handful of objects (e.g.,~\citealt{itoh:2003aa,bary:2002aa,bary:2008aa,carmona:2008aa,panic:2009aa}).
In addition to H$_2$ disk emission, H$_2$ emission may also originate in shock environments; in particular the near-IR lines and the mid-IR line at 12.28~$\mu$m are associated with shock densities of n$\sim10^{4-5}$~cm$^{-3}$ (\citealt{hollenbach:1989aa,hollenbach:1989ab}).

We have detected H$_2$ in six sources: one of the newly identified Taurus members SST~043905.2+233745, which is a Class~I source, the Class~II sources FS~Tau~A, FX~Tau, IQ~Tau, V710~Tau, and the Class~III HBC~366.  
FX~Tau and V710~Tau are both binary systems composed of a classical and a weak-lined T Tauri star.
These are the first detections of H$_2$ reported toward these sources. 

We have tested correlations between the total luminosity of the H$_2$ line with the following parameters: L$_*$, M$_*$, L$_{\rm IR}$, L$_{\rm X}$, \.M$_{\rm acc}$, and L$_{\rm [O~I]}$. 
The total luminosity of the H$_2$ line was calculated by adding the luminosities of the three lines accessible in the IRS wavelength range.
Our analysis does not reveal any correlation with the above parameters. Although correlations with H$_2$ were found for both the mid-IR and  [O~I] luminosities, this was only
the case for one test method out of three. We prefer, thus, to conclude that no correlation was found.

\begin{table*}
\caption{Correlations tested for the luminosity of the H$_2$ line.}
\label{table:5}
\centering
\begin{tabular}{lcccccc}
\hline\hline

\multicolumn{7}{c}{ \bf All infrared classes}\\
 & &\multicolumn{2}{c}{N$^\circ$ Upper L. }& \multicolumn{3}{c}{Probability } \\
\hline
Var. 	&	N$^\circ$ Points &Total  & H$_2$ &Cox &	Kendall  &	Spearman \\
\hline

M$_*$					&	53	&	48	& 48    & 0.97 (0.001)	 &	0.79 (-0.26) 	&	0.57 (-0.078) 	\\
L$_*$	        			&	57	&	52	& 52    & 0.17 (1.9)	 &	0.61	 (-0.51)    &	0.41 (-0.11)	 	\\
L$_{\rm X}$  			&	47	&	42	& 39 	& ...	 &	0.86 (-0.17) 	&	0.54 (-0.095)	 	\\
L$_{\rm IR}$  			&	64	&	58	& 58 	& {\bf0.03 (4.5)}	 &	0.09 (1.7)		&	0.09 (0.01)	\\
$\dot{\rm M}_{\rm acc}$  &	47	&   44 	& 40 	& ...	 &	0.22	 (-1.2)	&	0.43 (-0.15) 		\\
L$_{\rm[O I]}$   		&	27	&	26	& 26 	& ...	 &	0.99 (0.00)	&	{\bf 0.04} (0.40)	\\

\hline
\hline
\end{tabular}
\tablefoot{Probabilities (P) of a correlation achieved by chance are presented in columns 5, 6, and 7, where P~$\leq 5 \%$ are shown in bold fonts.
The value of the correlation parameters are shown between parentheses. 
}

\end{table*}

\begin{figure*}[!h]
   \centering
   	\begin{minipage}[!b]{1.\textwidth}
		\centering
			$\begin{array}{cc}
		    \includegraphics[height=6.2cm]{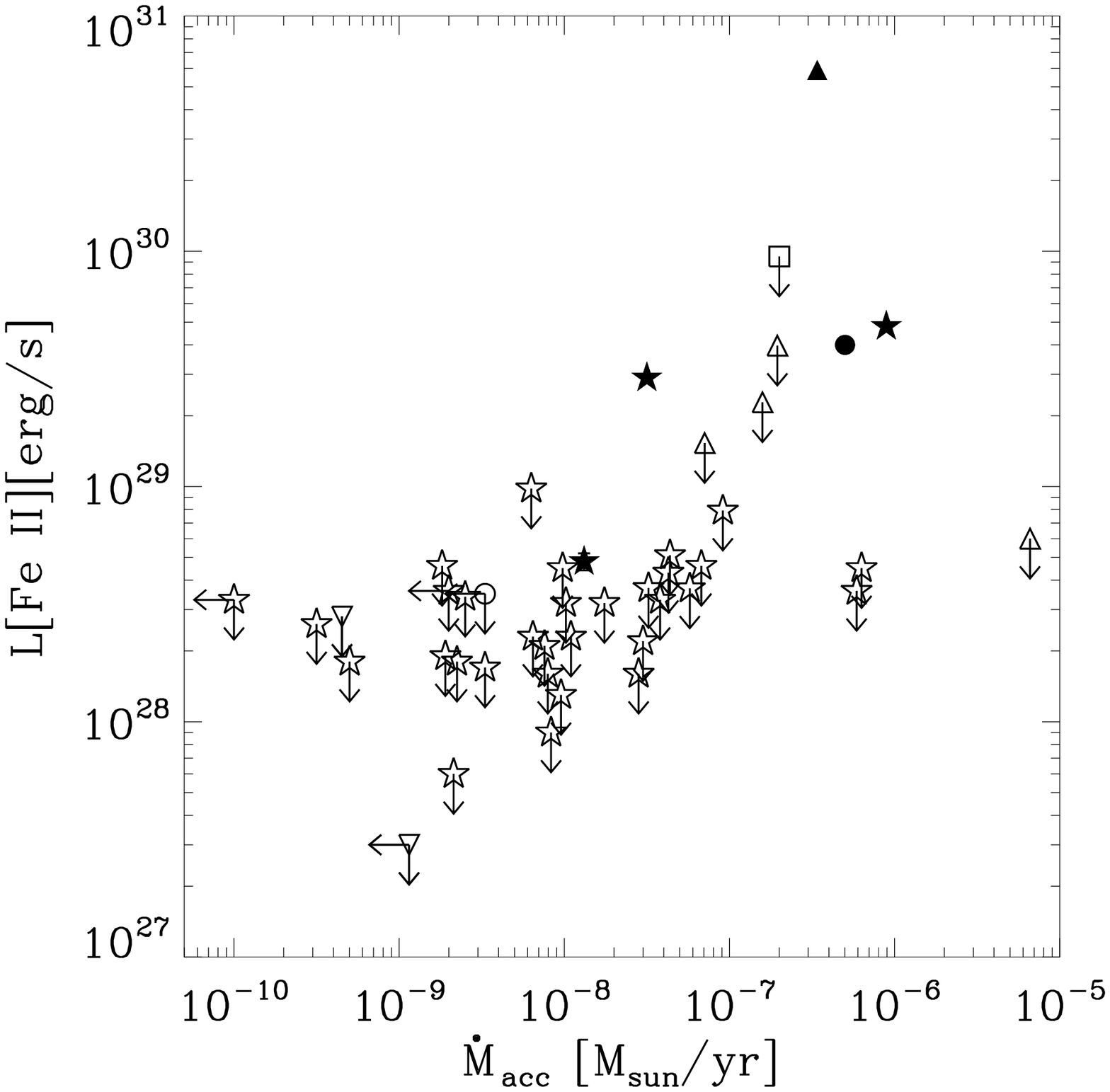} &
		    \includegraphics[height=6.2cm]{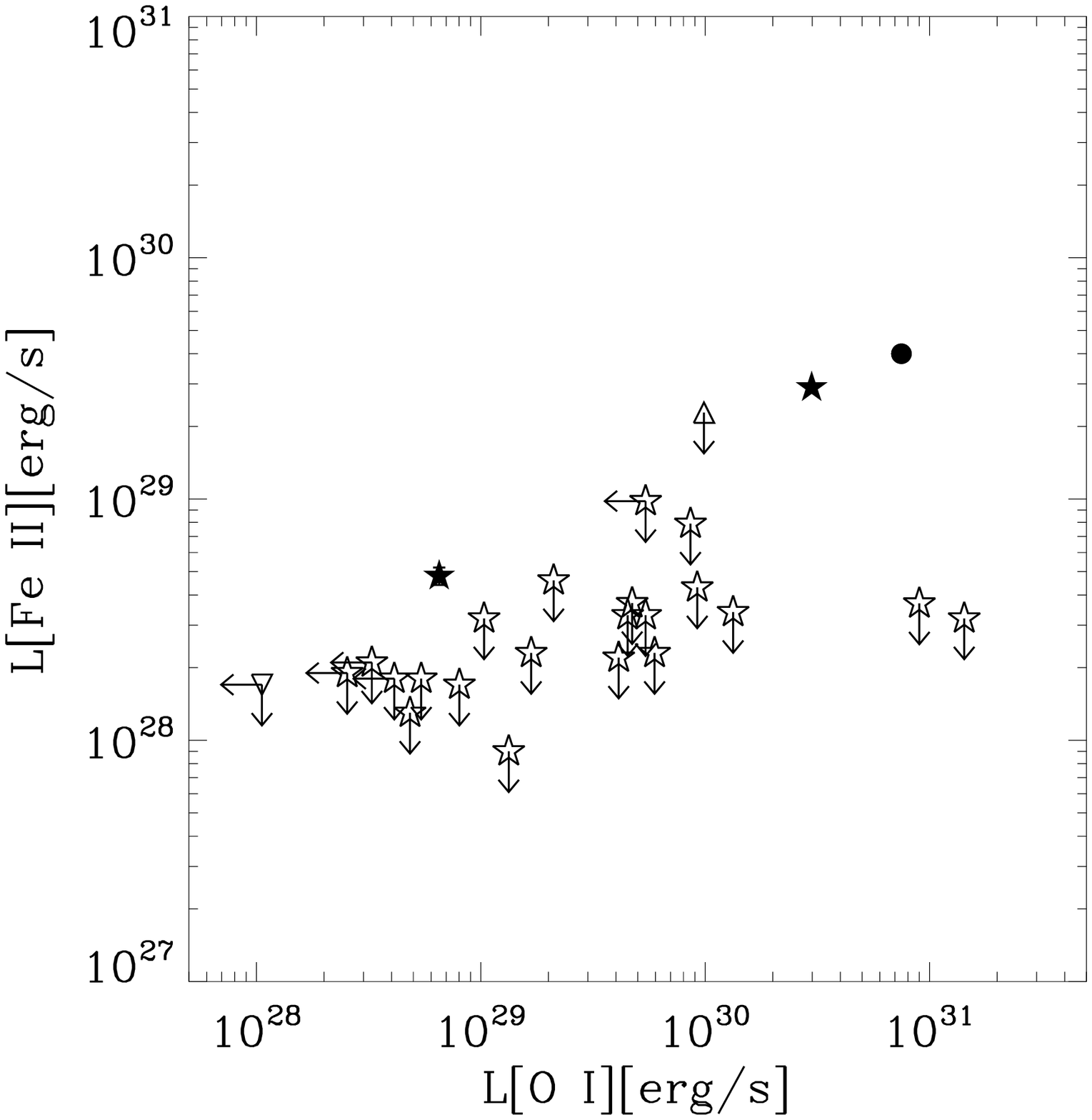}
		\end{array}$

	\end{minipage}
     \caption{ [Fe~II] luminosity versus mass accretion rate and luminosity of the [O~I] line. The symbol convention is the same used in Fig.\ref{Ne2corr}}
     \label{Fe2corr}
 \end{figure*}

\section{Summary and Conclusions}
\label{conclusions}

We have searched for gas in a large sample of pre-main sequence stars (64 objects) spanning a broad range of evolutionary stages. 
Our sample is dominated by Class~II objects but it also includes Class~I, III, and Herbig AeBe stars.
We have detected 36 gas emission lines; in particular, we have obtained 18 detections of the [Ne~II] line.
Nine out of the 18 [Ne~II] detections are reported here for the first time. 
For 4 targets there are previously reported detections of [Fe~II] based on near-IR observations, which we complement with the mid-IR information.
The six sources in which we detect H$_2$ emission do not have previously reported detections.

The [Ne~II] detections are concentrated towards Class~II sources; we have not detected [Ne~II] lines towards Class~III or Herbig AeBe stars.
Two Class~I sources show [Ne~II] line emission. 
These sources have higher accretion rates, and are more likely to show jet/outflow phenomena, but they also are more embedded in the circumstellar material. 
If neon line emission is produced by jets, we would expect to have a larger detection rate for Class~I objects.
We have attempted to disentangle sources with and without jets, but we cannot statistically separate the two populations.

We have tested correlations between the [Ne~II] line with parameters probing different stellar-disk properties; high-energy photon irradiation, accretion, and jets/outflows. 
We have found that the luminosity of the [Ne~II] line is dependent on the continuum mid-IR luminosity.
The dependence with L$_{\rm X}$ obtained in \citet{gudel:2010aa} is confirmed for the sub-sample of Class~II sources.
A good correlation is also found for the L$_{\rm [O~I]}$, echoing the finding by \citet{gudel:2010aa} and clearly supporting a trend with jet parameters.

In the jet/outflow scenario proposed by \citet{hollenbach:1989ab}, the [Ne~II] line is proposed as an effective coolant.
For high density shocks (n$\sim10^{5-6}$~cm$^{-3}$), the [Fe~II] line (25.99~$\mu$m) is more effective, with expected luminosities up to 2 orders of magnitude higher.
Only in lower density environments (n$\sim10^{3-4}$~cm$^{-3}$) and high velocities ($>80$~km s$^{-1}$) the luminosity of the [Ne~II] surpasses the luminosity of [Fe~II].
Among the sources with [Ne~II] detections, three also show [Fe~II] emission: DG~Tau, L1551~IRS~5 (both known to drive jets), and SST~041412+289837 (for which little information is available in the literature).
In these cases, a plausible explanation for the [Ne~II] emission is outflow activity.
The question arises for sources with [Ne~II] detections that do not show [Fe~II] emission:  
if [Ne~II] is excited in shocks produced by outflows, what is the explanation for the absence of [Fe~II] emission? 
This might be due to different reasons; perhaps lower signal-to-noise was obtained in the [Fe~II] wavelength range, or the conditions in the emitting region are such that the neon line is brighter, 
or simply for those sources, neon is not excited by shocks, but its line emission originates, e.g., from the disk atmosphere.

The [Fe~II] emission line has been detected in a smaller number of sources, in comparison to the [Ne~II] line.
The statistical analysis has shown that the luminosity of [Fe~II] is dependent on the mass accretion rate.
This indicates a link with outflow phenomena, since the outflow rate is proportional to the mass accretion rate. 
On the other hand, no correlation between  the [Fe II] luminosity and the luminosity of the outflow-tracing [O~I] (6300 \AA) line was found.

Furthermore, \emph{Spitzer}~IRS can access three H$_2$ emission lines. 
We have obtained few detections, and, despite the large number of upper limits, we have tested different correlations. 
Our results do not show any dependency on the parameters tested.

Our results show a trend towards accretion-related phenomena as the origin for the forbidden lines. 
Several mechanisms could be playing a role in the heating and excitation mechanism of gas in protoplanetary disks.
A sample with several Class~I, and Class~III objects would be more suitable to study possible trends with evolutionary state.
But no matter what mechanism produces the neon emission, it is likely originating very close to the star.
High spatial and spectral resolution observations are crucial in order to determine its origin.


\begin{acknowledgements}
This work is based on observations made with the \emph{Spitzer} Space Telescope, which is operated by the Jet Propulsion Laboratory, California Institute of Technology under a contract with NASA.
This research has made use of the SIMBAD database, operated at CDS, Strasbourg, France.
C. B-S., M.~A., and A.~C. acknowledge support from the Swiss National Science Foundation (grants PP002-110504 and PP00P2-130188).
We thank Fred Lahuis for useful discussion on the data reduction, Maria Suveges for insightful discussion on bootstrap techniques, and an anonymous referee for suggestions and comments that improved the manuscript.

\end{acknowledgements}

\bibliographystyle{aa}
\bibliography{15622}

\clearpage
\onecolumn

\Online

\setcounter{table}{0}
\begin{center}
\tablefirsthead{%
\hline
\hline
Source &	 \small{Class} & \small{Mass}& L & \small{EW(H$\alpha$)} &  $\log$\tiny{(\.{M}/M$_\odot$ yr$^{-1}$)} & L$_{\rm X}$ & L$_{\rm mid-IR}$ &log(L$\rm{_{[O~I]}}$/L$_\odot$) & Jet\tablefootmark{a} & Ref. \\
			&				& (M$_\odot$) 	& (L$_{\odot}$) & (\AA) &    & \tiny{($10^{30}$ erg/s)} & \tiny{($10^{32}$ erg/s)} & & \\ 
\hline}
\tablehead{%
\hline
\hline
Source &	 \small{Class} & \small{Mass}& L & EW(H$\alpha$) & $\log$\tiny{(\.{M}/M$_\odot$ yr$^{-1}$)}  & L$_{\rm X}$ & L$_{\rm mid-IR}$ & log(L$\rm{_{[O~I]}}$/L$_\odot$) & Jet\tablefootmark{a} & Ref. \\
			&				& (M$_\odot$) 	& (L$_{\odot}$) & (\AA) &    & \tiny{($10^{30}$ erg/s)} & \tiny{($10^{32}$ erg/s)} & & \\ 
\hline}
\topcaption{Stellar properties used in this study.}
\label{prop}
\begin{supertabular}{p{1.25in}p{0.15in}p{0.2in}p{0.2in}p{0.4in}p{0.77in}p{0.54in}p{0.44in}p{0.7in}p{0.2in}p{0.2in}}

\object{AA Tau} &II     & 0.76   & 0.80 &    37 &  -8.48	&    1.24 & 1.27 &-4.68 & $\surd$ &1,22,25\\
\object{BP Tau} &II     & 0.75   & 0.95 &    40 &  -7.88	&    1.36 & 1.80 &-4.77 & $\times$ &1,22\\
\object{CW Tau} &II     & 1.40   & 1.10 &   135 &  -7.99	&    2.84 & 4.62 &-2.43 & $\surd$ &1,23,26\\
\object{CZ Tau (A,B)}    &III    & 0.32   & 0.27 &     4 &  -9.35	&    2.24 & 3.11 &... & $\times$ &1,2\\
\object{DF Tau (A,B)}    &II     & 0.34   & 1.77 &    54 &  -7.96	&     ... & 2.62 &-3.81 & $\surd$ &2,3,26\\
\object{DG Tau} &I/II   & 0.91   & 1.70 &    63 &  -6.13	&    0.55 & 15.8 &-2.71 & $\surd$ & 1,4,7,22,27\\
\object{DG Tau B}  &I/II   & 0.72   & 5.50 &   276 &   ...     &    9.69 &	5.09 &... & $\surd$ & 1,31\\
\object{DK Tau} &II     & 0.74   & 1.30 &    40 &  -7.42	&    0.92 & 3.93 &-3.85 & $\times$ &1,22\\
\object{DL Tau} &II     & 0.75   & 0.85 &   105 &  -7.17	&    ...    & 2.87 &-4.26 & $\surd$ &1,2,5,25\\
\object{DM Tau} &II     & 0.47   & 0.30 &   139 &  -8.67	&    3.18 &  0.30 &... & $\times$ &1,5\\
\object{DN Tau} &II     & 0.56   & 1.00 &    20 &  -8.72	&    1.15 & 1.19 &$<-5.18$& $\times$ &1,2,22\\
\object{DO Tau} &II     & 0.70   & 1.05 &   100 &  -7.52	&    0.18 & 7.56 &-3.97 & $\surd$ &3,6,7,22,25\\
\object{DR Tau} &II     & 0.75   & 0.96 &   106 &  -7.49	&    ...  & 7.41 &-3.91& $\surd$ &2,3,6,22,25\\
\object{FS Tau A} &II     & 0.53   & 0.15 &    81 &  -9.50	&    3.22 & 3.24 &... & $\times$ &1,6\\
\object{FV Tau (A,B)}    &II     & 0.88   & 0.98 &    50 &  -6.23	&    0.53 & 5.93 &... & $\times$ &1\\
\object{FX Tau} & II     & 0.47   & 1.02 &    10-15 &  -8.65	&    0.50 & 1.27 &-4.85 & $\times$ &1,22\\
\object{FZ Tau} &II	   & 0.57   & 0.98 &	 181 & 	-7.69	& 	 0.64 & 2.81 &... & $\times$ &1,2\\ 
\object{GG Tau} &II     & 0.44   & 0.98 &    30 &  -7.76	&    ...  & 2.50 &-4.57 & $\times$ &3\\
\object{GI Tau} &II    & 0.75   & 1.00 &    20 &  -8.08	&    0.83 & 3.57 &-4.46 & $\times$ &1,23\\
\object{GK Tau (A,B)}    &II     & 0.74   & 1.40 &    25 &  -8.19	&    1.47 & 3.80 &-4.36 & $\times$ &1,22\\
\object{GV Tau (A,B) }   &I      & 0.68   & 1.82 &    86 &  -6.71	&    0.65 & 64.7 &... & $\times$ &1\\
\object{HL Tau} &I      & 1.20   & 1.53  &   49 &  -6.80	&    3.84 & 29.7 &-3.59 & $\surd$ &1,22,25\\
\object{HN Tau} &II     & 1.10   & 0.25  &  150 &  -8.60	&    0.15 & 2.67 &-3.46 & $\surd$ &1,2,23,25\\
\object{HP Tau} &II     & 1.39   & 1.40  &   28 &  ...      &    2.55 & 3.63 &... & $\surd$ &1,27\\
\object{HQ Tau} &III    &  ...   & ...  &    2 &  ...     	&    8.16 & 3.75 &... & $\times$ &1,8  \\
\object{IQ Tau} &II     & 0.51   & 0.88  &   13 &  -7.55	&    0.42 & 1.43 &... & $\times$ &1\\
\object{RY Tau} &II     & 2.37   & 7.60  &   13 &  -7.04	&    5.52 & 35.8 &-3.65 & $\surd$ &1,22\\
\object{T Tau} &	II & 2.41   & 8.90  &   41 &  -7.24    &    8.05 & 59.5 &-2.63 & $\surd$ &1,22,27 \\
\object{UX Tau A} &II    & 1.10    & 1.81  &  9.5 &  -9.30	&    0.95 & 1.27 &$<-4.97$& $\times$ &2,6,9,22\\
\object{UZ Tau} &II     & 0.47   & 0.49  &   73 &  -8.01	&    0.89 & 3.98 &... & $\times$ &1\\
\object{VY Tau} &III    & 0.57   & 0.55  &  4.9 &    ...    &    1.29 & 0.38  &$<-5.56 $& $\times$ &2,7,10,22\\
\object{XZ Tau (A,B)}    &II     & 0.34   &  0.33  &  62 &  -8.74	&    0.96 & 13.1 &... & $\surd$ &1,2,25\\
\object{Coku Tau-1 }  &II     & 0.70   &  0.15  &  70 &  -7.36	&   ...  & 3.90 &... & $\surd$ &1,31\\
\object{Coku Tau-3}   &II     & 0.46   &  0.98  &   5 &  ...	&    5.85 & 1.31 &... & $\times$ &1\\
\object{Coku Tau-4}   &II     & 0.43   &  0.68  & 1.8 &  ...  	&$< 0.21$ & 0.70 &... &  $\times$ &2,9,10\\
\object{AB Aur}     &HAe&  2.40  &  49.00   & 33 &  -6.85	&    0.35 & 64.6 &-3.36 & $\surd$ &1,11,28\\
\object{GM Aur}     &II   &  0.52  &  0.74   & 96 &  -8.02	&    1.68 & 0.68 &-4.90 &  $\times$ & 3,5,7,22\\
\object{RW Aur A}   &II &  1.34  &  3.37  & 76 &  -7.50	& 1.60 & 4.70 &-3.11 &  $\surd$ &4,23,25,32\\
\object{SU Aur }    &II   &  1.91  & 9.90    & 63 &  -8.20	&    9.46 & 15.0 &$<-3.85$ &  $\times$ &1,22\\
\object{UY Aur (A,B) } &II   &  0.57  &  1.25  & 36 &  -7.37	& $< 0.25$ & 10.6 &-3.62 &   $\surd$ &2,9,12,23,25\\
\object{Haro 6-28}   &II & 0.21   &  0.12   & 48-92 &  -8.10	&    0.25 & 0.47 &... & $\times$ &1,12\\
\object{Haro 6-37}   &II   & 0.91   & 1.74   &  22 &  -8.12	&  ...       &  2.35 &$<-5.07$& $\times$ &2,6,22\\
\object{L1551 IRS5}        &I    &  1.58  & 2.60  &   83 &  -6.47	&  0.021 &  62.0 &... &  $\surd$ &1,13,14,30\\
\object{MHO-1/2}       &I/II &  0.34  & 0.42 &    58 &  $<-8.48$	&  1.64 & 7.20 &... &  $\times$ &1\\
\object{V710 Tau}     &II   &  0.68  &  1.10  &  69 &  ...		&    1.38 & 0.75 &... &  $\times$& 1,6\\
\object{V773 Tau }    &II   &  0.58  &  5.60  &  2  & $< -8.6$	&    9.49 & 4.46 &-3.93 &  $\times$ &1,6,22\\
\object{V892 Tau}    &HAe&   2.89 & 77.00   &  7-13  &  ...   &    9.21 & 132 &... & $\surd$ &1,29\\
\object{HBC 366}    &III  & 0.40   &  2.60 &  1.3 & $< -8.94$ &    4.14 & 0.23 &... & $\times$ &1,16\\
\object{HD 31648}    &HAe& 2.30   &  23.7 & 18.3 &  -8.39 & ... & 34.8 &... & $\times$ &17\\
\object{HD 35187 }   &HAe& 2.30   &  1.37 & 5.1 &  -7.81 	& ... & 18.5 &... & $\surd$ &11,18 \\
\object{HD 36112}    &HAe&  2.20  &  1.45 & 14.5 &  -6.70	& ... & 17.4 &... &  $\times$ &18,19\\
\object{IC2087IR}      &II   & ... & ... &  22 &  -6.20	&    4.67 & 16.3 &... &  $\times$ &1\\
\object{IRAS 04016+2610 }   &I    &  1.45  & 3.70 &   41-56 &  -7.15 & 4.47 & 13.6 & ... &  $\surd$ &1,21,27\\
\object{IRAS 04101+3103}         &HAe& ... & ... & 11 &  ... & ... & 10.7 &... & $\times$ &15\\
\object{IRAS 04108+2803 B }      &I    &  0.36 & 0.4 &... & ...  	&    0.42 & 2.46 &... &  $\surd$ &1,27\\
\object{IRAS 04239+2436}         &I    & ... & 1.27 &... & ...	& ... & 8.76 &... &  $\surd$ &13,27 \\
\object{IRAS 04278+2253}         &I    & ... & 7.20  &  17 &  -5.18	& ... & 28.6 &...  & $\times$ &4,20,21 \\
\object{IRAS 04299+2915}     & II    &  ...  &...&  6.5 & ...	&  ...  & 0.47 &... & $\times$ &15\\
\object{IRAS 04303+2240 }       &II   &  0.51  & 2.20  &    67 &  -6.05	&    5.01 & 6.43 &...  & $\times$ &1\\
\object{IRAS 04361+2547}        &I    &... & 3.70 &... &...	&$< 0.30$ & 8.28 &... & $\surd$ &1,9,27  \\
\object{IRAS 04365+2535}      &I    &... & 2.20 &  ... &... &    1.99 & 6.13 &... & $\surd$ &1,27  \\
\object{SST~041412.2+280837}		& I & ... &... & ... & ... & ... & 1.73 &... & $\times$ & 33  \\
\object{SST~042936.0+243555}		& II & ... & ... & ... & ... & ... & 0.32 &... & $\times$ & 24,33 \\
\object{SST~043905.2+233745}		& I/II & ... & ... & ... & ... & ... &  0.23 &... & $\times$ & 24,33 \\	
\hline
\hline 
\end{supertabular}

\tablefoottext{a}{$\surd$ indicates presence of jet, $\times$ indicates that no jet has been reported.}\\
\tablefoottext{b}{``...'' indicates no information available.}
\tablebib{{\bf[1]}\citet{gudel:2007aa}; {\bf[2]}~\citet{watson:2009aa}; {\bf[3]} \citet{gullbring:1998aa}; {\bf[4]} \citet{white:2004aa}; {\bf[5]} \citet{herbig:1995aa}; {\bf[6]} \citet{white:2001aa}; {\bf[7]} G\"udel private communication 2009; {\bf[8]} \citet{nguyen:2009aa}; {\bf[9]} \citet{neuhaeuser:1995aa}; {\bf[10]} \citet{cohen:1979aa}; {\bf[11]} \citet{garcia-lopez:2006aa}; {\bf[12]} compiled in \citet{mccabe:2006aa}; {\bf[13]} \citet{prato:2009aa}; {\bf[14]} \citet{kenyon:1998aa}; {\bf[15]} \citet{kenyon:1990aa}; {\bf[16]} compiled in \citet{martin:1994aa}; {\bf[17]} \citet{mannings:1997aa}; {\bf[18]} \citet{manoj:2006aa}; {\bf[19]} \citet{isella:2008aa}; {\bf[20]} \citet{bitner:2008aa}; {\bf[21]} \citet{furlan:2008aa}, {\bf[22]} \citet{cabrit:1990aa}, {\bf[23]} \citet{hartigan:1995aa}, {\bf[24]} \citet{luhman:2010aa}; {\bf[25]} \citet{hirth:1997aa}; {\bf[26]} \citet{hartigan:2004aa}; {\bf[27]} \citet{moriarty-schieven:1992aa}, {\bf[28]} \citet{corcoran:1997aa}, {\bf[29]} \citet{maheswar:2002aa}, {\bf[30]} \citet{davis:2003aa},  {\bf[31]} \citet{eisloffel:1998aa}, {\bf[32]} \citet{gudel:2010aa}, {\bf[33]} \citet{rebull:2010aa}}
\end{center}

\setcounter{table}{2}
\addtocounter{table}{-1}
\begin{table*}
\caption{Line luminosities derived in this study.}
\centering
\begin{tabular}{lcccccccccc}

\hline\hline

		&	H$_2$	& [Ne~II] & [Ne~III] &  H$_2$ & [Fe~II] & [S~III] & [Fe~II] & H$_2$ & [S~III] \\
	Name 	&	12.2786 & 12.8135 & 15.5551 & 17.0348 & 17.9359  & 18.7130  & 25.9883 & 28.2188 & 33.4810\\	
\hline

\object{AA Tau} & 1.2 & {\bf 2.7 $\pm$ 0.3} & 2.2 & 2.8 & 1.7 & 2.2 & 1.7 & 1.9 & 1.9 \\
\object{BP Tau} & 2.1  & 1.0 & 1.3 & 3.5 & 2.3& 1.8 & {\bf4.8 $\pm$ 0.4}& 1.9 & 1.9\\
\object{CW Tau} & 11.2 & 4.9 & 9.5 & 8.2 & 10.9 & 4.9 & 3.2 & 6.2 & 8.8\\
\object{CZ Tau} (A,B)    & 4.2 & 3.8  & 3.0 & 3.9 & 6.5 & 5.4 & 2.8 & 3.2 & 3.6\\
\object{DF Tau} (A,B)    & 5.2 & 8.4 & 2.4 & 10.0 & 3.6 & 4.6 & 2.3  & 3.0 & 5.8 \\
\object{DG Tau} & 13.6 & {\bf 43.0 $\pm$ 4} & 9.5 & 9.8 & 10.3 & 10.6 & {\bf 40 $\pm$ 2} & 15.2 & 9.7 \\
\object{DG Tau} B  & 5.1 &{\bf 10 $\pm$ 2} & 12.3 & 11.6 & 8.5 & 10.1 & 6.1 & 3.8 & 12.9 \\
\object{DK Tau} & 5.7 & 5.6 & 6.7 & 7.9 & 6.5 & 2.7 & 3.3 & 3.1 & 7.4\\
\object{DL Tau} & 7.7 & 4.0 & 5.5 & 5.9 & 6.6 & 5.6 &4.7 & 2.0 & 8.9\\
\object{DM Tau} & 1.1 & {\bf 0.8 $\pm$ 0.2} & 1.2 & 0.8 & 1.2 & 1.0 & 0.6 & 1.5 & 0.3\\
\object{DN Tau} &1.3 & 1.2 & 0.8 & 0.9 & 0.8 & 0.9 & 1.9 & 1.4 & 2.0 \\
\object{DO Tau} & 4.6  & 4.5 & 3.7 & 4.3 & 3.2 & 4.6 & 2.2 & 3.9 & 8.0\\
\object{DR Tau} &12.6 & 6.5 & 12.9 & 18.0 & 8.8 & 8.8 & 3.7 & 3.3 & 14.2\\
\object{FS Tau} A &{\bf 10 $\pm$ 2}  & {\bf17 $\pm$ 2} & 4.9 & 5.2 & 7.9 & 7.0 & 2.6 & 4.1 & 4.0\\
\object{FV Tau} (A,B)    & 6.7 & 4.3  & 4.2 & 6.6 & 6.3 & 1.2 & 3.6 & 2.1 & 5.4\\
\object{FX Tau} & 1.5 & 1.1 & 3.8 & 2.5 & 3.3 & 5.3 & 1.8 & {\bf 2.3 $\pm$ 0.4} & 1.2 \\
\object{FZ Tau} & 8.2 & 4.7 & 5.2 & 4.7 & 8.0 & 7.0 & 2.6 & 6.6 & 5.8\\
\object{GG Tau} & 7.0 & 6.5 & 4.1 & 8.2 & 5.1 & 3.9 & 3.2 & 3.8 & 8.9\\
\object{GI Tau} & 2.0 & 2.0 & 2.9 & 3.2 & 2.8 & 2.1 & 0.9 & 2.2 & 3.1\\
\object{GK Tau} & 3.0 & 5.0 & 2.5 & 2.8 & 3.4 & 2.1 & 2.3 & 3.9 & 2.9\\
\object{GV Tau (A,B) }   & 50.5 & 40.0 & 41.6 & 21.4 & 76.4 & 30.2 & 39.7 & 56.8 & 51.6 \\
\object{HL Tau} & 40.0 & 14.1 & 14.5 & 19.8 & 14.7 & 13.9 & 22.8 & 17.1 & 35.2\\
\object{HN Tau} & 6.3 & 2.1 & 4.1 & 3.3 & 5.6 & 4.0 & 3.4 & 2.3 & 6.0\\
\object{HP Tau} & 6.8 & 4.6 & 4.8 & 3.9 & 8.8 & 6.5 & 5.0 & 2.6 & 6.9\\
\object{HQ Tau} & 5.4 & 5.2 & 6.4 & 5.9 & 10.7 & 3.1 & 4.0 & 3.0 & 8.4\\
\object{IQ Tau} & 1.5 & {\bf 1.2 $\pm$ 0.4} & 1.7 & 1.6 & 1.4 & 2.5 & 1.6 & {\bf 3.9 $\pm$ 0.6} & 1.7\\
\object{RY Tau} & 36.6 & 13.7 & 33.7 & 25.2 & 17.0 & 23.3 & 7.9 & 7.7 & 40.9\\
\object{T Tau(N,S)} & 49.5 & {\bf 400 $\pm$ 9} & 21.9 & 54.6 & 31.2 & 25.4 & 36.9 & 109 & 119 \\
\object{UX Tau} A  & 6.9 & 5.6 & 2.5 & 2.9 & 7.5 & 7.4 & 1.8 & 1.8 & 1.3\\
\object{UZ Tau} & 9.1 & 5.1  & 4.5 & 7.3 & 8.6 & 9.1 & 4.5 & 9.7 & 8.8\\
\object{VY Tau }& 0.3 & 0.4 & 0.3 & 0.2 & 0.5 & 0.5 & 1.7 & 1.3 & 1.7\\
\object{XZ Tau (A,B) }   & 9.5 & {\bf 7.2 $\pm$ 0.9} & 5.4 & 9.2 & 15.9 & 9.9 & 4.6 & 9.6 & 9.6\\
\object{Coku Tau-1 }  & 4.1 & {\bf 27 $\pm$ 1} & 3.9 & 3.7 & 5.7 & 5.2 & 5.1 & 7.1 & 6.0 \\
\object{Coku Tau-3 } & 1.5 & 1.5 & 0.5 & 0.8 & 1.5 & 1.3 & 1.9 & 1.6 & 1.3 \\
\object{Coku Tau-4 } & 1.3 & {\bf 3.2 $\pm$ 0.2} & 0.8 & 0.9 & 0.9 & 1.6 & 1.4 & 1.1 & 2.1  \\
\object{AB Aur }   & 29.7 & 16.8 & 18.5 & 25.1 & 18.2 & 34.9 & 44.3 & 25.7 & 51.8\\
\object{GM Aur }   & 2.6 & {\bf 1.5 $\pm$ 0.3} & 1.0 & 1.0 & 1.4 & 2.2 & 1.3 & 1.1 & 4.0\\
\object{RW Aur }  & 14.6 & 3.8 & 7.9 & 12.9  & {\bf 10 $\pm$ 2} & 6.6 & {\bf29 $\pm$ 1} & 8.1 & 5.0\\
\object{SU Aur}    & 25.2 & 6.5 & 6.8 & 7.8 & 18.6 & 17.4 & 9.8 & 11.2 & 10.3 \\
\object{UY Aur (A,B)}    & 14.1 & {\bf 11 $\pm$ 2} & 9.9 & 8.7 & 10.0 & 5.4 & 4.3 & 8.3 & 13 \\
\object{Haro 6-28}   & 4.4 & 4.4 & 10.5 & 3.7 & 3.7 & 5.5 & 1.6 & 1.4 & 3.9\\
\object{Haro 6-37}   & 5.4 & 4.0 & 2.2 & 4.9 & 5.5 & 5.7 & 2.1 & 3.0 & 4.0\\
\object{L1551 IRS5}      & 30.4 & {\bf 129 $\pm$ 9} & 75.8 & 23.3 &{\bf 200 $\pm$ 19}  & 49.8 &{\bf 597 $\pm$22 } & 20.1 & 13.1\\
\object{MHO-1/2}   & 8.4 & {\bf 10 $\pm$ 2}& 4.7 & 6.9 & 19.7 & 12.8 & 3.5 & 7.7 & 11.3\\
\object{V710 Tau A }  & 0.9 & 0.4 & 0.6 & {\bf 1.0 $\pm$ 0.3} & 0.5 & 1.1 & 1.0 & {\bf 2.1$\pm$ 0.4} & 3.5 \\
\object{V773 Tau }   & 3.6 & {\bf 34 $\pm$ 2} & 4.7 & 8.9 & 6.8 & 4.1 & 3.3 & 3.6 & 6.4\\
\object{V892 Tau}    & 100.4 & 40.1 & 141.4 & 53.7 & 79.9 & 56.9 & 84.3 & 64.3 & 166.8\\
\object{HBC 366}   & 0.4 & 0.2 & 0.1 & {\bf 0.5 $\pm$ 0.1} & 0.3 & 0.2 &   0.3 & 0.5 & 1.0\\
\object{HD 31648}    & 45.3 & 7.5 & 9.9 & 15.4 & 35.1 & 15.5 & 7.1 & 5.3 & 10.4 \\
\object{HD 35187}    & 11.6 & 11.4 & 10.8 & 9.7 & 16.9 & 13.5 & 8.4 & 8.3 & 11.1\\
\object{HD 36112}    & 21.0 & 11.1 & 11.2 & 7.7 & 15.8 & 18.3 & 9.5 & 5.4 & 1.8\\
\object{IC2087IR}   & 20.9 & 11.3 & 16.3 & 11.8 & 7.8 & 7.2 & 4.5 & 6.4 & 5.9\\
\object{IRAS 04016+2610}      & 9.7 & 5.6 & 15.9 & 9.1 & 8.5 & 11.4 & 15.3 & 8.0 & 21.1\\ 
\object{IRAS 04101+3103 }     & 4.8 & 6.3 & 6.1 & 3.8 & 9.5 & 7.1 & 6.2 & 8.7 & 4.9\\ 
\object{IRAS 04108+2803 B}    & 3.1 & 3.9 & 4.6 & 4.6 & 6.5 & 4.5 & 8.0 & 6.6 & 10.0 \\ 
\object{IRAS 04239+2436}      & 4.6 & 4.4 & 6.0 & 7.3 & {\bf 24 $\pm$ 3} & 8.3 & {\bf 83 $\pm$ 3} & 16.4 & 38.8\\ 
\object{IRAS 04278+2253}      & 24.2 & 17.1 & 19.1 & 5.7 & 21.9 & 13.4 & 6.0 & 9.7 & 17.8 \\ 
\object{IRAS 04299+2915 }    & 3.6 & 3.2 & 3.3 & 3.6 & 4.6 & 2.0 & 2.9 & 2.7 & 6.4\\ 
\object{IRAS 04303+2240 }     & 14.8 & 8.2 & 11.5 & 13.9 & {\bf 21 $\pm$ 2} & 9.8 & {\bf 48 $\pm$ 1} & 16.9 & 11.1\\ 
\object{IRAS 04361+2547}      & 7.5 & 3.0 & 7.8 & 7.0 & 8.9 & 6.6 & 6.8 & 34.3 & 26.4\\ 
\object{IRAS 04365+2535}      & 9.4 & 4.6 & 8.4 & 3.7 & 10.5 & 6.5 & 11.9 & 12.5 & 19.0\\ 
\object{SST~041412.2+280837}		& 4.0 & {\bf 7.5 $\pm$ 0.6} & 7.1 & 3.2 & {\bf 5.3 $\pm$ 0.7} & 2.8 & {\bf 3.5 $\pm$ 0.5} & 4.7 & 6.6\\
\object{SST~042936.0+243555}		& 0.9 & {\bf 2.5 $\pm$ 0.2} & 0.7 & 0.8 & 0.5 & 0.8 & 0.6 & 0.8 & 1.0\\
\object{SST~043905.2+233745}		& {\bf 0.8$\pm$0.2} & {\bf 1.6 $\pm$ 0.1} & 0.4 & {\bf 1.2 $\pm$ 0.1} & 0.8 & 0.09 & 0.5 & 1.9 & 2.6\\
\hline
\hline
\end{tabular}
\tablefoot{Detections plus 1$\sigma$ errors are in bold characters, 3$\sigma$ upper limits are presented in normal font. All values are given in units of $10^{28}$ erg s$^{-1}$. The wavelength of the emission lines are given in $\mu$m.}
\label{linelum}
\end{table*}

\twocolumn

\begin{appendix}

\section{Results for Individual Objects}
\label{ind_objects}

We start with \object{L1551~IRS~5} because it is a good example to illustrate the procedure used to study the line emission.
The rest of the objects are ordered as in Table~\ref{prop}. 

\noindent{\bf L1551 IRS 5} is a Class I protostar driving a molecular outflow. 
High angular resolution observations in the radio have revealed a binary with separation of 0$^{\prime\prime}$.3 \citep{bieging:1985aa}. 
One of the binary components is now known to be a binary itself with separation of 0$^{\prime\prime}$.09 \citep{lim:2006aa}.
We do not have any further information on the nature of the components of the system; \emph{Spitzer} observations do not resolve the components (its spatial resolution is $\sim 4^{\prime\prime}$ at 13~$\mu$m).

\begin{figure}[h]
	\centering
		\begin{minipage}[!b]{.5\textwidth}	
     		$\begin{array}{cc}
	    			\includegraphics[width=4.5cm]{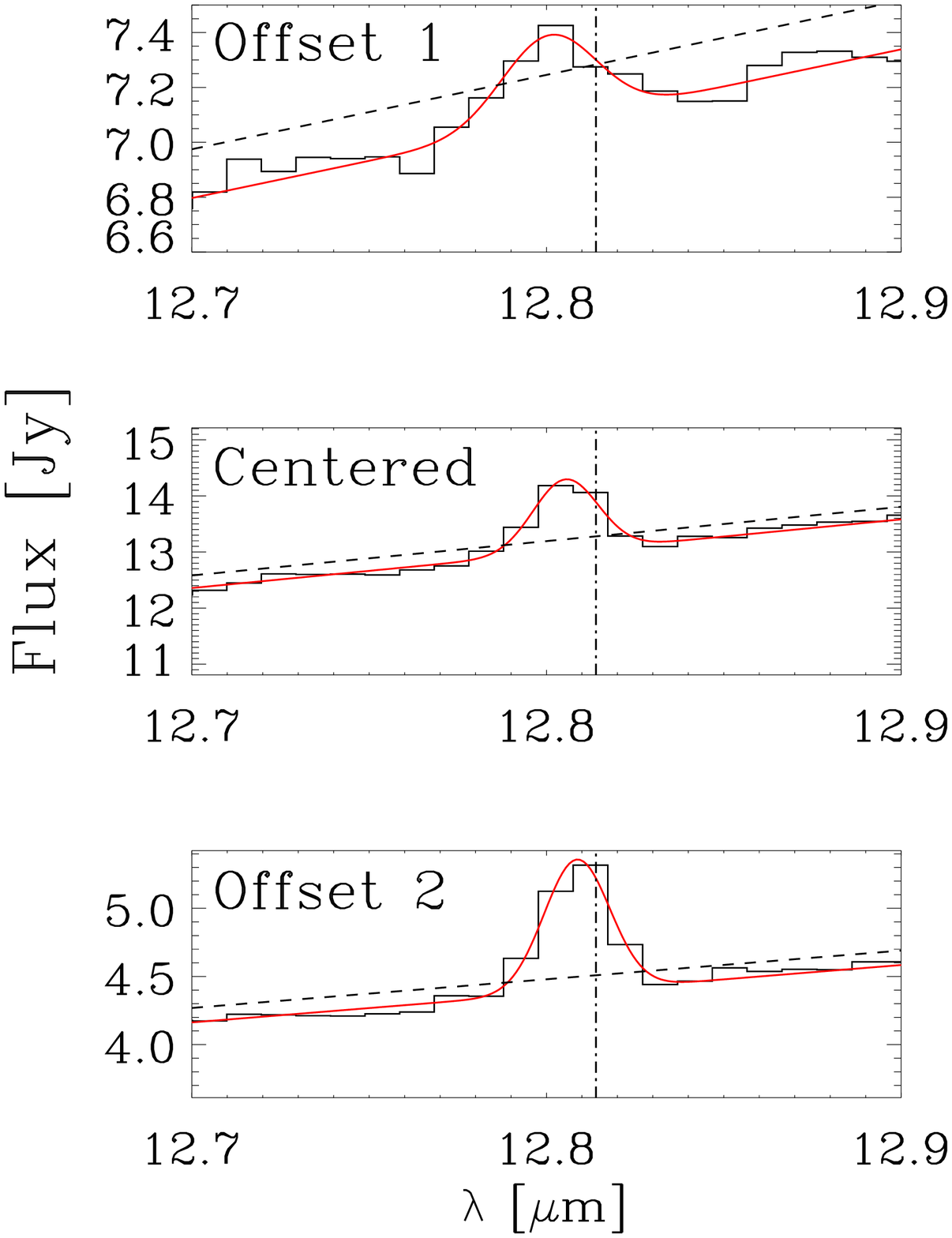} &
	    			\includegraphics[width=4.5cm]{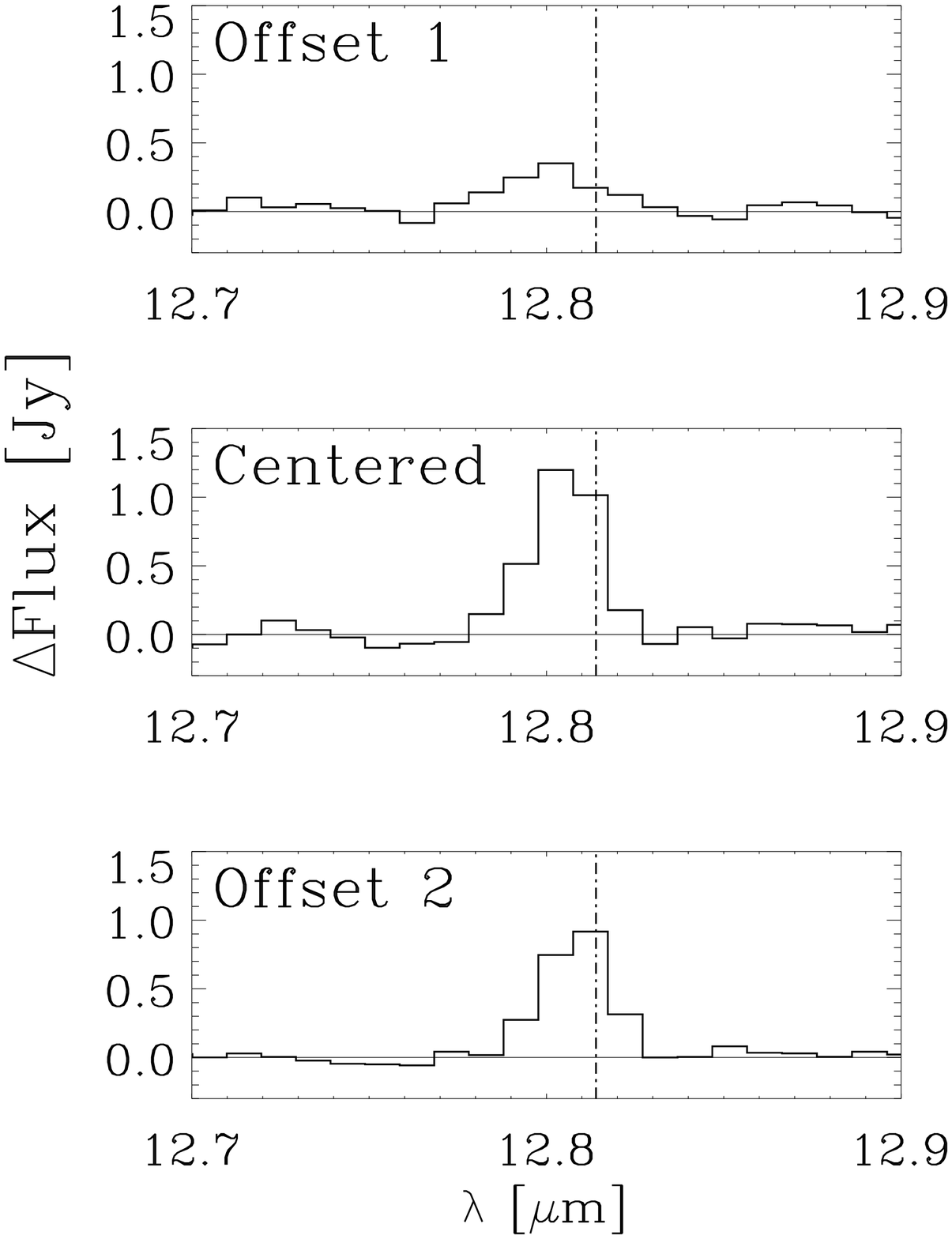}
	  		\end{array}$  
		\end{minipage}
     \caption{Spectrum of L1551 IRS 5 surrounding the position of the [Ne~II] line at 12.8~$\mu$m. The plots on the left panel show the average of 2 pointings at each position. We have labeled the positions as Offset 1, Centered, and Offset 2. The spectrum obtained for the position ``Centered" has the highest flux in the continuum so it was used for the line analysis. The expected position of the [Ne~II] line is overplotted with a black dash-dotted line, we note that the line is shifted. The right panel shows the difference between the line flux and the continuum flux in order to show the real strength of the line.}        				
     \label{L15point}
 \end{figure}

This target is part of a GTO program. 
The observations were done with 6 different pointings offset in spatial direction. 
The pointings correspond to three sets of standard nod observations. Therefore we could obtain three average spectra located at three positions; one centered at the star position, and the two others at the offset positions.
We can use these data to compare the continuum-subtracted flux at the different positions in order to have a rough estimate for the contribution of the extended emission, whenever the line is detected at multiple positions. 
In Fig.~\ref{L15point} we show the spectrum of L1551 IRS 5 around the position of the [Ne~II] line. 
Each plot in the left panel represents the average of the spectra between 2 nods at a given position. 
In this specific case, we have three different positions identified as ``Offset 1", ``Offset 2", and ``Centered". 
We have labeled each position in the upper left side of each plot. 
The maximum flux in the continuum is achieved for the ``Centered" position. 
The line is detected in all three sets of pointings and we point out that it is shifted by 0.1~$\mu$m with respect to the expected wavelength of the [Ne~II] line. Such a wavelength shift corresponds to a velocity shift of $\sim230$~km s$^{-1}$. 
This result is consistent with the velocity of the [Fe~II] jet (285~km s$^{-1}$) derived by \citet{davis:2003aa} from near-IR observations.
\citet{pyo:2009aa} have observed the [Fe~II] jet in the near-IR and derived a position angle (P.~A.) of $260^{\circ}$ for the northern component and $235^{\circ}$ for the southern component.
~\emph{Spitzer} observations have been obtained using a slit P.~A. of $-56.7 ^{\circ} $ and $-141.5 ^{\circ}$ for the SH and LH modules respectively.  

In the right panel of Fig. \ref{L15point} we have plotted the continuum-subtracted flux. In this way we could verify that although there is an extended component in the emission, the maximum,~i.~e., the highest line peak, corresponds to the position centered on the object. 
But we recall that extended emission close to the star can be present and not be resolved by \emph{Spitzer}.

\begin{figure}[h]
	\centering
		\begin{minipage}[!b]{.5\textwidth}	
     		$\begin{array}{cc}
	    			\includegraphics[width=4.4cm]{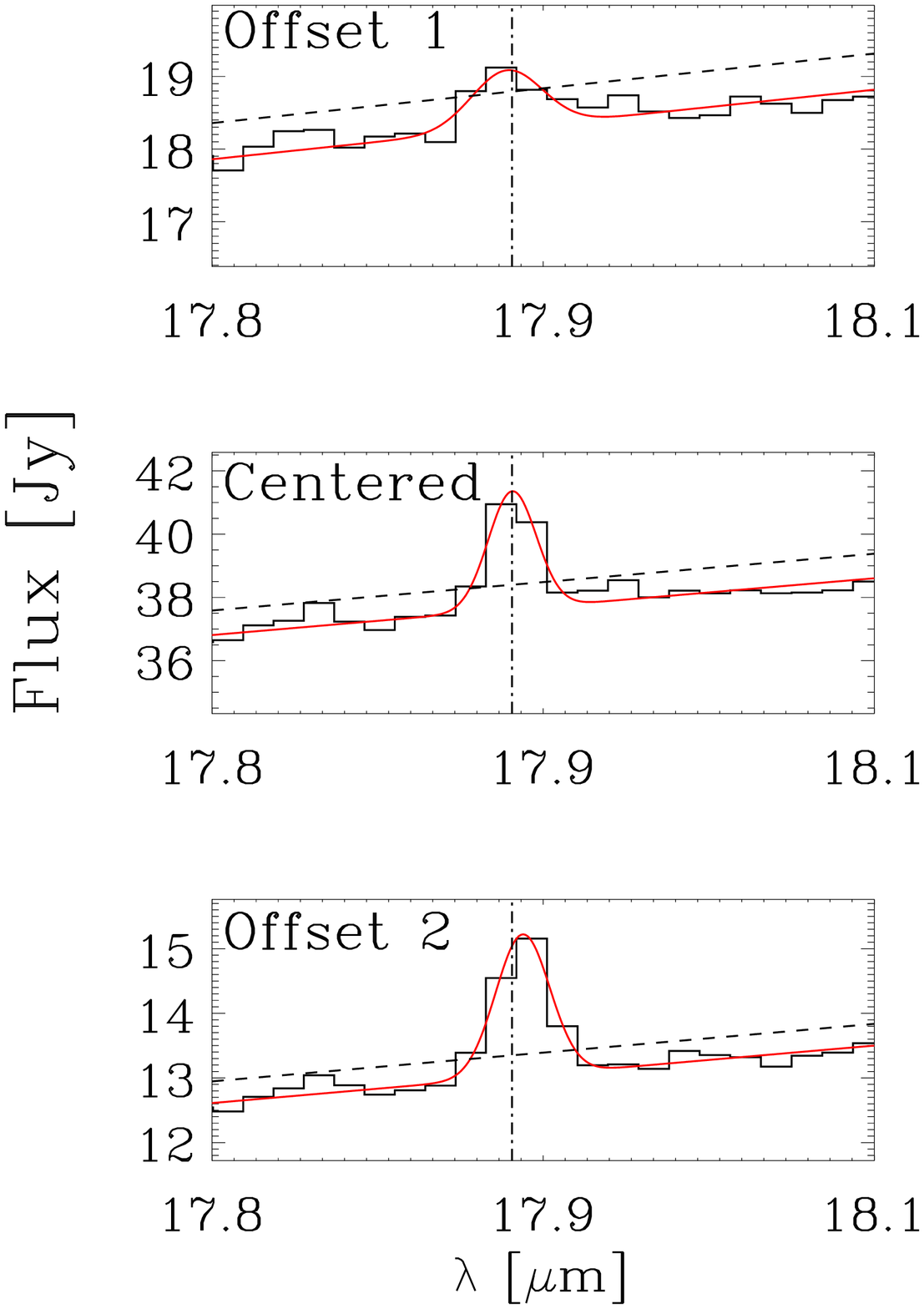} &
	    			\includegraphics[width=4.4cm]{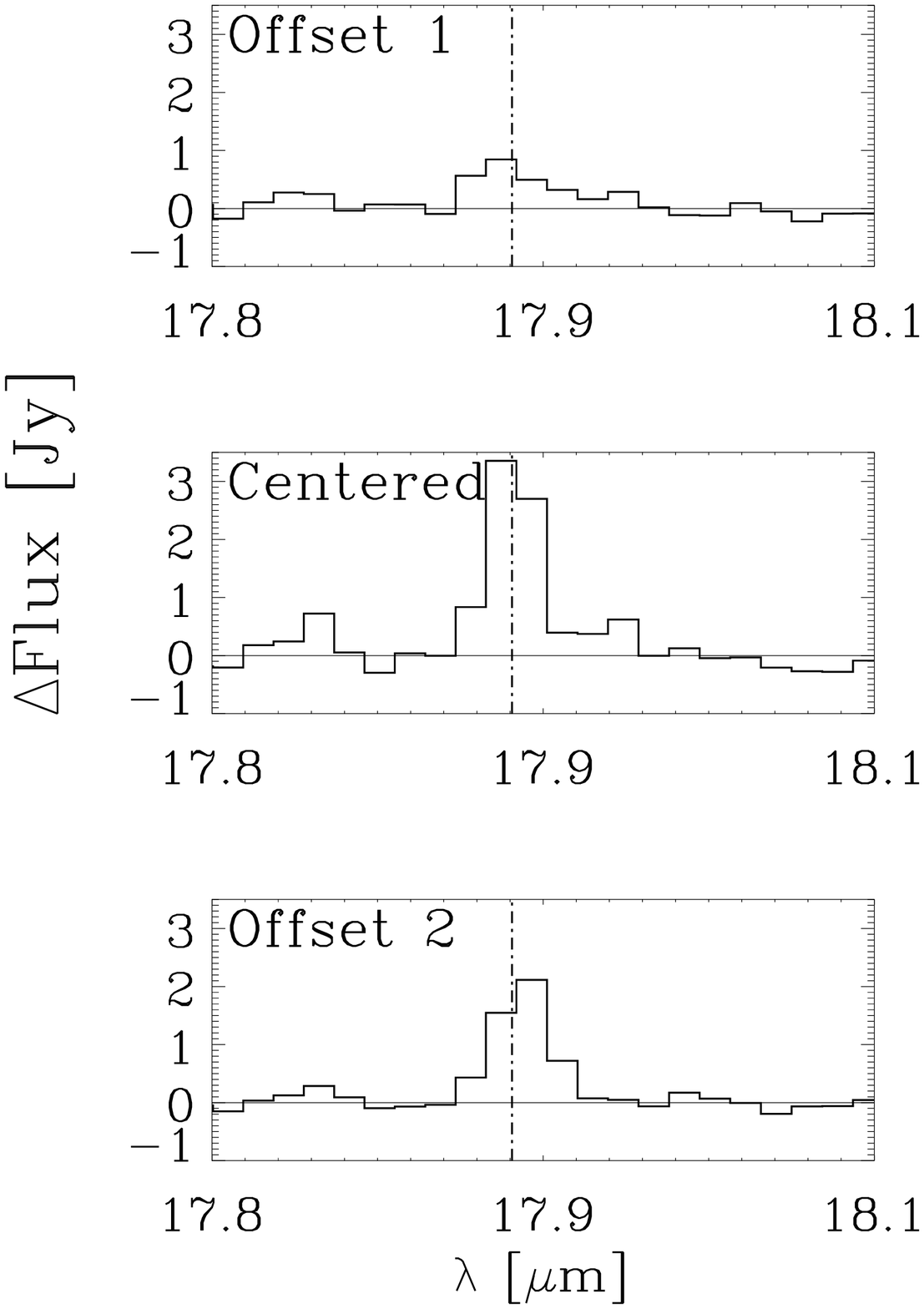} \\
	    			\includegraphics[width=4.2cm]{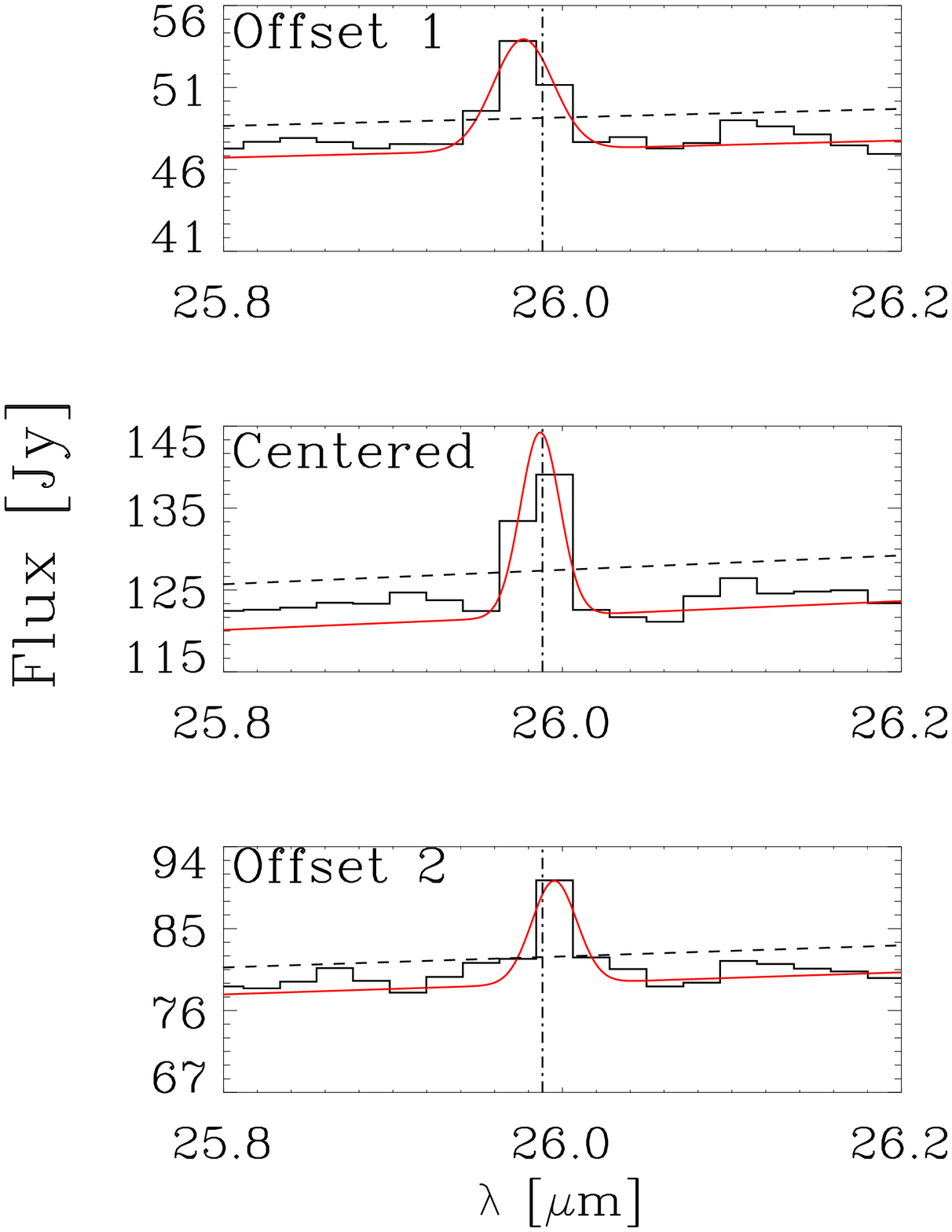} &
	    			\includegraphics[width=4.2cm]{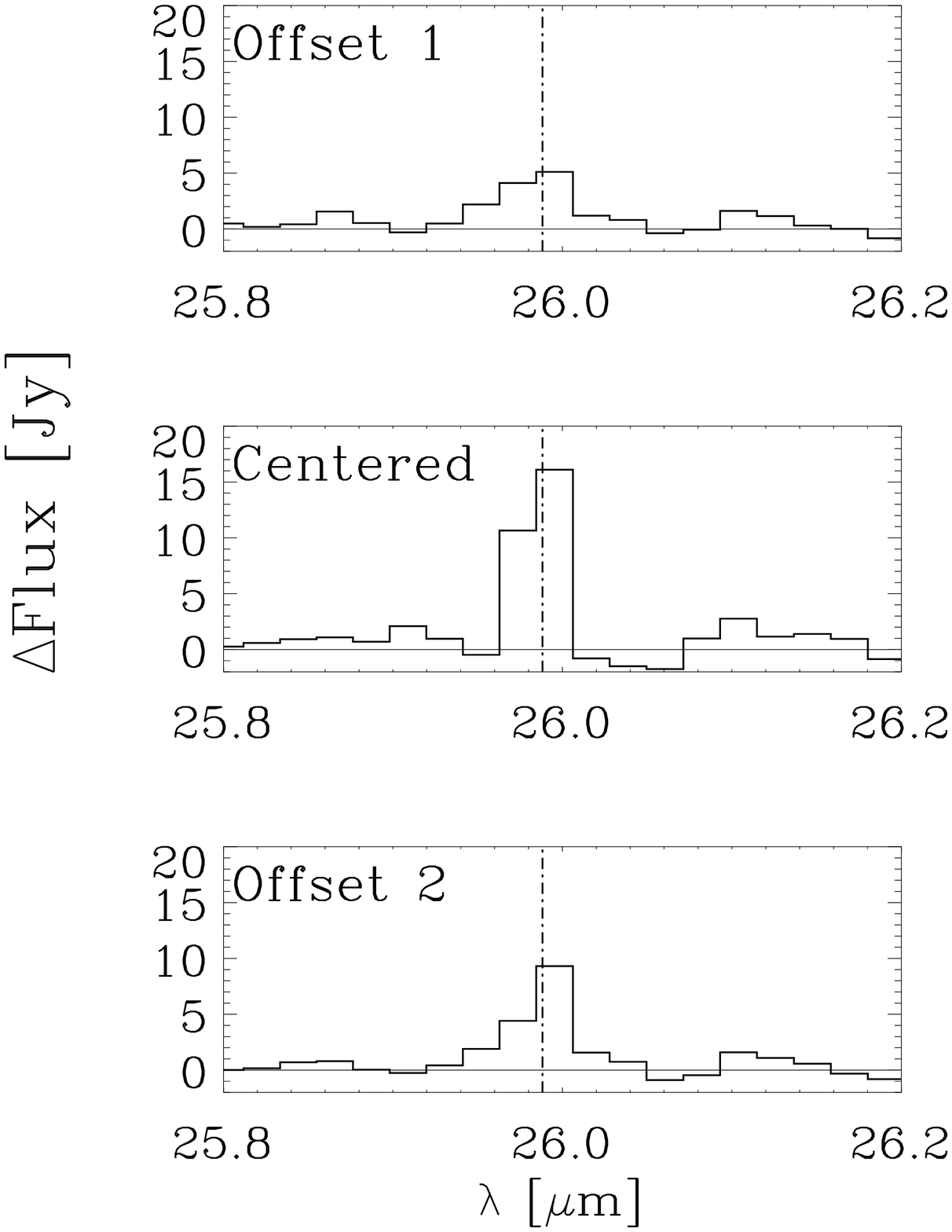}				
	  		\end{array}$  
		\end{minipage}
     \caption{Spectrum of L1551 IRS 5 surrounding the position of the [Fe~II] line at 17.9 and 25.9~$\mu$m. The right panel shows 		 the difference between the line flux and the continuum flux in order to show the real strength of the line. 
     The plots on the left panel show the average of 2 pointings at each position. 
     Labels follow the convention used for Fig. \ref{L15point}}
          \label{L15pointFe}
 \end{figure}
 
In the case of [Fe~II] at 17.9 and 25.9~$\mu$m, the lines are detected in all three pointings (Fig.~\ref{L15pointFe}).
Both lines are centered at the expected wavelength, in contrast to near-IR observations \citep{davis:2003aa} and the [Ne~II] line. 
In both cases, there is an important contribution from extended emission inferred from the high peak of the continuum-subtracted line in the position labeled Offset~2, suggesting diffuse emission as the main contributor to the line luminosity. 
However, the highest luminosity is still obtained from the pointing centered at the position of the star. 

\citet{davis:2003aa} have reported a detection of an H$_2$ emission line at 2.12~$\mu$m, but we do not detect H$_2$ in the mid-IR spectra obtained with \emph{Spitzer}.

The emission lines from L1551 IRS 5 are the brightest of the sample, with luminosities on the order of $10^{30}$~erg~s$^{-1}$.
We have obtained VISIR/VLT observations in order to study the kinematics and profile of the [Ne~II] line. 
We have not detected [Ne~II] emission at the position of the star, which might suggest a jet origin for the [Ne~II] emission, although we have detected the system in imaging mode using the [Ne~II] filter (Baldovin-Saavedra et al. 2011, in preparation).

\noindent{\bf \object{AA~Tau}} is a Class II source for which [Ne~II] is the only line we detected (of the lines we studied). 
However, \citet{carr:2008aa} have presented the spectrum of AA~Tau showing multiple emission lines from organic molecular species (HCN, C$_2$H$_2$, and CO$_2$), water vapor, OH, and [Ne~II].
Follow-up observations (at R$\sim$ 80~000, \citealt{najita:2009aa}) have confirmed that the [Ne~II] emission is centered at the stellar velocity, which means that its origin lies in the disk rather than in a jet or outflow. 
Nevertheless, due to the high inclination of the system ($\sim 75 ^{\circ}$, \citealt{bouvier:1999aa}), the absence of velocity shifts in the emission line does not rule out a jet origin for the emission.
\citet{cox:2005aa} have indeed reported a microjet detected using coronographic imaging with the Hubble Space Telescope (HST).

\noindent{\bf \object{BP Tau}} is a Classical T Tauri star (CTTS) that was observed together with background observations. 
We detect [Fe~II] at 25.99~$\mu$m, with a marginal detection at 17.93~$\mu$m. 
There is no record in the literature of previous [Fe~II] detections. 
We have plotted in Fig.~\ref{fig:BP+AA} the SH spectra of AA~Tau together with BP~Tau and DK~Tau.
Both spectra show spectral features similar to the ones observed in AA~Tau and reported by \citet{carr:2008aa}. 
The presence of molecular line emission in the spectra of young stars were recently reported in \citet{pontoppidan:2010aa}.
Although the spectra of BP~Tau and DK~Tau are quite similar to the spectrum of AA~Tau, we do not detect [Ne~II] emission from these two stars.
\citet{gudel:2010aa} detected [Ne~II] emission in BP~Tau using the same data set as this study, although the line is weak ($\sim 10^{27}$~erg~s$^{-1}$). 
We observe an excess at the position of the [Ne~II] line (see Fig.~\ref{fig:BP+AA}), but below our 3$\sigma$ detection threshold.
 A similar situation occurs for the H$_2$ line at 12.27 $\mu$m (see Fig. \ref{fig:BP+AA}). The presence of numerous faint emission lines likely makes our detection threshold slightly too high. 
 
 \begin{figure*}[h!]
   \centering
   		\includegraphics[height=14.cm,angle=90]{15622fga3.eps}
      \caption{Archival SH spectra of DK~Tau and BP~Tau showing a variety of emission lines from water and organic molecules. We present for comparison the spectrum of AA~Tau previously presented in \citet{carr:2008aa}. The empty diamonds show the position of some water lines (see \citealt{pontoppidan:2010aa}). The spectra were background subtracted and shifted to allow comparison.} 
     \label{fig:BP+AA}
 \end{figure*}

\noindent{\bf \object{DG~Tau}} is classified as both Class I and II in the literature. 
This young star drives a collimated Herbig-Haro (HH) jet observed in the near-IR.
The \emph{Spitzer} observations were made in the same way as L1551~IRS5, with 3 sets of 2 nod pointings. 
The [Ne~II] line is detected in all of them. 
The observations were made using a slit P.A. of $-57.1^\circ$ and $-141.9^\circ$ for the SH and LH modules respectively.
The jet P.~A. is $-138 ^\circ$ \citep{lavalley:1997aa}, almost coincident with the LH module P.~A.
When comparing the difference between the line and the continuum flux for each set of pointings, we note that the bulk of the emission comes from the position centered on the coordinates of the star, with the ratio being roughly 2:1 between the centered and the offset positions.
A detection of [Ne~II] has been previously reported in \citet{gudel:2010aa}, based on a different data set than this study.
We detect [Fe~II] emission at 25.99~$\mu$m; 
this line is detected in all pointings, again with a higher contribution from the position centered on the coordinates of the star.
Emission from [Fe~II] in the near-IR has been previously reported (e.g., \citealt{bacciotti:2002aa,davis:2003aa}).
In particular,~\citet{davis:2003aa} do not detect any [Fe~II] emission within 1$^{\prime\prime}$ from the source, suggesting a jet origin for the [Fe~II] emission.

\noindent{\bf \object{DG~Tau~B}} has an edge-on disk and a bipolar jet detected in the optical (\citealt{mundt:1983aa,eisloffel:1998aa}) and radio \citep{rodriguez:1995aa}. The jet is oriented perpendicular to the disk. 
The IRS spectrum has been obtained at only one position. We have detected [Ne~II] emission towards this object.

\noindent{\bf DM~Tau} is a Class~II object with [Ne~II] line emission. 
The IRS observations were done with a background observation, so we are confident that the detected line comes from the source itself.
Based on \emph{Spitzer} observations made with the low-resolution modules (SL and LL), \citet{espaillat:2007ab} have derived a luminosity of the [Ne~II] line of $1.3 \times 10^{28}$~erg~s$^{-1}$. 
In contrast, we have obtained a luminosity of $(8.0~\pm~2.0)\times~10^{27}$~erg~s$^{-1}$, which is 40\% lower than what is presented in \citet{espaillat:2007ab}.
This object is particularly interesting because it belongs to the category of transitional disks \citep{calvet:2002aa}. 
These objects show a lack of emission at wavelengths shorter than 10 $\mu$m, which is sometimes interpreted as an evidence of an inner hole and planet growth.

\noindent{\bf \object{FS~Tau A}} is a close binary with separation of 0$^{\prime\prime}$.25 \citep{krist:1998aa}. 
The system is surrounded by a filamentary nebula. 
\emph{Spitzer} observations have been done in 3 sets of 2 nod pointing; the [Ne~II] line is detected in all of them.
By analyzing the flux difference between the continuum and the line, we have found that the height of the line in the centered position is comparable to the height of the line at one of the offset positions, indicating the emission is likely extended.
In addition we have detected H$_2$ at 12.28 $\mu$m, while there is no detection of the lines at 17.03 and 28.22~$\mu$m.
At 28.22~$\mu$m, there could be a marginal detection, but only below the detection threshold of 3$\sigma$. 
In the case of the 12.28 $\mu$m line, it has been detected in 2 out of the 3 sets of pointings.
Comparing the continuum-subtracted line flux at the different positions, we determine that the line flux is higher for a position shifted with respect to the star position, indicating an important contribution from extended emission. 

\noindent{\bf \object{FX~Tau}} is a binary with 0.$^{\prime\prime}$89 separation composed of a classical and a weak-lined T Tauri star \citep{white:2001aa}. Assuming that the mid-IR emission is dominated by the CTTS, we have included this system in the Class~II objects for our analysis.
We have detected H$_2$ emission at 28~$\mu$m towards this object, and it is the only line detected. 
No previous detections have been reported.

\noindent{\bf \object{IQ~Tau}} is a CTTS showing [Ne~II] line emission.
There is no information in the literature on possible jet or outflow emission associated with this object.
Carmona et al.\ 2011 (in preparation) detected [O~I] emission towards this object employing optical spectroscopy (R$\sim$ 40 000), observing a velocity shift of the line which indicates the presence of an outflow.

\noindent{\bf \object{T~Tau}} is a triple system visible in the optical and bright in X-rays. 
T~Tau~S is an infrared companion located 0$^{\prime\prime}$.6 away \citep{dyck:1982aa} not detected in X-rays.
The southern component is a binary itself with a projected separation of $\sim 0^{\prime\prime}.13$ \citep{kohler:2008aa}.
We have detected the [Ne~II] line towards the T~Tau system, consistent with previous studies (e.g., \citealt{van-den-ancker:1999aa,gudel:2010aa}). 
Based on the analysis of the continuum-subtracted line, there is a contribution from extended emission, in agreement with previous results.
The first detections of [Ne~II], [Fe~II], and [O~I], among other gas tracers, come from \citet{van-den-ancker:1999aa}, based on Infrared Space Observatory (ISO) observations.
However, we do not detect [Fe~II] emission, which would be expected to be present in an outflow environment. 
Van Boekel et al.\ (2009)  studied this system (R$\sim$30 000) and have determined that the emission is extended and associated with an outflow from the North and South components. 
Only a small fraction of the emission is related to the X-ray bright T~Tau~N.

\noindent{\bf \object{XZ~Tau}} is a close binary system with separation of $0.^{\prime\prime}3$. 
Both stars have collimated, bipolar jets \citep{krist:2008aa}.
Due to the close separation, we cannot resolve the system with \emph{Spitzer}.
Observations have been made in 3 sets of 2 nod pointings; we detect [Ne~II] only at the pointing with the highest flux in the continuum, centered at the position of the stars. This is the first detection of [Ne~II] reported toward this source.

\noindent{\bf \object{CoKu~Tau 1}} is a Class~II object. It is a binary with separation of $0^{\prime\prime}.24$ \citep{robitaille:2007aa} in which we have detected [Ne~II] emission.
Previous observations in H$\alpha$ show evidence of an outflow \citep{eisloffel:1998aa}, but we do not detect the [Fe~II] emission usually observed in outflows.  

\noindent{\bf \object{CoKu Tau 4}} is a Class~II object known to have a large excess at $\sim 30~\mu$m and no excess at $<8~\mu$m \citep{forrest:2004aa}. This was interpreted as a clear signature of a transitional disk, but recent observations \citep{ireland:2008aa} have shown that the clearing of the disk at shorter wavelengths is actually due to the presence of a stellar-mass companion. 
The binary projected separation is calculated to be $\sim 7.8$~AU. 
We have detetected [Ne~II] emission towards this binary.

\noindent{\bf \object{GM Aur}} is a Class II source with [Ne~II] emission; this object is also a transitional disk \citep{calvet:2005aa}.
We have dedicated background observations, so we can discard the possibility of the emission coming from the background or a nearby source. 
The first detection of [Ne~II] in the \emph{Spitzer} spectrum was reported by \citet{carr:2008aa}. 
Follow-up observations of the [Ne~II] line at R$\sim$80 000 \citep{najita:2009aa} have confirmed that the emission is from a disk rather than an outflow.
\citet{espaillat:2007ab} have derived a 5$\sigma$ upper limit for the luminosity of the [Ne~II] line of $7.4 \times 10^{28}$~erg~s$^{-1}$. This result is based on observations made with \emph{Spitzer} IRS low-resolution modules (R$\sim$ 60--120).  
Our calculations based on the high resolution module give a luminosity of $(1.5~\pm~0.3)~\times 10^{28}$~erg~s$^{-1}$.

\noindent{\bf \object{RW~Aur}} is an optically bright T Tauri star; \citet{petrov:2001aa} suggested that this is a binary system. 
The secondary, RW~Aur~B, is located $1^{\prime\prime}.2$ away from the primary.   
We have detected both [Fe~II] lines (17.93 and 25.99 $\mu$m) in the spectrum of this system, which is not spatially resolved by \emph{Spitzer}.
The observations were made in 3 sets of 2 nod pointings.
The line at 17.93~$\mu$m is detected in one of the offset positions, indicating that the emission is likely to be extended. 
The [Fe~II] line at 25.99~$\mu$m is detected in all the three positions: centered on the position of the star, and two in positions displaced from the star position. 
The contribution from extended emission is important, but the line flux is higher at the position of the source.
Previous observations have shown emission of [Fe~II] in the near-IR from the jet associated with this system \citep{hartigan:2009aa}.

\noindent{\bf \object{UY~Aur~(A,~B)}} is a binary system with separation $0^{\prime\prime}.88$ \citep{hirth:1997aa}.
Observations were made in 3 sets of 3 nod pointings. 
We have detected [Ne~II] emission at the 3$\sigma$ limit. 
The line is detected only in the pointing centered in the position of the system. 
We note that in this case the line is broad, with a FWHM 1.6 times higher than the IRS instrumental FWHM.
This system is known to drive a high velocity ($\sim300$~km s$^{-1}$) [O~I] bipolar outflow \citep{hirth:1997aa}. 
We do not observe any shift of the position of the line; the broadening could indicate either a strong wind or a jet with a blue and red-shifted component. 

\noindent{\bf \object{MHO~1/2}} is a binary with a separation of 3$^{\prime\prime}$ \citep{briceno:1998aa}. 
Since the \emph{Spitzer} observations do not resolve the system, we cannot know in which star the emission originates.
The system has been observed in 3 sets of 2 nod pointings.  
The [Ne~II] line is detected only at one of the three positions.

\noindent{\bf \object{V773 Tau}} is a multiple system. V773~Tau~AB is a WTTS spectroscopic binary with a separation of 0$^{\prime\prime}$.003, i.e.,$\sim0.37$~AU at the distance of Taurus \citep{welty:1995aa}. V773~Tau~C is a visual companion at 0$^{\prime\prime}$.17 from V773~Tau~AB, classified as a CTTS. A fourth member of this system, V773~Tau~D, has an angular separation of $\sim 0^{\prime\prime}.22$ with respect to V773~Tau~A \citep{woitas:2003aa} and it has been classified as an infrared companion due to the shape of its SED \citep{duchene:2003aa}.
These objects are believed to be T Tauri stars embedded in dusty envelopes and undergoing episodes of intense accretion. 
But their true nature is not yet understood.
The \emph{Spitzer} observations do not resolve the components of the system, but according to the SED presented in \citet{duchene:2003aa} we assume that V773~Tau~D is responsible for the bulk of the mid-IR emission and therefore for the level of continuum observed.
However, we cannot be sure that the [Ne~II] emission comes from it. 
In fact, it could also be produced in shocks from one of the components or in the interstellar environment in a similar way as the case of T~Tau.  
The stellar properties for the system are compiled in Table~\ref{prop}.
The H$\alpha$ emission and X-ray luminosity probably come from V773~Tau~AB, the WTTS binary, while the mass accretion rate value adopted is the one derived for V773~Tau~C, which is likely similar to V773~Tau~D.
We assume that in the infrared regime, the emission is dominated by the CTTS, so we have included the system in the group of Class~II sources.
In the correlation tests presented in Section~\ref{discussion}, we have not used the H$\alpha$ equivalent width, which is derived for the WTTS binary.

\noindent{\bf \object{HBC~366}} is a WTTS. We have detected one of the H$_2$ lines (17.03~$\mu$m) in this spectrum. 
No other gas lines have been detected. 

\noindent{\bf \object{IRAS~04239+2436}} is a binary of 0$^{\prime\prime}$.3 separation, driving a highly collimated near-IR [Fe~II] jet \citep{reipurth:2000aa}. 
The observations were done in three sets of two nod pointings.
We detect both iron lines (17.93 and 25.99~$\mu$m) in the IRS spectrum.
The two lines are detected in all three sets of pointings, but when analyzing the continuum-subtracted line, the highest contribution comes from the position centered on the star. 
The luminosity of the line at 25.99~$\mu$m is more than a factor of 3 higher than the luminosity at 17.93~$\mu$m.
Previous observations (\citealt{davis:2001aa,davis:2003aa}) have detected emission of H$_2$ (2.122 $\mu$m) and [Fe~II] (1.644 $\mu$m). 
The \emph{Spitzer} spectrum of this source does not show any evidence of H$_2$ emission. 

\noindent{\bf \object{IRAS~04303+2240}} is a CTTS. 
We have detected both [Fe~II] lines (at 17.93 and 25.99~$\mu$m), with the line at 17.93~$\mu$m detected at the 3$\sigma$ level. 
Although there is an excess at the wavelength of the [Ne~II] line, it is formally not detected (at the 3$\sigma$ detection limit).
The ratio between the luminosity of both [Fe~II] lines is 1:2.
Observations were done in 3 sets of 2 nod pointings; the line at 17.93~$\mu$m is detected only in the position centered at the star.
The line at 25.99~$\mu$m is detected in two out of three positions; the contributions of the centered position and the offset position are similar.  
Both lines are broadened by a factor of two compared to the intrinsic line width of the instrument. 
There is no information in the literature about previous detections of [Fe~II] towards this source or about jets or outflows that could explain the [Fe~II] emission, or winds explaining the broadening of the lines. 

\noindent{\bf \object{V710~Tau}} is a binary system formed by a CTTS and a WTTS. 
Observations in the X-rays made with \emph{Chandra} \citep{shukla:2008aa} and in the near-IR \citep{white:2001aa} have shown that the separation of the binary is 3$^{\prime\prime}$.2. 
We detect H$_2$ at 17.03 and 28.22~$\mu$m.
Considering that the star responsible for the continuum at mid-IR wavelengths is the CTTS, V710~Tau~A,
we have used the H$\alpha$ equivalent width and X-ray luminosity of that star in the present study, and included the system in the Class~II group, but the lines could also be produced by the WTTS.

\noindent{\bf \object{SST~041412.2+280837}} was included in a sample of the low-mass star population of Taurus for which we have obtained  \emph{Spitzer}-IRS spectroscopy. It was previously known as IRAS 04111+2800G and is classified as a Class~I object \citep{rebull:2010aa}.			 
This is the only object in our sample for which we have used low-resolution spectra (SL and LL) since we do not have the SH and LH spectra.  
We have detected [Ne~II] and [Fe~II] at 17.9 and 25.9~$\mu$m towards this object.

\noindent{\bf\object{SST~042936.0+243555}} is a newly discovered member of Taurus \citep{rebull:2010aa}. It belongs to the sample of low luminosity sources for which we have obtained IRS spectra. \citet{rebull:2010aa} and \citet{luhman:2010aa} both classify this object as a Class~II. The spectrum shows [Ne~II] emission. 

\noindent{\bf \object{SST~043905.2+233745}} also belongs to the sample of newly discovered Taurus members \citep{rebull:2010aa}. It is a Class~I object as classified by \citet{luhman:2010aa}, but a Class II object according to \citet{rebull:2010aa}. We have detected emission lines from [Ne~II] and H$_2$ at 17.0~$\mu$m.\\ 

\section{[Ne II] non-detections.}
\label{neon_non_det}
\vspace{-2cm}
\begin{figure*}[h!]
	\vspace{20pt}
	\begin{center}
	\includegraphics[width=16cm]{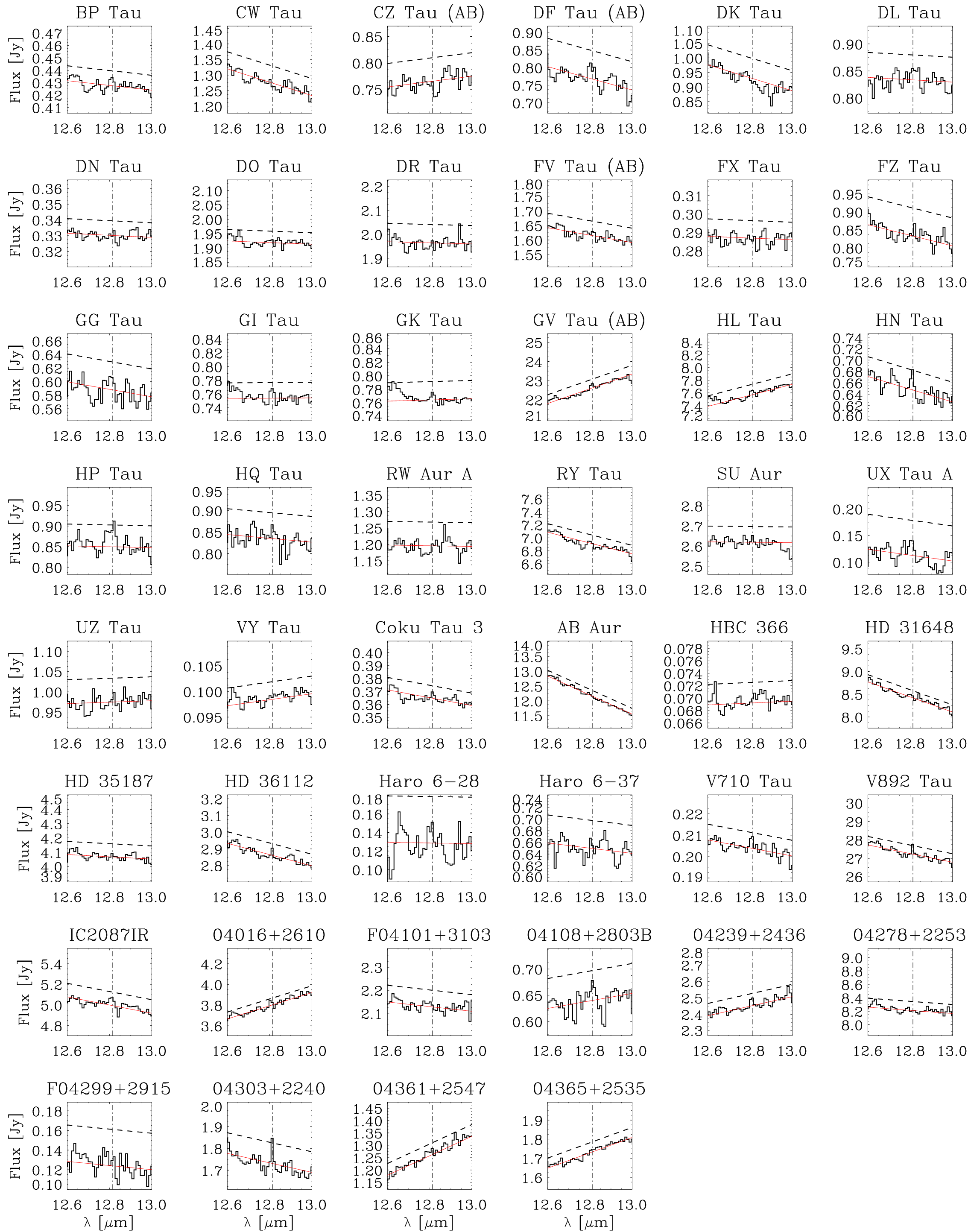}
\end{center}
	\caption{Spectra around the [Ne~II] line of the targets in which we did not detect the line. The position of the line is plotted in dashed-dotted line and the 3$\sigma$ noise level is plotted in dashed line. The name of each star is labeled in the top of each plot. }
	\label{Nenondet}
	\end{figure*}

\end{appendix}

\end{document}